\documentclass[%
 aps,
 prx,
 amsmath,amssymb,
 reprint,%
floatfix,
superscriptaddress,
longbibliography
]{revtex4-2}

\usepackage{silence}  %
\usepackage{graphicx}%
\usepackage{lmodern}
\usepackage{dcolumn}%
\usepackage{bm}%
\usepackage[utf8]{inputenc}
\usepackage[T1]{fontenc}
\usepackage{mathptmx}
\usepackage{etoolbox}
\usepackage{xcolor} %
\usepackage{xparse} %
\usepackage{braket} %
\usepackage{amsmath}
\usepackage{siunitx}
\usepackage{tabularx}
\usepackage{hyperref}
\usepackage[at]{easylist}
\usepackage{mleftright}  %
\mleftright  %
\usepackage{orcidlink}
\usepackage{float}
\usepackage{enumitem}
\usepackage{placeins}  %
\usepackage{booktabs}
\usepackage{layouts}

\usepackage{algorithm}
\usepackage{algpseudocode}

\definecolor{table_qnn_color}{HTML}{ff9202}   %
\colorlet{table_qnn_color_bg}{table_qnn_color!25}          %
\definecolor{table_svm_color}{HTML}{039cff}   %
\colorlet{table_svm_color_bg}{table_svm_color!25}          %
\colorlet{table_special_case_color}{black!60}

\usepackage{rotating}
\usepackage{multirow}
\usepackage{adjustbox}
\usepackage{nicematrix}
\usepackage{etoolbox}

\usepackage{soul,xcolor}
\setstcolor{red} %

\makeatletter
\DeclareDocumentCommand\todo{g}{%
\def\@message{\IfNoValueTF{#1}{TODO}{TODO: #1}}
\textbf{\textcolor[HTML]{FF8811}{\@message}}
\@latex@warning{\@message}{}{}}
\makeatother

\NewDocumentCommand{\expval}{m g}{%
  \IfNoValueTF{#2}
    {\Braket{#1}}             %
    {\Braket{#2 | #1 | #2}}   %
}

\DeclareMathOperator*{\argmin}{arg\,min}

\DeclareMathOperator*{\minimize}{minimize}

\definecolor{tab_1_blue}{HTML}{1f77b4}
\definecolor{tab_2_orange}{HTML}{ff7f0e}
\definecolor{tab_3_red}{HTML}{d62728}
\definecolor{tab_4_purple}{HTML}{9467bd}
\definecolor{plt_lightgreen}{HTML}{90ee90}
\definecolor{plt_tab_green}{HTML}{2ca02c}
\definecolor{plt_darkgreen}{HTML}{006400}

\definecolor{color_quantum}{RGB}{145,0,159}
\definecolor{color_classical}{RGB}{0,109,61}

\makeatletter
\def\@email#1#2{%
 \endgroup
 \patchcmd{\titleblock@produce}
  {\frontmatter@RRAPformat}
  {\frontmatter@RRAPformat{\produce@RRAP{*#1\href{mailto:#2}{#2}}}\frontmatter@RRAPformat}
  {}{}
}%
\makeatother

\renewcommand{\selectlanguage}[1]{}  %

\newcommand{\Qbadge}[1][Q]{{%
    \setlength{\fboxsep}{0pt}%
    \smash{\colorbox{color_quantum}{%
        \makebox[1.2em][c]{%
            \ifstrempty{#1}%
                {\rule[-0.2\baselineskip]{0pt}{1.2\baselineskip}}%
                {\rule[-0.2\baselineskip]{0pt}{0.95\baselineskip}\textcolor{white}{\textsf{#1}}}%
        }%
    }}%
}}

\newcommand{\Cbadge}[1][C]{{%
    \setlength{\fboxsep}{0pt}%
    \smash{\colorbox{color_classical}{%
        \makebox[1.2em][c]{%
            \ifstrempty{#1}%
                {\rule[-0.2\baselineskip]{0pt}{1.2\baselineskip}}%
                {\rule[-0.2\baselineskip]{0pt}{0.95\baselineskip}\textcolor{white}{\textsf{#1}}}%
        }%
    }}%
}}

\newcommand{\QState}[1][Q]{\State \Qbadge[#1]\hspace{2pt}\ignorespaces}
\newcommand{\CState}[1][C]{\State \Cbadge[#1]\hspace{2pt}\ignorespaces}

\begin{document}

\preprint{AIP/123-QED}

\title{Quantum feature-map learning with reduced resource overhead}

\author{Jonas Jäger\,\orcidlink{0000-0001-7631-8689}}
 \thanks{Both authors contributed equally to this work.\scriptsize\\\url{jojaeger@cs.ubc.ca}, \url{philipp.elsaesser@physik.uni-freiburg.de}}
 \affiliation{Department of Computer Science and Institute of Applied Mathematics, University of British Columbia (UBC), Vancouver, B.C. V6T 1Z4, Canada}
 \affiliation{Stewart Blusson Quantum Matter Institute (QMI), Vancouver, B.C. V6T 1Z4, Canada}

\author{Philipp Elsässer\,\orcidlink{0000-0002-5692-0709}}%
 \thanks{Both authors contributed equally to this work.\scriptsize\\\url{jojaeger@cs.ubc.ca}, \url{philipp.elsaesser@physik.uni-freiburg.de}}
 \affiliation{Institute of Physics, University of Freiburg, Freiburg (Breisgau), 79104, Germany}
 \affiliation{Department of Chemistry, University of British Columbia (UBC), Vancouver, B.C. V6T 1Z1, Canada}

\author{Elham Torabian\,\orcidlink{0000-0002-3097-2519}}
 \affiliation{Department of Chemistry, University of British Columbia (UBC), Vancouver, B.C. V6T 1Z1, Canada}
 \affiliation{Stewart Blusson Quantum Matter Institute (QMI), Vancouver, B.C. V6T 1Z4, Canada}

\date{\today}%

\begin{abstract}
Current quantum computers require algorithms that use limited resources economically. 
In quantum machine learning, success hinges on quantum feature-maps, which embed classical data into the state space of qubits. 
We introduce Quantum Feature-Map Learning via Analytic Iterative Reconstructions (Q-FLAIR), an algorithm that reduces quantum resource overhead in iterative feature-map circuit construction. 
It shifts workloads to a classical computer via partial analytic reconstructions of the quantum model, using only a few evaluations.
For each probed gate addition to the ansatz, the simultaneous selection and optimization of the data feature and weight parameter is then entirely classical.
Integrated into quantum neural network and quantum kernel support vector classifiers, Q-FLAIR shows state-of-the-art benchmark performance.
Since resource overhead decouples from feature dimension, we train a quantum model on a real IBM device in only four hours, surpassing \SI{90}{\percent} accuracy on the full-resolution MNIST dataset (784 features, digits 3 vs 5).
Such results were previously unattainable, as the feature dimension prohibitively drives hardware demands for fixed and search costs for adaptive ansätze. 
Furthermore, Q-FLAIR demonstrates de-quantization robustness against direct classical modeling, satisfying a benchmark rare in the literature and a necessary condition for potential quantum advantage.
By rethinking feature-map learning beyond black-box optimization, this work takes a concrete step toward enabling quantum machine learning for real-world problems and near-term quantum computers.
\end{abstract}

\maketitle

\section{Introduction}

Quantum machine learning (QML) aims to harness the computational and representational power of quantum computing and quantum information systems to enhance data-driven tasks in machine learning \cite{biamonte_QuantumMachineLearning_2017,dunjko_MachineLearningArtificial_2018,cerezo_ChallengesOpportunitiesQuantum_2022}. 
While quantum speed-ups over classical algorithms have traditionally been a central topic in QML \cite{biamonte_QuantumMachineLearning_2017,dunjko_MachineLearningArtificial_2018}, the focus of research has recently shifted toward representational advantages of quantum models \cite{havlicek_SupervisedLearningQuantumenhanced_2019,schuld_QuantumMachineLearning_2019,cerezo_ChallengesOpportunitiesQuantum_2022}. Quantum models may provide access to richer model classes, potentially enhancing generalization and capturing complex data patterns and structure. Beyond practical motivations \cite{huang2021quantum, sakka2025automating, riste2017demonstration, cho2024machine}, this research draws on the theoretical premise that relevant datasets may stem from classically hard problems \cite{liu_RigorousRobustQuantum_2021,jager_UniversalExpressivenessVariational_2023}. For recent comprehensive primers on QML, see  Refs.~\cite{chang_PrimerQuantumMachine_2025, pira2026fundamentalsquantummachinelearning}.

However, the realization of such QML algorithms necessarily requires a mapping from the data (feature) space $\mathcal{X} = \mathbb{R}^d$ into the exponentially large Hilbert space $\mathcal{H} = \mathbb{C}^{2^n}$ of $n$ qubits. Such maps are known as \emph{quantum feature-maps} \cite{havlicek_SupervisedLearningQuantumenhanced_2019,schuld_QuantumMachineLearning_2019}. 
Concretely, a unitary transformation $U(\bm{x}, \bm{\theta})$ embeds classical data points $\bm{x} \in \mathcal{X}$ into $\mathcal{H}$ as quantum states of the form
\begin{equation}\label{eq:quantum_feature_map}
    \ket{\psi(\bm{x}, \bm{\theta})} = U(\bm{x}, \bm{\theta}) \ket{\psi_0}
    .
\end{equation}
They are prepared based on common parameters $\bm{\theta}$ and initial state $\ket{\psi_0}$.
This work adopts the prominent gate-based quantum computing paradigm.
Thus, the transformation $U(\bm{x}, \bm{\theta})$ is realized by a (variational or parameterized) quantum circuit, formed of multiple individual quantum gates 
($U_1, U_2, \ldots$) as
\begin{equation}\label{eq:gate_decomp}
    U(\bm{x}, \bm{\theta}) = \ldots U_2(\theta_2, x_{k_2})U_1(\theta_1, x_{k_1})
    .
\end{equation}
The exact arrangement of gates is referred to as the \emph{ansatz}.

Despite their central role in QML, there is, in general, no principled approach for designing suitable quantum feature-maps, making this one of the most significant open challenges in the field \cite{cerezo_ChallengesOpportunitiesQuantum_2022}.
The common practice of using fixed, often generic and problem-agnostic, circuit ansätze and only aligning the free (continuous) parameters $\bm{\theta}$ to the training data \cite{schuld_EffectDataEncoding_2021,mitarai_QuantumCircuitLearning_2018,watabe2019quantum,qi2023qtn,terashi2021event,chen2020hybrid,blance2021quantum,kwak2021introduction,sierra2020dementia,chen2020variational,lloyd2020quantum,hubregtsen_TrainingQuantumEmbedding_2022,henry2021quantum} often results in trainability issues and overly cumbersome models, which are inadequate for current hardware. The trainability issues arise particularly from the absence of inductive biases \cite{larocca_BarrenPlateausVariational_2025}.
It is the lack of concrete prior knowledge (inductive bias) for the ansatz construction and related shortcomings of fixed ansatz approaches that motivate a fully data-driven quantum feature-map design, including the ansatz.

\begin{figure*}[tbp]
	\includegraphics[width=\linewidth]{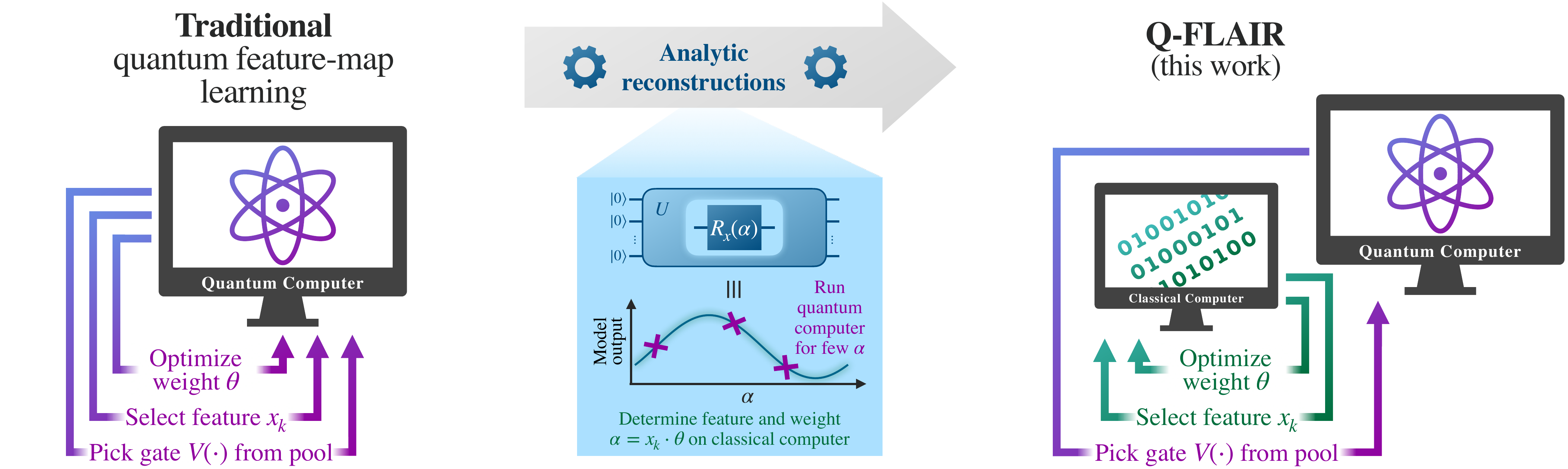}
\caption{Schematic comparison of traditional quantum feature-map learning and Q-FLAIR. 
The figure illustrates the distribution of computational workloads between quantum (violet) and classical (green) computers. Multiple subroutines that are repeatedly executed on a quantum computer in traditional approaches are shifted to classical computers in Q-FLAIR.
By retrieving partial analytic reconstructions of the quantum model from only a few quantum evaluations (blue box), Q-FLAIR reduces the resource overhead of quantum feature-map learning.
}
\label{fig:graphical_abstract}
\end{figure*}

Recent efforts have therefore shifted toward automated and adaptive ansatz design, which iteratively modifies the ansatz to maximize relevant metrics.
This approach achieves task-specific ansätze tailored for the dataset at hand, rather than optimizing the parameters alone, to balance expressivity, trainability, and hardware compatibility.
Successful realizations of this \emph{quantum architecture search} range in complexity from simple searches \cite{du2022quantum,torabian_CompositionalOptimizationQuantum_2023,zahid2024unlocking} over optimization guided by reinforcement learning \cite{fosel2021quantum}, graph theory \cite{chakraborty2020hybrid} or evolutionary and genetic algorithms \cite{lu2021markovian,wang2022quantumnas,altares2021automatic,pellow2024hybrid}, to graph neural network surrogate modeling \cite{liu2025hardware}.
However, the inherent compositional search over the (discrete) space of possible ansätze might quickly entail a combinatorial \emph{blow-up}, leading to detrimental runtimes. Especially considering the resource overhead scaling with the feature dimension $d$, their applicability to relevant high-dimensional datasets becomes impractical.
Hence, learning quantum feature-maps from scratch must be rethought from a resource perspective to make QML applicable in real-world scenarios in the near-term NISQ or early fault-tolerant era.

We devise an algorithm, Q-FLAIR, to learn quantum feature-maps with the clear design goal of reducing quantum resource overhead. We view this reduction as two-fold: 
First, Q-FLAIR conserves hardware resources, such as the number of gates and qubits required to implement the feature-map. This controlled usage is achieved by a bottom-up design:
Starting from an empty ansatz, Q-FLAIR incrementally increases complexity by allocating quantum resources only when required to improve model performance. Hence, feature-map circuits remain as shallow as possible, and qubits are entangled selectively. The latter allows to respect restrictive qubit topologies of current devices.
A key characteristic of Q-FLAIR is that each growth step (iteration) involves the \emph{simultaneous} (and greedy) optimization over three choices:
\begin{itemize}
    \item[\textit{(i)}] quantum gate choice (from a pool of $M$ possibilities),
    \item[\textit{(ii)}] feature selection (of $d$ features), 
    \item[\textit{(iii)}] weight optimization (with $T_{\rm max}$ loss evaluation bound). 
\end{itemize}
Second, Q-FLAIR reduces the number of circuit evaluations required for the actual feature-map learning process. 
Instead of naively querying the quantum model to determine the feature \textit{(ii)} and weight \textit{(iii)} for every gate candidate, we employ \emph{analytic reconstructions} of the model output as a function of the gate rotation angle. 
While analytic reconstructions have been utilized in fixed-ansatz gradient-free optimization \cite{ostaszewski2021structure,nakanishi2020sequential,parrish_JacobiDiagonalizationAnderson_2019,vidal_CalculusParameterizedQuantum_2018}, and related ideas for adaptive ansätze in quantum simulation, e.g., in selection heuristics \cite{feniou2025greedy,jager_FastGradientfreeOptimization_2025} for ADAPT-VQE \cite{grimsley_AdaptiveVariationalAlgorithm_2019}, Q-FLAIR introduces this paradigm to QML to efficiently address its unique challenge of encoding classical input data. 
Specifically, this additional data dependence must be introduced carefully: A naive adaptation would compromise scalability with a quantum resource overhead in the data dimension $d$, preventing high-dimensional applications.
Crucially, the fact that Q-FLAIR employs classical reconstructions of local gate effects does not render the resulting quantum circuit efficiently simulable. Necessarily, the use of universal gate sets generally rules out efficient exact classical simulation \cite{nielsen_QuantumComputationQuantum_2010}.

The analytic reconstructions are processed on a classical computer, efficiently offloading choices \textit{(ii)} and \textit{(iii)}.
Choice \textit{(i)} requires a constant overhead per gate candidate only to obtain a full set of reconstructions for a fixed-size batch of training data:
one quantum circuit evaluation more than querying the candidate model output alone.
This constant overhead stems from the functional class of sine curves induced by the gates studied here, although Q-FLAIR is not limited to such gates in principle.
As a result, the scaling of circuit evaluations on the quantum computer reduces from $\mathcal{O}(M d T_{\rm max})$ to $\mathcal{O}(M)$ per iteration, shifting the $\mathcal{O}(d T_{\rm max})$ dependence entirely to classical processing.
Figure~\ref{fig:graphical_abstract} illustrates this shift: Whereas traditional architecture search repeatedly queries the quantum computer for all three choices, Q-FLAIR sweeps and reconstructs the gate pool once per iteration, thereby avoiding the aforementioned combinatorial blow-up.

Q-FLAIR decouples quantum resource overhead from the feature dimension $d$ through analytic reconstructions, a crucial advantage that avoids the detrimental twofold scaling overhead of prior quantum feature-map approaches on high-dimensional datasets. In fixed-ansatz feature-maps, either the number of gates or qubits typically grows at least linearly with $d$ \cite{schuld_EffectDataEncoding_2021, mitarai_QuantumCircuitLearning_2018, watabe2019quantum}, while adaptive ansatz approaches require circuit evaluation counts increasing with $d$ \cite{lu2021markovian, du2022quantum,torabian_CompositionalOptimizationQuantum_2023} (cf. traditional learning in Fig.~\ref{fig:graphical_abstract}).
\emph{Feature selection} plays an implicit yet central role in Q-FLAIR. 
Feature selection techniques in classical machine learning can be applied to QML, as in prior works \cite{grossi2022mixed,albino2023evolutionary,wang2023quantum,mucke2023feature,zahid2024unlocking}. 
Some methods act as pre-processing, however, ignoring feedback from the downstream model (mutual information \cite{battiti1994using} or variance \cite{fida2021variance} filtering) or the supervision signal (dimensionality reduction such as principal component analysis or autoencoders).
Other methods include the model in the loop (recursive feature elimination \cite{guyon2002gene} or LASSO regularization \cite{tibshirani1996regression}), reintroducing feature-dimension-dependent quantum resource scaling.
In contrast, Q-FLAIR performs \emph{quantum-aware} feature selection, meaning that it explicitly evaluates the precise (analytic) impact of candidate features on the quantum model output during training.

We utilize Q-FLAIR to learn quantum feature-maps for the prominent quantum neural network (QNN) and quantum support vector machine (QSVM) classifiers (defined in Sec.~\ref{sec:background} and connected to Q-FLAIR in Sec.~\ref{sec:method}). In Sec.~\ref{sec:experiments}, we evaluate Q-FLAIR using established benchmark datasets \cite{bowles_BetterClassicalSubtle_2024}, both in numerical simulations and on real noisy intermediate-scale quantum (NISQ) hardware. Hereby, we demonstrate its practical applicability on the full-resolution MNIST dataset, achieving over \SI{90}{\percent} accuracy in distinguishing handwritten digits 3 and 5 on NISQ devices for the first time. We also supplement experiments for extended analyses of Q-FLAIR. 
Beyond verifying practical hardware scalability, we conclude with empirical evidence for potential quantum advantage. Using an established benchmark \cite{schreiber_ClassicalSurrogatesQuantum_2023}, we rule out that, for most datasets, Q-FLAIR models could be readily de-quantized. This algorithmic advantage contrasts both the data-restricted regime of the to us only known prior positive benchmark outcome \cite{stein_BenchmarkingQuantumSurrogate_2024} and the readily de-quantized fixed-ansatz models benchmarked originally \cite{schreiber_ClassicalSurrogatesQuantum_2023}.
The broader relevance of Q-FLAIR and the results are discussed in Sec.~\ref {sec:discussion}.

\section{Background}\label{sec:background}

    To learn a (binary) classification model from data, a training dataset $\mathcal{D} = \left\lbrace \bm{x}_i, y_i \right\rbrace_{i=1}^N$ is provided, consisting of $N$ pairs of input data points $\bm{x}_i$ of dimension $d$ and true classification labels $y_i \in \lbrace -1, +1\rbrace$. The data dimension $d$ is also referred to as the number of input features.
    The two QML methods employed in this work, which integrate quantum feature-maps to construct classification models, are briefly presented here.

    \subsection{Quantum neural networks (QNNs)}

        Quantum neural networks (QNNs) \cite{farhi_ClassificationQuantumNeural_2018,schuld_CircuitcentricQuantumClassifiers_2020} are a widely adopted analogue of classical (deep) neural networks, where trainable quantum gates replace artificial neurons. To clarify the various definitions of QNNs, variational quantum classifier \cite{havlicek_SupervisedLearningQuantumenhanced_2019,schuld_CircuitcentricQuantumClassifiers_2020} is a more specific name for the QNN paradigm used in the present work, and is defined as follows.
        The QNN output is obtained from an observable $O$ as the expectation value when measured after the data point $\bm{x}$ has been embedded via the quantum feature-map,
        \begin{equation}\label{eq:qnn_output}
            m(\bm{x}; \bm{\theta}) 
            = \Braket{\psi_0 | U^\dagger(\bm{x}, \bm{\theta}) \; O \; U(\bm{x}, \bm{\theta}) | \psi_0}
            .
        \end{equation}
        The choice of the observable $O$ should be subject to $\left\lVert O\right\rVert_{\mathrm{op}} = 1$ to bound the model output to the interval $[-1, 1]$. %
        Thresholding this output, defined as
        \begin{equation}\label{eq:qnn_output_sign}
            m_{\pm}(\bm{x}; \bm{\theta}) 
            = \operatorname{sign}\left[ m(\bm{x}; \bm{\theta}) - b \right]
            ,
        \end{equation}
        provides the binary classification prediction for a data point $\bm{x}$. The threshold parameter $b \in [-1,1]$ is also learnable.

        To train the QNN, a loss function is defined to assess the fit of the model to the dataset $\mathcal{D}$.
        The standard choice in classification tasks is the \emph{negative log likelihood (NLL) loss} \footnote{This loss function draws motivation from statistics and information theory, also known as log loss, logistic loss or cross-entropy loss, and has been established as the standard choice in the classical machine learning literature -- including extensions to multi-class -- classification. \cite{bishop_PatternRecognitionMachine_2006,hastie2009elements,murphy_MachineLearningProbabilistic_2012}}, or solely \emph{log loss}, between a true label $y$ and model prediction $\hat{y}$
        \begin{equation}\label{eq:QNNlog}
            l_{\mathrm{log}}\left(y, \hat{y}\right)
            = -\log\left( 0.5 + y\hat{y} / 2 \right)
            .
        \end{equation}
        Rescaling inside the $\log$ interprets the model output $\hat{y} = m(\bm{x}; \bm{\theta})$ as the probability that the data point $\bm{x}$ is of class $\pm1$:
        \begin{equation}
            \Pr(y=\pm1|\bm{x}) = {0.5 \pm m(\bm{x}; \bm{\theta})/2}
            .
        \end{equation}
        Compared to a linear loss function \cite{farhi_ClassificationQuantumNeural_2018}, low-probability predictions for the true class $y$, leading to a misclassification, dominate the log loss non-linearly.

        For generalization to the entire data distribution, the model should ideally learn to minimize the \emph{expected} loss
        \begin{equation}
            \mathcal{L}_{\log}
            = \mathbb{E}_{\bm{x}, y} \left[ l_{\log} \left(y,  m(\bm{x}; \bm{\theta}) \right) \right] 
            .
        \end{equation}
        Given that this data distribution is presumed to be inaccessible in applications, the expected loss is estimated from finite samples (training data $\bm{x}_i, y_i \in \mathcal{D}$) and predictions $\hat{y}_i = m(\bm{x}_i; \bm{\theta})$
        \begin{equation}
            L_{\log}(\bm{y}, \bm{\hat{y}}) 
            = \frac{1}{N} \sum_{i=1}^N l_{\log} \left(y_i, \hat{y}_i \right)
            \approx \mathcal{L}_{\log}
            ,\label{eq:empirical_loss}
        \end{equation}
        which is known as the \emph{empirical} loss.

    \subsection{Quantum kernel support vector machines (QSVMs)}\label{sec:background:qsvm}

        Quantum kernel support vector machines (QSVMs) or, more specifically, quantum kernel support vector \emph{classifiers}, quantum-enhance the classical support vector machine (SVM) algorithm \cite{havlicek_SupervisedLearningQuantumenhanced_2019,schuld_QuantumMachineLearning_2019,hubregtsen_TrainingQuantumEmbedding_2022}. This is achieved by providing access to a quantum kernel function, which is thus evaluated on a quantum computer and constructed from quantum feature-maps as
        \begin{align}
            \kappa(\bm{x},\bm{x}';\bm{\theta})
            &= \left\lvert \Braket{\psi\left( \bm{x}', \bm{\theta} \right) | \psi\left( \bm{x}, \bm{\theta} \right)} \right\rvert^2 \\
            &=
            \left\lvert \Braket{\psi_0 | U^\dagger\left( \bm{x}', \bm{\theta} \right) U\left( \bm{x}, \bm{\theta} \right)  | \psi_0} \right\rvert^2 \label{eq:kernel}
            .
        \end{align}
        Note that other quantum kernel formulations exist besides this quantum \emph{fidelity} kernel definition \cite{huang_PowerDataQuantum_2021}. 
        This QSVM paradigm leverages the kernel trick concept in SVMs to perform a linear classification in a potentially high-dimensional feature space, such as the quantum state space in the case of quantum kernels. The `trick' is that the feature space does not have to be accessed explicitly, but only kernel values $\kappa(\bm{x},\bm{x}';\bm{\theta})$ of pairs of data points $\bm{x},\bm{x}'$ have to be known for both learning and inference. Learning such a linear SVM classification is efficient via quadratic optimization with desirable generalization guarantees, such as the maximum margin property \cite{boser_TrainingAlgorithmOptimal_1992}. At the same time, it is capable of modelling complex decision boundaries in the original data space.
        The QSVM model provides the binary classification prediction for a data point $\bm{x}$ as
        \begin{equation}\label{eq:SVMclassification}
            m_{\pm} = \operatorname{sign}\left[ \sum_{i=1}^N a_i y_i \kappa\left(\bm{x}_i, \bm{x}; \bm{\theta} \right)  + b\right]
            .
        \end{equation}
        This decision is based on the \emph{support vectors} among the training data points $\bm{x}_i$ corresponding to a non-zero parameter $a_i$. QSVM training yields these parameters $\bm{a}$ (and $b$), which is classical given the fixed kernel matrix $K = [\kappa(\bm{x}_i,\bm{x}_j,\bm{\theta})]_{i,j=1}^N$, with details omitted here \footnote{Note that given the evaluated quantum kernel values, commonly aggregated in the kernel matrix (also referred to as Gram matrix), the training of the QSVM does not differ from that of classical SVMs \cite{bishop_PatternRecognitionMachine_2006,hastie2009elements,murphy_MachineLearningProbabilistic_2012}.}. 
        Importantly, the parameters $\bm{\theta}$ of the quantum kernel/feature-map are unaffected.

        For the QSVM to be successful, it is crucial to choose a suitable quantum feature-map $U(\bm{x}, \bm{\theta})$ to construct the quantum kernel $\kappa(\bm{x}_i,\bm{x}_j,\bm{\theta})$. However, this is usually unknown a priori. Therefore, kernel target alignment (TA) formulates an (empirical) loss \cite{cristianini2001kernel} that can be leveraged to learn the quantum kernel parameters $\bm{\theta}$ and/or construction $U$ from data:
        \begin{equation}\label{eq:L_TA}
            L_{\rm TA}(\bm{y}, K)=-\frac{\bm{y}^\top K \bm{y}}{N\lVert K\rVert_{F}}
        \end{equation}
        This formulation is motivated by the fact that non-trivial quantum feature-maps should achieve a low fidelity for $\ket{\psi(\bm{x}_i,\bm{\theta})}$ and $\ket{\psi(\bm{x}_j,\bm{\theta})}$ when the data points $\bm{x}_i, \bm{x}_j$ belong to different classes, reflected in a low kernel value. Conversely, data points $\bm{x}_i, \bm{x}_j$ belonging to the same class should yield a higher kernel value by bringing their quantum feature states closer together. 
        The TA problem formulation can also be viewed as a separate classification problem in its own right. Concretely, with the input data consisting of data point tuples $(\bm{x}_i, \bm{x}_j)$ along with $\pm1$ class labels $y_i  y_j$.
        Ideally, the quantum kernel could be aligned with a target kernel that is one for data points in the same class ($y_i y_j = 1$) and zero otherwise ($y_i y_j = -1$).

\section{Method}\label{sec:method}

    The proposed algorithm, Q-FLAIR, is based on the fundamental connection that the expectation value (of an observable) as a function of the angle in a rotation gate in the quantum circuit has a sinusoidal form \cite{ostaszewski2021structure,nakanishi2020sequential,parrish_JacobiDiagonalizationAnderson_2019,vidal_CalculusParameterizedQuantum_2018}. Formally, consider any unitary gate being generated by a Hermitian operator $A$ as
    \begin{equation}\label{eq:def_rotation_gate}
        R(\alpha) = \exp\left({-\frac{i}{2} \alpha A}\right) \quad\text{with}\quad A^2 = I
        .
    \end{equation}
    When acting with $R(\alpha)$ on some state $\ket{\phi}$, the expectation value of any observable $\hat{M}$ obeys an analytic form with respect to the (independent) rotation angle $\alpha$, reading as
    \begin{equation}
        \expval{R^\dagger(\alpha) \hat{M} R(\alpha)}{\phi} %
        = a \sin\left(\alpha - b\right) + c 
        .
        \label{eq:analytic_form_rotation_gate}
    \end{equation}
    Solely estimations of the expectation value at three different rotation angle positions $\alpha$ on the quantum computer can determine the coefficients $a,b,c\in\mathbb{R}$, and, hence, be utilized to obtain a reconstruction $f(\alpha)$. 
    Concretely, given expectation values $z_0, z_+, z_-$ for the angles $\alpha_0, \alpha_0 + \pi/2, \alpha_0 -\pi/2$, respectively, Ref.~\cite{ostaszewski2021structure} provides direct formulae for the coefficients
    \begin{align}
        a &= \tfrac{1}{2}{\sqrt{(2z_0 - z_+ - z_-)^2 + (z_+ - z_-)^2}} \label{eq:coeffs:a}\\
        b &= \alpha_0 -\mathrm{arctan2}\left( 2z_0 - z_+ - z_-, z_+ - z_- \right) \label{eq:coeffs:b}\\
        c &= \tfrac{1}{2} ({z_+ + z_-}) 
        .\label{eq:coeffs:c}
    \end{align}
    Appendix~\ref{app:reconstruction} details the reconstruction and implementation.

\begin{figure}[tb]
    \centering
    \includegraphics[width=\linewidth]{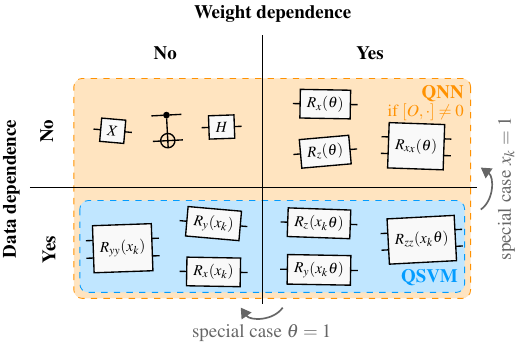}
    \caption{Types of variational dependence in gates. Gates may depend on a data feature $x_k$, a weight parameter $\theta$, both, or neither (fixed, non-variational). Colors indicate which gate types contribute in Q-FLAIR for QNN (orange) and QSVM (blue). Arrows denote reductions to special cases by setting variational parameters to one.}
    \label{tab:param_vs_data_dep}
\end{figure}

    Before introducing further details of Q-FLAIR, we define the different types of gates used. Figure~\ref{tab:param_vs_data_dep} provides a compact summary. In many variational quantum algorithms, the gates depend on a weight parameter $\theta$ or are fixed entirely. To map the data to quantum feature states in QML, however, gates may also depend on a (classical) data feature, i.e., one of the $d$ components $x_k$ of the data points $\bm{x}$. 
    Therefore, four types of gates are possible: fixed gates, purely weight-dependent gates as defined in Eq.~\eqref{eq:def_rotation_gate}, purely data-dependent gates by setting the rotation angle $\alpha$ in Eq.~\eqref{eq:def_rotation_gate} to a data feature $x_k$, and both weight-data-dependent gates. The latter type is of the form
    \begin{equation}\label{eq:def_parameter_data_gate}
        R(\theta, x_{k}) = \exp\left(-\frac{i\theta x_{k}}{2} A_j\right), \quad \text{with} \quad A^2 = I 
        .
    \end{equation}
    Due to the $2\pi$ gate periodicity, we assume features to be normalized to $[-\pi, \pi]$, where rescaling can be inferred from the training data $\mathcal{D}$ in advance. Weights are constrained to $[-1, 1]$.

    Q-FLAIR learns the quantum feature-map of a QML method with an adaptive circuit ansatz, rather than optimizing circuit parameters alone.
    Initialized with the empty ansatz $U^{(1)}(\bm{x}, \bm{\theta}^{(1)}) = I$, Q-FLAIR grows the ansatz iteratively, probing possible candidates from a pool $\mathcal{V}$ of $M$ gates. When selecting the \textit{(i)} gate that yields the largest immediate loss reduction, analytic reconstructions (Eq.~\eqref{eq:analytic_form_rotation_gate}) yield the optimal \textit{(ii)} feature and \textit{(iii)} weight at no extra quantum cost. 
    Concretely, in the $\ell$-th iteration of the algorithm, the $M$ possible candidates $V_m \in \mathcal{V}$ are appended to the current circuit $U^{(\ell-1)}(\bm{x}_i; \bm{\theta})$
    \begin{equation}\label{eq:candidate_feature_map}
        V_{m}(\alpha) U^{(\ell)}(\bm{x}_i; \bm{\theta}) \Ket{\psi_0} 
        = V_{m}(\alpha) \Ket{\psi^{(\ell)}\left(\bm{x}_i;\bm{\theta}\right)} 
        .
    \end{equation}
    To evaluate these candidate quantum feature-maps, a case distinction follows, based on the type of gate $V_m$ (see Fig.~\ref{tab:param_vs_data_dep}).
    
    If gate $V_m$ is fixed, evaluate the loss function under this $m$-th candidate feature-map, denoted by $L_m^*$, which involves quantum circuit evaluations for \emph{each} training data point $\bm{x}_i$. 
    If gate $V_m$ is instead variational in $\alpha$, reconstruct ($\equiv$) the model output under this $m$-th candidate
    for \emph{each} training data point $\bm{x}_i$
    \begin{equation} \label{eq:candidate_reconstruction}
        f_{m,i}(\alpha) \equiv \expval{V_{m}^\dagger(\alpha) \hat{M} V_{m}(\alpha)}{\psi^{(\ell)}\left(\bm{x}_i;\bm{\theta}\right)}
        .
    \end{equation}
    This reconstruction requires two additional estimates (of the expectation value) of the candidate quantum circuit under shifting $\alpha$, described in Appendix~\ref{app:reconstruction}. 
    
    The remaining evaluation of this $m$-th gate candidate relies solely on \emph{classical} processing of the reconstructions $f_{m,i}(\alpha)$, requiring no further quantum hardware queries.
    If the gate $V_m$ is both weight- and data-dependent, we substitute $\alpha = \theta x_{i,k}$ in $f_{m,i}(\alpha)$ to classically infer the model output reconstruction for each feature index $k$ as a function of $\theta$. We denote this as 
    $f_{m,i,k}(\theta) = f_{m,i}(\theta x_{i,k})$.
    Conversely, if $V_m$ is weight- but not data-dependent, we simply set $\alpha = \theta$ to obtain $f_{m,i}(\theta)$.

    Consequently, the (empirical) loss $L(\bm{y}, \bm{\hat{y}})$ can be formulated entirely in closed form.
    By relating the candidate reconstructions, which vary in $\theta$ (and $k$), to the true training labels $\bm{y}$, we construct the $m$-th candidate loss function $L_{m}$ as
    \begin{equation}
        L_{{m}}(\theta, k) = L\left(\bm{y}, \left[ f_{m,i,k}(\theta) \right]_{i=1}^N \right).
    \end{equation}
    For data-independent gates, the feature index $k$ is naturally omitted, yielding $L_{m}(\theta)$.
    The optimization problem for \textit{(ii)} feature and \textit{(iii)} weight choice is \emph{classical} as
    \begin{equation}
        \minimize_{k\in\{1,\ldots, d\}}
        \Big(
        \minimize_{\theta \in [-1,1]} \;
                L_{{m}}(\theta, k)
        \Big)
        ,
    \end{equation}
    where $\theta_m^*$ (and $k_m^*$) denote optimal solutions with value $L_m^*$. Since the inner optimization generally has no closed-form solution, it can be solved numerically or by fine sampling with modest classical effort due to its one-dimensional bounded domain. The outer optimization, relevant for data-dependent gates only, selects the feature index $k$ corresponding to the lowest inner-optimal value. Together with the fixed gate loss evaluations $L_m^*$, the optimal \textit{(i)} gate choice minimizes the loss
    \begin{equation}        
        m^* \in \argmin\left\{ 
        L_m^*
        \right\}_{m=1}^{M}
        .
    \end{equation}
    We then append the selected gate $V_{m^*}$ to the current feature-map ansatz $U^{(\ell)}$, record the feature index $k_m^*$ and extend the parameter vector $\bm{\theta}$ with $\theta_m^*$ if applicable. This process repeats, adding candidates from the pool until convergence, i.e., until the loss improvement falls below a threshold $\Delta L$.

    To apply Q-FLAIR to a QML method, it must be ensured that appending gate candidates (with a generator $A^2=I$) to the feature-map circuit obeys the form of a measurement expectation value after a state was evolved by this gate as in Eq.~\eqref{eq:analytic_form_rotation_gate}. We will provide such verification along with implementation details for the QNN and QSVM in the following two sections. Figure~\ref{tab:param_vs_data_dep} highlights differences in terms of gate types.

\subsection{Q-FLAIR for quantum neural networks}

    The immediate applicability of Q-FLAIR to the QNN method is evident by comparing the model output in Eq.~\eqref{eq:qnn_output} to the required form in Eq.~\eqref{eq:analytic_form_rotation_gate}. They coincide when appending a rotation gate $R(\alpha)$ to the end of an (arbitrary but fixed) quantum feature-map circuit $U(\bm{x}, \bm{\theta})$:
    \begin{equation}\label{eq:qnn_derivation}
        m(\bm{x}; \bm{\theta}, \alpha) 
            = \Braket{\psi(\bm{x}, \bm{\theta}) | R^\dagger(\alpha) \, O \, R(\alpha) | \psi(\bm{x}, \bm{\theta})}
        .
    \end{equation}
    with $\ket{\psi(\bm{x}, \bm{\theta})} = \ket{\phi}$ and $O=\hat{M}$ between Eqs.~\eqref{eq:qnn_output} and \eqref{eq:analytic_form_rotation_gate}.
    Any loss based on the QNN output $m(\bm{x}; \bm{\theta})$, such as the log loss, can then be assessed classically via analytic reconstructions of the form in Eq.~\eqref{eq:analytic_form_rotation_gate} for the QNN output.

    Since Q-FLAIR adds gates exclusively to the end of the quantum feature-map circuit, any gate $V_m$ in the pool $\mathcal{V}$ must not commute with the observable $O$, $[V_m, O] \neq 0$. Otherwise, such gates cannot impact the model output. Therefore, it is essential to carefully balance both the pool design and observable choice. 
    The gate pool used in our experiments comprises
    \begin{equation}
        \begin{aligned}
        \mathcal{V} = \lbrace
            &R_{x}(\theta), R_{y}(\theta),
            R_{x}(\theta, x_k), R_{y}(\theta, x_k),\\
            &R_{xx}(\theta), R_{yy}(\theta),
            R_{xx}(\theta, x_k), R_{yy}(\theta, x_k), H
        \rbrace
        .
        \end{aligned}
        \label{eq:gateSetQNN}
    \end{equation}
The two-qubit gates in this pool always act on two neighboring qubits. While all gates exhibit sine curve reconstructions, the gate set universality \cite{nielsen_QuantumComputationQuantum_2010} is crucial to preclude efficient exact classical simulation of the full learned quantum feature-maps in general.
    The observable chosen is the all-zero-state projector $O = (\ket{0}\!\bra{0})^{\otimes n}$, which realizes the probability of observing all $n$ qubits in zero upon a computational basis measurement.

\subsection{Q-FLAIR for quantum kernel support vector machines}

    To confirm the applicability of Q-FLAIR to the QSVM method, we verify that the quantum kernel function as defined in Eq.~\eqref{eq:kernel} obeys the required form as specified in Eq.~\eqref{eq:analytic_form_rotation_gate}. Compared to the QNN model output, we always encode pairs of data points $\bm{x}_i, \bm{x}_{i'}$ in the quantum feature-map $U(\bm{x}, \bm{\theta})$. Therefore, we append rotation gates (with identical generator), but with rotation angles that may differ, i.e., $R(\alpha_i)$ and $R(\alpha_{i'})$, to encode $x_i$ and $x_{i'}$ into their respective quantum feature states. The required form of Eq.~\eqref{eq:analytic_form_rotation_gate} is then recovered as follows:
    \begin{widetext}
    \begin{align}
        \kappa(\bm{x}_i,\bm{x}_{i'};\bm{\theta}, \alpha_i, \alpha_{i'})
        &=
        \left\lvert \Braket{\psi\left(\bm{x}_{i'};\bm{\theta}\right) | 
        R^\dagger\left(\alpha_{i'}\right) R\left(\alpha_i\right) 
        | \psi\left(\bm{x}_i;\bm{\theta}\right)} \right\rvert^2 \label{eq:qsvm_derivation_add_rotation}  \\
        &= \left\lvert \Braket{\psi\left(\bm{x}_{i'};\bm{\theta}\right) | 
        R\left(\alpha_i - \alpha_{i'}\right)
        | \psi\left(\bm{x}_i;\bm{\theta}\right)} \right\rvert^2  \label{eq:qsvm_derivation_combine_rotations}\\
        &= \underbrace{\Bra{\psi\left(\bm{x}_i;\bm{\theta}\right) } }_{=~\Bra{\phi}}
        R^\dagger(\underbrace{\alpha_i - \alpha_{i'}}_{=~\alpha})
        \underbrace{
        \Ket{ \psi\left(\bm{x}_{i'};\bm{\theta}\right)}
        \!\Bra{\psi\left(\bm{x}_{i'};\bm{\theta}\right) }
        }_{=~\hat{M}}
        R(\underbrace{\alpha_i - \alpha_{i'}}_{=~\alpha})
        \underbrace{\Ket{ \psi\left(\bm{x}_i;\bm{\theta}\right)}}_{=~\Ket{\phi}} \label{eq:qsvm_derivation_complex_conjugate}
        .
    \end{align}
    \end{widetext}
    Two properties of rotation gates are applied in Eq.~\eqref{eq:qsvm_derivation_combine_rotations}. First, the adjoint flips the rotation direction $R^\dagger(\varphi) = R(-\varphi)$. Second, two subsequent rotation gates with matching generators combine into a single rotation as $R(\alpha)R(\beta) = R(\alpha + \beta)$. Eq.~\eqref{eq:qsvm_derivation_complex_conjugate} uses the property $\lvert z \rvert^2 = zz^*$ for any $z\in\mathbb{C}$.
    
    Compared to the QNN derivation in Eq.~\eqref{eq:qnn_derivation}, not only does the state depend on a data point $\bm{x}_i$, but the observable also depends on a second data point $\bm{x}_{i'}$. The application of Q-FLAIR to QSVMs becomes more transparent when the kernel target alignment setting is viewed as a classification problem, as explained in Sec.~\ref{sec:background:qsvm}. Thus, per probed gate candidate, an analytic reconstruction of the quantum kernel is required for each \emph{pair} of data points (as opposed to a single data point). 
    For weight-data-dependent gates as in Eq.~\eqref{eq:def_parameter_data_gate}, the appended rotation gate then depends on the feature difference of the two data points weighted by a parameter $\theta(x_{i,k} - x_{i',k})$, which is what the rotation angle $\alpha$ is substituted by when processing the analytic reconstruction classically for such gates.

    A well-known phenomenon, sometimes referred to as \textit{gate erasure bug} \cite{paine_QuantumKernelMethods_2023,salmenpera2024impact}, occurs when constructing quantum (fidelity) kernels directly from quantum feature-maps $U(\bm{x}, \bm{\theta})$.
    Precisely, data-independent gates at the end of the quantum feature-map circuit %
    cancel out, due to the adjoint construction 
    in the quantum kernel, as it becomes evident when setting $\alpha_i = \alpha_{i'}$ in Eq.~\eqref{eq:qsvm_derivation_add_rotation}.
    Since we use Q-FLAIR to grow the quantum feature-map circuits by adding gates from the pool at the end, we therefore restrict the gate pool to gates that always exhibit data dependence. Data dependence is achieved by either weight-data-dependent gates as in Eq.~\eqref{eq:def_parameter_data_gate} or by fixed gates preceding a (weight-) data-dependent gate \footnote{The required form to apply Q-FLAIR is still maintained when appending a fixed gate along with a (weight-) data-dependent gate because the fixed gate is absorbed by the state and observable in Eq.~\eqref{eq:qsvm_derivation_complex_conjugate}}. 
    Eventually, the gate pool used in our experiments comprises
    \begin{equation}
        \begin{aligned}
        \mathcal{V} = \{
            & R_z(\theta, x_{k}), (CZ,H,R_z(\theta, x_{k})) \\
            &R_{xx}(\theta, x_{k}),R_{yy}(\theta, x_{k}),
            R_{zz}(\theta, x_{k}),
        \}
        .
        \end{aligned}
    \end{equation}
    Like in the QNN gate pool, the two-qubit gates only act on neighboring qubits, except the $CZ$ gate combination, which connects qubits with up to two qubits between them. Furthermore, all gates possess sine curve reconstructions and constitute a universal gate set \cite{nielsen_QuantumComputationQuantum_2010} to prevent exact classical simulability of the learned feature-map circuits.

\section{Experimental results}\label{sec:experiments}

    We present experimental results and analyses for Q-FLAIR, 
    starting with demonstrating the algorithm's ability to learn high-quality QNN and QSVM models. For QNNs, we then illustrate the scalability of Q-FLAIR to high-dimensional, real-world datasets and conduct an ablation study to analyze the impact of individual components within its simultaneous optimization mechanism. We finally report benchmarking results of Q-FLAIR on a real IBM quantum computer for both QNN and QSVM. See Appendix~\ref{app:hyperparams} for hyperparameter settings.

    \begin{table}[tb]
        \centering
        \begin{tabular}{c c c c c}
            \toprule
            Dataset & \# Training data & \# Test data & Dimension $d$ & Color \\
            \midrule
            \textit{linearly separable} & 1000 & 1000 & 10 & \textcolor{tab_2_orange}{$\blacksquare$}\\
            \textit{two-curves} & 1000 & 1000 & 10 & \textcolor{tab_1_blue}{$\blacksquare$} \\
            \textit{bars \& stripes} & 1000 & 200 & 16 & \textcolor{tab_3_red}{$\blacksquare$}\\
            \textit{MNIST PCA} & 11552/1000 & 1902/1000 & 10 & \textcolor{tab_4_purple}{$\blacksquare$}\\
            \textit{MNIST $7 \times 7$} & 1000 & 1000 & 49 & \textcolor{plt_lightgreen}{$\blacksquare$}\\
            \textit{MNIST $14 \times 14$} & 1000 & 1000 & 196 & \textcolor{plt_tab_green}{$\blacksquare$}\\
            \textit{MNIST $28 \times 28$} & 1000 & 1000 & 784 & \textcolor{plt_darkgreen}{$\blacksquare$}\\
            \bottomrule
        \end{tabular}
        \caption{Details of datasets employed in this work. Reported are the number (\#) of training and test samples, the feature dimension $d$, and the color used in plots for each dataset.
        For \textit{MNIST PCA}, the QNN is trained and tested on the full datasets, while the QSVM uses smaller subsets. The data is based on Pennylane datasets \cite{bowles2024subtledata}, except for \textit{MNIST $28 \times 28$} and its down-scaled variants (\textit{MNIST $7 \times 7$, MNIST $14 \times 14$}), which are based on the raw MNIST data \cite{lecun1998}. All MNIST datasets are restricted to two classes (digits 3 and 5).}
        \label{tab:num_datasets}
    \end{table}
    
    The performance and flexibility of Q-FLAIR are evaluated on a variety of datasets, summarized in Tab.~\ref{tab:num_datasets}.
    Four of these are well-established classification benchmarks \cite{bowles_BetterClassicalSubtle_2024}: \textit{linearly separable} is easily solved with classical models, but has been reported to pose challenges for QML approaches \cite{bowles_BetterClassicalSubtle_2024}. \textit{bars \& stripes} consists of images designed explicitly to test translational invariance in models such as convolutional architectures. \textit{two-curves} is based on a work in which the influence of the curvature of two curves on the accuracy of a neural network has been investigated \cite{buchanan2020deep}. \textit{MNIST PCA} is a dimensionality-reduced version, using principal component analysis (PCA), of the MNIST handwritten digit data \cite{lecun1998} restricted to the digits 3 and 5. This classification task is considered one of the most challenging digit pairings \cite{bowles_BetterClassicalSubtle_2024}.
    We further extend the analysis of Q-FLAIR to the MNIST in full $28 \times 28$-pixel resolution \cite{lecun1998}. To specifically examine scaling behaviors with the feature dimension, we study down-scaled resolutions labeled as \textit{MNIST $7 \times 7$} and \textit{MNIST $14 \times 14$}. This dataset not only extends our study to real-world data, as opposed to the synthetic or pre-processed datasets, but data of significantly higher dimensionality (up to $d=784$) than the other benchmarking datasets $d \leq 16$.

    We report accuracies using \emph{average} (\emph{balanced}) accuracy,
    \begin{equation}\label{eq:averAcc}
        \overline{A}=\tfrac{1}{2}(t_{\mathrm{p}}+t_{\mathrm{n}}), 
    \end{equation}
    averaging over the ratio of correct predictions $t_{\mathrm{p}}$ and $t_{\mathrm{n}}$ for the positive and negative class, respectively. 
    Since the datasets we use deviate up to \SI{6}{\percent} from perfect balance, this measure avoids the bias that standard accuracy would introduce.

\subsection{Q-FLAIR performance benchmark for QNN and QSVM}\label{ssec:performacne_qsvm_qnn}  
    \begin{figure*}%
        \centering
        \includegraphics[width=\linewidth]{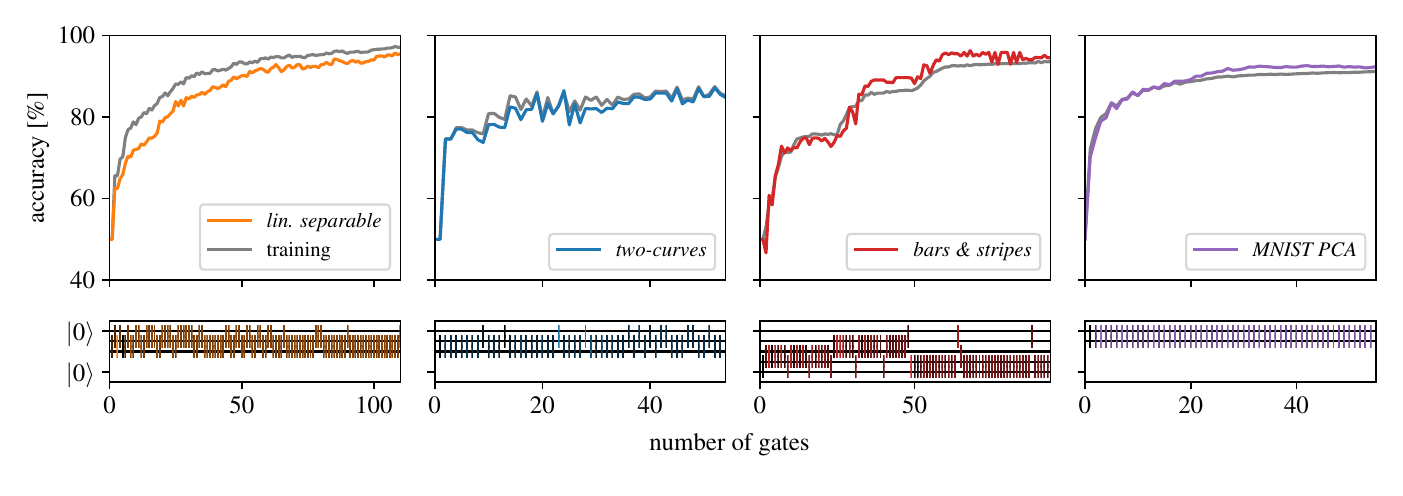}
        \caption{QNN performance benchmark on four different datasets. 
        \textit{Top panels:} 
        Accuracy over the number of gates appended to the feature-map circuit by Q-FLAIR, where each point corresponds to the QNN with the intermediate circuit obtained after each gate addition.
        Gray curves show training accuracy, colored curves show test accuracy.
        \textit{Bottom panels:}
        The corresponding feature-map circuits built gate by gate with Q-FLAIR. Each rectangle denotes a gate (single- or two-qubit), with coloring indicating its parameter value between \num{-1} (black) and \num{1} (color).
        }
        \label{fig:ann_acc}
    \end{figure*}
    \begin{figure*}%
        \centering
        \includegraphics[width=\linewidth]{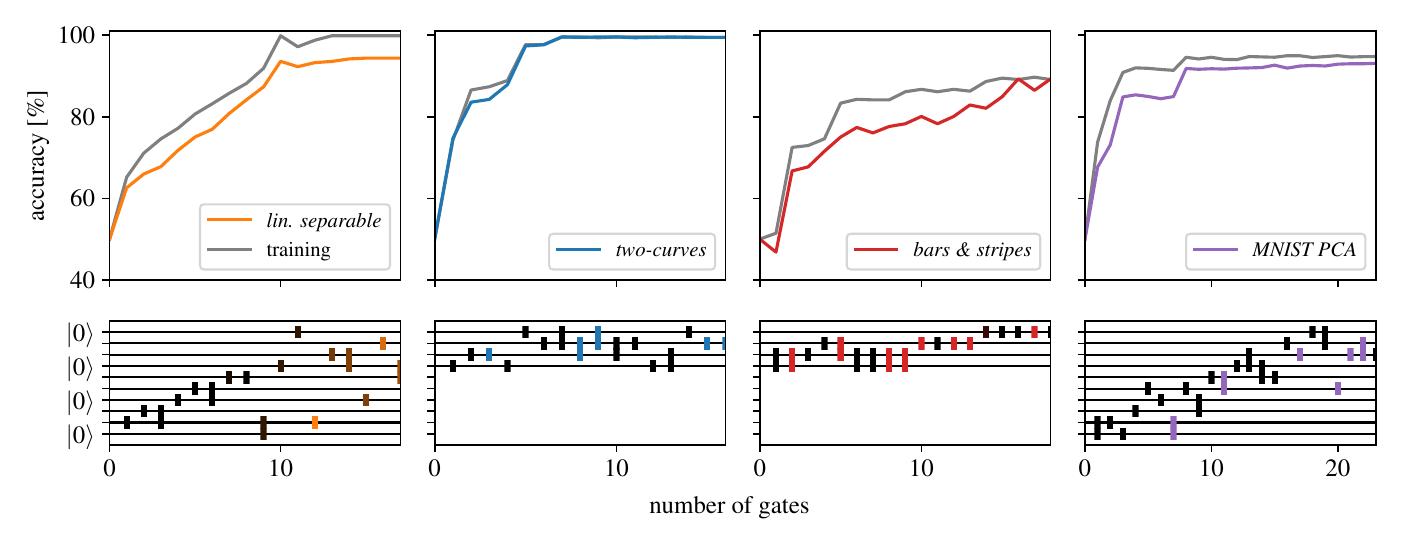}
        \caption{QSVM performance benchmark on four different datasets. 
        \textit{Top panels:} 
        Accuracy over the number of gates appended to the feature-map circuit by Q-FLAIR, where each point corresponds to the QSVM with the intermediate circuit obtained after each gate addition.
        Gray curves show training accuracy, colored curves show test accuracy.
        \textit{Bottom panels:}
        The corresponding feature-map circuits built gate by gate with Q-FLAIR. Each rectangle denotes a gate (single- or two-qubit), with coloring indicating its parameter value between \num{-1} (black) and \num{1} (color).}
        \label{fig:ta_acc}
    \end{figure*}

    For the four benchmarking datasets, Figs.~\ref{fig:ann_acc} and~\ref{fig:ta_acc} show the accuracies achieved with Q-FLAIR for QNNs and QSVMs, together with the gate-by-gate construction of the learned quantum feature-maps.
    With QNNs, test accuracies above \SI{92}{\percent} are reached in all tasks except for \textit{two-curves}, which saturates at \SI{85}{\percent}.  The best result is obtained on \textit{linearly separable}, with a test accuracy of \SI{96}{\percent}. Overfitting is negligible across datasets as the gap between final training and test accuracies never exceeds \SI{1.5}{\percent}.
    With QSVMs, test accuracies above \SI{89}{\percent} are achieved on all datasets. The best accuracy on \textit{two-curves}, with \SI{99}{\percent}, marks a sharp improvement over the QNN setting, where this dataset had the lowest accuracy, and even surpasses the QNNs on any dataset. The weakest QSVM result is on \textit{bars \& stripes}. Unlike QNNs, QSVMs show slight overfitting (train-test gap of up to \SI{5}{\percent} on \textit{linearly separable}).

    Both methods achieve similar accuracies, with QSVM performing better on \textit{two-curves} and QNN more reliably capturing \textit{bars \& stripes}. The strong QSVM performance on \textit{two-curves} is notable given its reported difficulty, though this complexity measure \cite{lorena2018data} should be interpreted with caution \cite{bowles_BetterClassicalSubtle_2024}. For instance, despite the good performance of Q-FLAIR for both models, \textit{linearly separable} poses difficulties for some QML methods \cite{bowles_BetterClassicalSubtle_2024}, but can be readily learned by most classical models. Appendix~\ref{app:re-optimization} shows that re-optimizing weights offers negligible gains, indicating near-optimal weights.

    Comparing the quantum circuits learned for the QNN and QSVM models (cf. bottom panels in Figs.~\ref{fig:ann_acc} and \ref{fig:ta_acc}) reveals clear differences.
    On the one hand, the circuit \emph{width} (i.e., number of qubits) is typically higher for QSVM. Two QSVM circuits (Fig.~\ref{fig:ta_acc}) use \num{10} qubits, reaching the maximum available in the Q-FLAIR simulations, whereas the QNN circuits (Fig.~\ref{fig:ann_acc}) never exceed \num{5} qubits, as observed for \textit{bars \& stripes}. In this dataset, the QNN makes particular use of entangling qubits, with each qubit addition corresponding to a noticeable increase in accuracy.
    On the other hand, the QSVM circuit \emph{depth} (i.e., number of gates) is yet lower across all datasets. None of the QSVM circuits use more than \num{25} gates, while the QNN circuit for \textit{linearly separable} exceeds \num{100} gates before the loss function change falls below the convergence threshold. 
    Two subtleties must be considered, however, when comparing circuit depths in this context. First, the convergence threshold itself drives the circuit length, and because it is applied to different loss functions for QNN and QSVM, the depth is not directly comparable. To address this, Appendix~\ref{app:accuracy_depth_comp} analyzes the minimum circuit depth needed to reach certain accuracy levels, confirming that QSVM circuits are shallower in the higher accuracy regime (above \SI{75}{\percent}).
    Second, the effective depth of QSVM circuits is doubled because each kernel entry corresponds to preparing two distinct states, whereas QNNs use only one state per data point.
    Another difference is their scaling with training data, since evaluating the full $N \times N$ kernel matrix requires many more circuit evaluations for QSVMs throughout training.
    Overall, the combination of lower measurement and qubit requirements makes QNNs more suitable for near-term quantum hardware, particularly when qubits are limited. 

    \subsection{Scaling with feature dimension} \label{ssec:featureDim}

    \begin{figure}
        \centering
        \includegraphics[width=\linewidth]{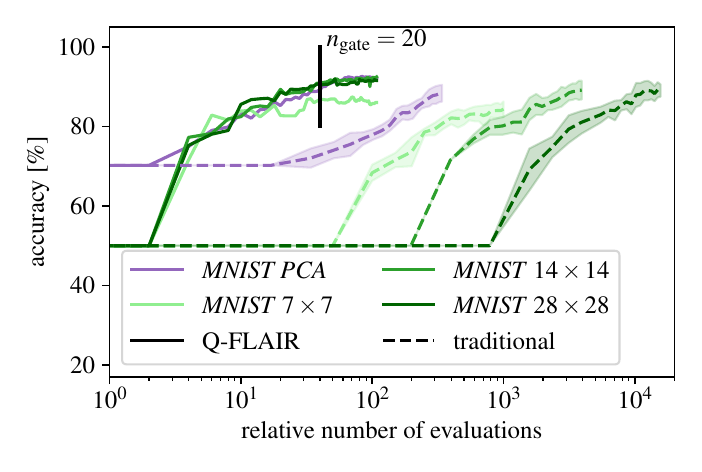}
        \caption{Comparison of Q-FLAIR with traditional feature-map learning on MNIST (digits 3 vs 5). Accuracies are shown for \textit{MNIST PCA} (purple) and pixel (light to dark green) variants. Solid lines show Q-FLAIR results, while dashed lines correspond to the traditional feature-map learning without analytic reconstruction. Here, training is terminated after \num{20} gates (Q-FLAIR curves are marked at \num{20} gates for comparison). A random parameter is drawn for each gate–feature combination. Each curve represents the mean over ten independent runs, with shaded bands indicating standard deviations. Traditional learning scales quantum evaluation overhead with feature dimension, exceeding Q-FLAIR by more than two orders of magnitude on \textit{MNIST $28\times28$}.
        }
        \label{fig:MNIST-MNISTPCA}
    \end{figure}
    
    We investigate the effect of scaling the feature dimension $d$ by incorporating, in addition to the PCA-reduced version ($d=10$), raw pixel variants of MNIST ($7 \times 7$, $14 \times 14$, $28 \times 28$) for digits 3 vs 5. Varying the pixel resolution allows us to study scaling on non-synthetic data across nearly two orders of magnitude in feature dimension (see Tab.~\ref{tab:num_datasets}).
    In these experiments, only QNNs were trained with Q-FLAIR.

    Figure~\ref{fig:MNIST-MNISTPCA} shows that Q-FLAIR maintains stable convergence and accuracy across all pixel resolutions compared to the PCA variant, with a slight drop for \textit{MNIST $7\times7$}. We hypothesize that this arises from limited visual information in these most down-scaled images rather than from the learning algorithm itself. 
    Matching the \textit{MNIST PCA} accuracies demonstrates that advanced pre-processing, such as PCA, is not necessary: Q-FLAIR integrates feature selection into the feature-map circuit composition and can directly process raw, high-dimensional data. In fact, accuracy initially increases faster for the pixel variants at higher resolutions (\textit{MNIST $14\times14$}, \textit{MNIST $28\times28$}) than for PCA-reduced data.
    While accuracies above \SI{90}{\percent} are reached for the \num{3}-\num{5} digit classification task on the higher-resolution datasets, Q-FLAIR achieves \SI{99}{\percent} and \SI{97}{\percent} accuracy on the simpler \num{0}-\num{1} and \num{0}-\num{2} tasks, respectively, with a single data-dependent gate targeting a central pixel. This underscores the effectiveness of Q-FLAIR's inherent feature selection in high-dimensional spaces while also highlighting the relative simplicity of the \num{0}-\num{1} task \cite{bowles_BetterClassicalSubtle_2024}, motivating our focus on the more challenging \num{3}-\num{5} scenario. At the same time, the result for \num{0}-\num{2} demonstrates that our algorithm is competitive with other established methods \cite{pan2023experimental}.
    Besides the high accuracy of higher-resolution datasets, it is essential to note that the quantum resources overhead is unchanged for all MNIST variants. Although \textit{MNIST PCA} is 16-dimensional, while MNIST $28\times28$ has 784 dimensions, the number of circuit evaluations is independent of the feature dimension, and the number of qubits is unchanged.

    In contrast, traditional feature-map learning, where the model is queried for probing individual features, scales unfavorably, because circuit evaluations are directly coupled to the number of features. To demonstrate this, we trained QNNs with up to 20 gates without reconstructions, repeating each run ten times with randomly drawn weight parameters. Because random parameters are drawn for each gate–feature combination, redundant features and similar gates effectively provide finer sampling of the parameter space, so that only a negligible decrease in accuracy is expected in comparison to the full optimization. This ensures that the number of evaluations is determined by the features rather than by weight optimization, resulting in a fair comparison with Q-FLAIR. 
    As presented in Fig.~\ref{fig:MNIST-MNISTPCA}, quantum circuit evaluation overhead grows rapidly with feature dimension, exceeding Q-FLAIR by more than two orders of magnitude for \textit{MNIST $28\times28$} to construct quantum feature-maps for similar accuracy and depth. In turn, the constant reconstruction overhead in Q-FLAIR is negligible relative to the high feature dimensions present.   

    \subsection{Ablation study of simultaneous optimization in Q-FLAIR}

\begin{figure}%
        \centering
        \includegraphics[width=\linewidth]{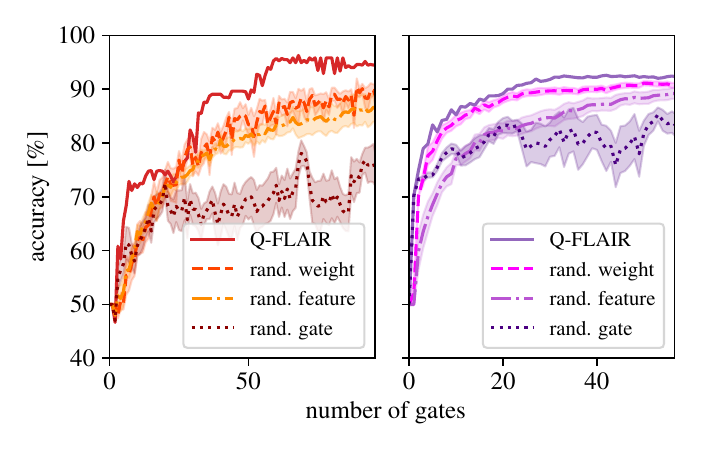}
        \caption{
        Ablation study of Q-FLAIR optimizations for QNNs on (left) \textit{bars \& stripes} and (right) \textit{MNIST PCA}. Solid curves (labeled `Q-FLAIR') correspond to the full algorithm comparison (same as in Fig.~\ref{fig:ann_acc}). Dotted, dash-dotted, and dashed curves show the effect of replacing gate, feature, or parameter optimization, respectively, by random choices instead of including them in the simultaneous optimization. Each curve represents the mean over ten independent runs, with shaded bands indicating standard deviations.
        }
        \label{fig:ann_rnd}
    \end{figure}

    We analyze the role of the three simultaneous optimizations in each Q-FLAIR iteration over gates, features, and weight parameters. Hence, we conduct ablation experiments in which each optimization is separately replaced with random choices. This design isolates the relative contribution of each optimization. We focus on QNNs trained on the \textit{bars \& stripes} dataset, which has the highest feature count, and on \textit{MNIST PCA}, the only non-synthetic dataset in our benchmarks. Training is terminated once circuits reach the same number of gates as in the unmodified Q-FLAIR runs. To account for stochasticity introduced by randomization, each ablation is repeated ten times, and Fig.~\ref{fig:ann_rnd} reports the resulting mean accuracies with standard deviations.

    In the first ablation (Fig.~\ref{fig:ann_rnd}, dotted), gate selection is no longer optimized but randomized. The best feature and parameter are then optimized as usual for this random but fixed gate. 
    To ensure the gate has an effect, the choice is restricted to those that modify the loss function and thus the model output. Learning behavior remains visible by improving beyond random guessing ($\sim\SI{50}{\percent}$), but accuracy quickly saturates at about \SI{70}{\percent} for \textit{bars \& stripes} and \SI{80}{\percent} for \textit{MNIST PCA}, well below the full Q-FLAIR performance.

    In the second ablation (Fig.~\ref{fig:ann_rnd}, dashed), the weight parameter is no longer optimized but randomly drawn (uniformly from $[-1,1]$).
    The best gate and feature are then optimized as usual for this random but fixed weight. Note that this ablation differs from the traditional learning scheme in the previous study (Sec.~\ref{ssec:featureDim}), where repeated resampling of weights across gate–feature combinations could implicitly optimize parameters through, e.g., redundant features.
    For \textit{MNIST PCA}, this leads to a consistent but modest decrease in accuracy across all (intermediate) circuit depths, whereas for \textit{bars \& stripes} the drop is more pronounced. The latter is evident in both slower convergence (more gates for the same accuracy) and substantially lower final accuracy.

    In the third ablation (Fig.~\ref{fig:ann_rnd}, dash-dotted), the feature is no longer optimized but randomly selected. The best gate and weight are then determined for this random but fixed feature. The impact of random feature selection is for \textit{bars \& stripes} similar to the random-parameter ablation, which means that the final accuracy is nearly $\Delta\overline{A}=\SI{10}{\percent}$ lower compared to full Q-FLAIR. For \textit{MNIST PCA}, the effect is initially stronger, with poor performance for shallow circuits, but the gap narrows at larger depths, eventually approaching the random-parameter ablation.

    Overall, these experiments demonstrate that all three optimizations make meaningful contributions to Q-FLAIR. Optimal gate selection has the most significant impact, but feature and parameter optimization also yield clear and non-negligible improvements. Importantly, Q-FLAIR consistently outperforms all ablation variants, establishing an effective upper bound on expected accuracies. It should also be emphasized that none of the ablation schemes could be implemented to reduce the quantum resource overhead in a substantial way, highlighting the information efficiency of Q-FLAIR through analytic reconstructions.

    \subsection{Benchmarking on IBM quantum computers}

    \begin{figure*}
        \centering
        \includegraphics[width=1\linewidth]{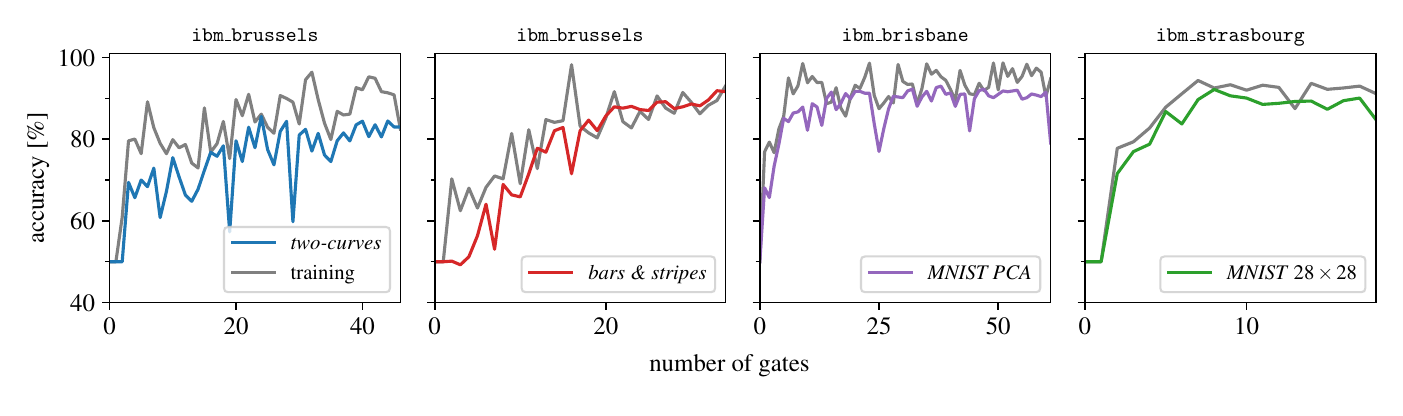}
        \caption{NISQ benchmark of QNN performance on four datasets, including full-resolution \textit{MNIST $28\times28$}. All results stem entirely from IBM quantum hardware (device names above panels), without numerical simulation. Mini-batch training is used. Curves are interpreted as in Fig.~\ref{fig:ann_acc}.
        }
        \label{fig:acc_ibm}
    \end{figure*}

    \begin{figure*}
        \centering
        \includegraphics[width=1\linewidth]{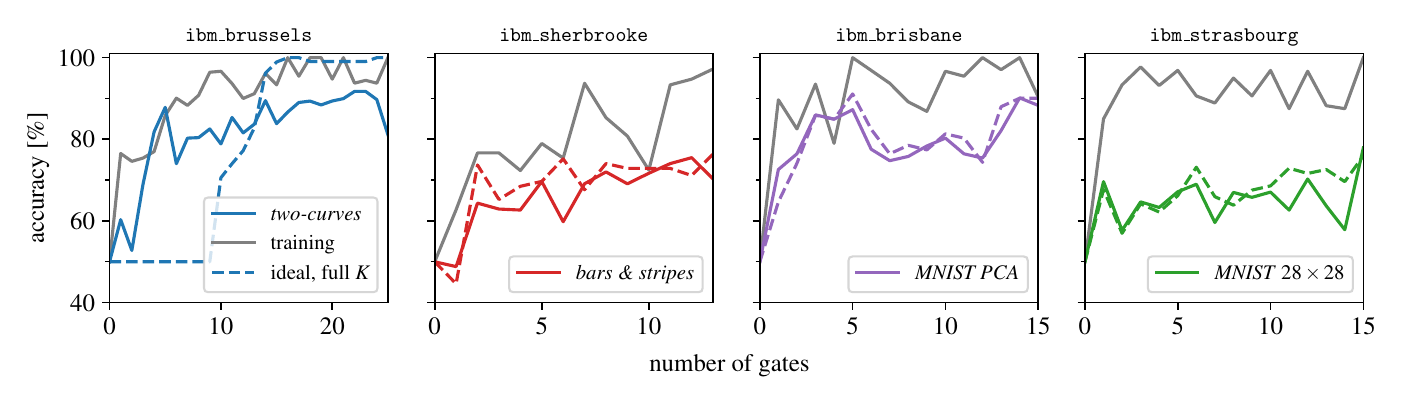}
        \caption{
        NISQ benchmark of QSVM performance on four datasets, including full-resolution \textit{MNIST $28\times28$}. Solid curves show results obtained entirely on IBM quantum hardware (device names above panels), without numerical simulation, using mini-batch kernels for training. Dashed curves correspond to the SVM trained on the full kernel matrix in ideal numerical simulations. Curves are interpreted as in Fig.~\ref{fig:ann_acc}.
        }
        \label{fig:acc_ta_ibm}
    \end{figure*}

    To conclude the experimental study, we transition from numerical simulations to benchmarks on real noisy intermediate-scale quantum (NISQ) hardware, running Q-FLAIR for both QNN and QSVM.  Experiments were conducted on IBM Quantum System One devices (\texttt{ibm\_brussels}, \texttt{ibm\_sherbrooke}, \texttt{ibm\_brisbane}, \texttt{ibm\_strasbourg}), featuring the Eagle architecture with up to 127 qubits. Q-FLAIR was executed with only minor modifications from the numerically simulated setup, and all quantum calculations, including training and testing, were performed on the IBM devices, without any numerical simulations. We switch from full training sets in simulation to \emph{mini-batching} \cite{murphy_MachineLearningProbabilistic_2012} on hardware to process a random subset at each iteration.
    Implementation details are provided in Appendix~\ref{app:specifics_ibm}. To increase practical relevance, we replace the simple synthetic \textit{linearly separable} with the full-resolution $28 \times 28$ MNIST dataset (digits 3 vs 5).

    As Fig.~\ref{fig:acc_ibm} shows, the QNN achieves test accuracies above \SI{90}{\percent} on the IBM device across all datasets, most notably surpassing \SI{92}{\percent} on full-resolution \textit{MNIST $28\times28$} with only eight gates. This performance is consistent with, and in the case of \textit{MNIST $28\times28$} even exceeds, the numerical simulations, where similar accuracies required deeper circuits. This seemingly counterintuitive result is likely due to stochasticity from mini-batching and noise in loss function estimates, as discussed in Sec.~\ref{sec:discussion}. 
    Compared to prior full-resolution MNIST benchmarks on real quantum hardware, to our knowledge, this is the first demonstration of MNIST classification with feature-maps learned from scratch on real hardware without heavy pre-processing. Importantly, no substantial overfitting occurs, and ideal simulations with the learned feature-maps reproduce the IBM results.

    Figure~\ref{fig:acc_ta_ibm} presents the QSVM results. However, the performance on the IBM hardware is consistently lower than in numerical simulation, yet it achieves a test accuracy of at least \SI{80}{\percent} across all tested datasets. This discrepancy can be attributed primarily to the mini-batching approach, as reported in Ref.~\cite{coelho_QuantumEfficientKernelTarget_2025}, which indicates that the resulting $32 \times 32$ training mini-kernels lead to overfitted SVM models. This phenomenon is also evident in the test–training accuracy gaps observed (Fig.~\ref{fig:acc_ta_ibm}). Larger kernels or approximations (e.g., Nyström method \cite{coelho_QuantumEfficientKernelTarget_2025}) could mitigate the overfitting.
    Moreover, the target alignment loss mini-batch estimation itself appears insufficient for learning effective quantum feature-maps, as opposed to observations in Ref.~\cite{coelho_QuantumEfficientKernelTarget_2025}. This is evident from the poor SVM performance using the quantum feature-maps learned on the IBM device, even for kernels incorporating the full training dataset evaluated in ideal numerical simulation (Fig.~\ref{fig:acc_ta_ibm}, dashed curves). Otherwise, these models should approach the high test accuracies of the simulated experiments in Sec.~\ref{ssec:performacne_qsvm_qnn}, which is only the case for, e.g., \textit{two-curves}.

    Overall, the IBM benchmarks demonstrate Q-FLAIR’s applicability to real NISQ hardware. This success stems from two main factors: the minimal impact of hardware noise, largely mitigated by the quantum resource efficiency of Q-FLAIR in learning lightweight circuits with few gates and qubits that map naturally to the hardware’s native gate set. The other factor is the low runtime, achieved by offloading subroutines to classical computation, a mechanism crucial for avoiding severe resource overhead with high-dimensional data. Notably, Q-FLAIR learns the QNN feature-map from scratch and achieves over \SI{90}{\percent} test accuracy entirely on a real IBM quantum computer within roughly four hours of total quantum computation, including test evaluations at each iteration, all in a single IBM runtime session.
    By circumventing the high resource overhead of traditional methods, which scale directly with the data dimension $d$ (cf.~Sec.~\ref{ssec:featureDim}), Q-FLAIR bypasses what would otherwise need an estimated four months to run.

    \subsection{Benchmarking quantum advantage via classical surrogates}\label{sec:classical_surrogate_bench}

        To conclude the empirical analysis of Q-FLAIR, we shift our focus from demonstrating practical (hardware) scalability to establishing concrete evidence of potential quantum advantage.
        While rigorous proofs of quantum advantage rely on theoretical frameworks that often lack direct applicability to relevant real-world data \cite{liu_RigorousRobustQuantum_2021,jager_UniversalExpressivenessVariational_2023}, an established empirical benchmarking technique to bridge this gap was proposed by \citet{schreiber_ClassicalSurrogatesQuantum_2023}.
        It relies on multivariate truncated Fourier series as \emph{classical surrogate models} to capture the input-output relation of quantum circuits, such as those learned by Q-FLAIR. The data-dependent quantum gates induce a frequency spectrum that the classical surrogate uses to match the accessible frequencies of the quantum model, allowing for Fourier coefficients that match the classical surrogate with the \emph{learned} quantum model.
        
        While the classical surrogate with such matched Fourier coefficients could replace the learned quantum model at inference time to make predictions for new data points $\bm{x}$, it is most intriguing to ask whether there exists a choice of Fourier coefficients such that the classical surrogate outperforms the quantum model. Given a training dataset, the globally optimal Fourier coefficients can be computed classically, entirely \textit{independent} of the learned quantum model. Hence, such training of the classical surrogate models ``provide a natural benchmark~[...]~for any claim of `quantum advantage'''~\cite{schreiber_ClassicalSurrogatesQuantum_2023} of a quantum model.
        We conduct classical surrogate benchmarking on Q-FLAIR QNNs. Details on the benchmark and its adaptation from Ref.~\cite{schreiber_ClassicalSurrogatesQuantum_2023} for the Q-FLAIR framework are provided in Appendix~\ref{app:classical_surrogate_bench}.

        \begin{figure}[t]
            \centering
            \includegraphics[width=1\linewidth]{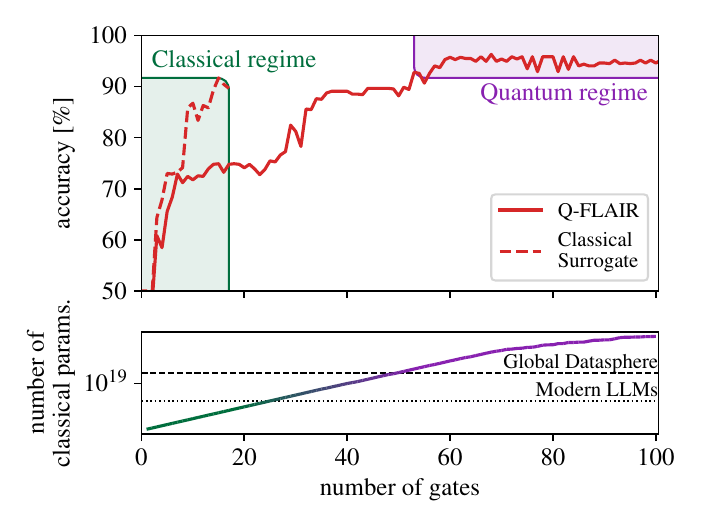}
            \caption{Classical surrogate benchmark with \textit{bars \& stripes} dataset. \textit{Top:} Comparison of the test accuracies of Q-FLAIR QNNs (solid) and the trained classical surrogate models (dashed). For each Q-FLAIR iteration (added gate), the reported surrogate model test accuracy corresponds to the highest-performing coefficients from \num{1000}-epoch training, evaluated on the training dataset every \num{100} epochs. 
            The classical regime (green) indicates where surrogates are tractable and can outperform intermediate Q-FLAIR models. The quantum regime (purple) begins at 53 gates, where Q-FLAIR surpasses the highest accuracy achieved by any classical surrogate.
            \textit{Bottom:} Parameter count required for the classical surrogate model to match the quantum circuit at each Q-FLAIR iteration. For reference, we indicate the scale of trillion-parameter state-of-the-art large language models~\cite{fedus2022switch} and the projected 2025 global datasphere of 175 zettabytes~\cite{reinsel2018digitization} (i.e., a model with $\approx 10^{23}$ parameters would match the total data stored globally, assuming single-byte precision).}
            \label{fig:classical_surrogate_bench}
        \end{figure}

        Figure~\ref{fig:classical_surrogate_bench} presents the benchmark for \textit{bars \& stripes} with a classical surrogate model trained for the QNN ansatz at each Q-FLAIR iteration. The test accuracy is initially similar but separates when Q-FLAIR reaches its first performance plateau. This performance gap (about \SI{90}{\percent} vs \SI{75}{\percent}) marks a clear classical regime because the classical surrogate achieves better performance from the same frequencies that the Q-FLAIR model has access to, and highlights the limitation of the greediness of Q-FLAIR where choices made in previous iterations cannot be reverted. In contrast, coefficient learning in the classical surrogates corresponds to the freedom of adjusting gate types (e.g., rotation axes) and placements in the quantum circuit. The frequency spectrum, which is dictated by the selected features and corresponding weights via Q-FLAIR, is fixed for both. Overall, the benchmark implies that the shallow quantum circuits ($\leq 17$ gates) at the beginning of the Q-FLAIR algorithm cannot exhibit a quantum advantage and ``enter a `classical regime' where one could just equivalently train the classical surrogate'' \cite{schreiber_ClassicalSurrogatesQuantum_2023}.

        However, classical surrogates exhibit a scaling limitation stemming from the number of feature encodings involved. Since the frequency spectrum is obtained from Minkowski sums of the spectra per encoding gate, its size generally grows exponentially in this gate count~\cite{caro_EncodingdependentGeneralizationBounds_2021}. 
        Crucially, because Q-FLAIR learns continuous weights, it inherently induces incommensurable frequencies (i.e., frequencies with irrational ratios).
        Unlike standard (unweighted) gates with integer frequencies \cite{schreiber_ClassicalSurrogatesQuantum_2023}, these incommensurate spectra prevent the frequency collisions that would otherwise collapse the spectrum to a polynomial size in the repeated encodings per feature.
        Because the number of Fourier coefficients scales with the frequency spectrum, classical surrogate models become intractably large as Q-FLAIR produces deeper circuits.
        While Fig.~\ref{fig:classical_surrogate_bench} indicates a classical regime only up to the point where we conducted surrogate training (for models with at most \num{e8} frequencies, tractable on a high-performance computer), it shows context for how this exponential trend rapidly exceeds any realistic computational requirements.

        For deeper feature-map circuits later in Q-FLAIR training, Q-FLAIR passes the benchmark. In this `quantum regime', test accuracies surpass that of classical surrogates at earlier stages, when the model was classically tractable. Due to the prohibitive parameter counts of the classical surrogate of deeper Q-FLAIR circuits, this ``impracticality of the classical surrogate can indicate a possible regime of advantage'' \cite{schreiber_ClassicalSurrogatesQuantum_2023}.

        To rule out that inferior performance was merely due to incomplete features at early Q-FLAIR stages, a surrogate that uses all 16 features simultaneously (one frequency/rotation gate per feature), despite slight but overfit improvements, still falls short of deeper Q-FLAIR QNNs.
        Crucially, the benchmark does not overall imply that no classical model exists that could match Q-FLAIR's performance on these datasets. 
        Rather, it highlights intractability within the specific function class native to the quantum models, i.e., incommensurable Fourier models for the frequencies determined by Q-FLAIR. 
        Nevertheless, from a practical learning perspective, it remains an open question to what extent this specific incommensurable complexity is an essential resource for learning, since classical approximations with commensurable frequencies may still yield sufficient learning outcomes.
        
        \begin{table}[t]
        \begin{tabular}{ccccc}
        \toprule
         \multirow{2}{*}[-0.25em]{Dataset} & \multicolumn{2}{c}{Classical surrogate} & \multicolumn{2}{c}{Q-FLAIR QNN} \\
        \cmidrule(lr){2-3} \cmidrule(lr){4-5}
         & Train. & Test & Train. & Test \\
        \midrule
        \textit{linearly separable} & 90.3\% & 84.4\% & 98.1\% & \textbf{97.3\%} \\
        \textit{bars \& stripes} & 95.0\% & 92.1\% & 94.0\% & \textbf{96.3\%} \\
        \textit{two-curves} & 98.3\% & \textbf{98.4\%} & 87.6\% & 87.6\% \\
        \textit{MNIST PCA} & 94.0\% & \textbf{94.5\%} & 92.5\% & 93.6\% \\
        \textit{MNIST $28 \times 28$} & 99.6\% & 91.6\% & 98.4\% & \textbf{92.9\%} \\
        \bottomrule
        \end{tabular}
        \caption{
        Comparison of classical surrogate and Q-FLAIR QNN accuracies. Maximum training and test accuracies achieved for each dataset. Bold values highlight the highest test accuracy.
        }
        \label{tab:classical_surrogate_bench}
        \end{table}
        
        For the remaining datasets, benchmarking results in Tab.~\ref{tab:classical_surrogate_bench} show that Q-FLAIR repeatedly produces QNNs that cannot be de-quantized via classical surrogates. However, as evidenced by slightly higher surrogate test accuracies on some datasets, this restriction to classically tractable surrogate models does not universally imply a deficiency in learning capability. According to Ref.~\cite{schreiber_ClassicalSurrogatesQuantum_2023}, claiming a quantum advantage over its surrogate requires the quantum model to exhibit superiority in trainability, generalization, or expressivity. The Q-FLAIR benchmarking results align with these criteria: First, Q-FLAIR exhibits a clear \emph{trainability} advantage, as surrogate models become intractable for deeper QNNs. Second, a slight \emph{generalization} advantage is consistently reflected in a smaller overfitting gap. Finally, Q-FLAIR provides a distinct structural advantage by building an ansatz that captures the dataset's inductive bias, effectively structuring the inherently vast spectrum of non-canceling incommensurable frequencies. Since classical surrogates fail to replicate this mechanism by unconstrained coefficient training, they cannot access this rich function class, granting Q-FLAIR a distinct \emph{expressivity} advantage.

        The benchmarks provide a concrete demonstration of a quantum regime driven by incommensurate frequencies. To successfully exploit this spectral richness, Q-FLAIR's adaptive learning is crucial compared to fixed quantum models. Traditionally, immense expressivity poses a severe trainability hazard \cite{mcclean_BarrenPlateausQuantum_2018,holmes_ConnectingAnsatzExpressibility_2022} for standard fixed-ansatz feature-maps that generate a generic, complex spectrum. Q-FLAIR overcomes this by dynamically tailoring the inductive bias of the ansatz to the specific dataset, selectively utilizing expressivity only where necessary. As verified in Appendix~\ref{sec:AccExpressZZFeature}, this adaptive approach maintains a tractable, low effective dimension (measured via the quantum Fisher information matrix (QFIM) rank) and remains trainable as the data dimension $d$ is increased. In contrast, the fixed feature-maps saturate at full QFIM rank and become untrainable at larger dimension $d$.

        While most benchmarks in the literature report that quantum models fail to outperform their classical surrogates \cite{schreiber_ClassicalSurrogatesQuantum_2023,Mittal_ExperimentalValidationDequantization_2025,hernicht_EnhancingScalabilityClassical_2025} or similar classical baselines \cite{bowles_BetterClassicalSubtle_2024}, we are aware of only one prior instance where QNNs successfully surpass classical surrogates, specifically in scenarios with scarce and noisy data \cite{stein_BenchmarkingQuantumSurrogate_2024}. Because their scenario of restricted data access is strictly complementary to our demonstration of a rich, quantum-trainable but classically intractable model class, the positive benchmark outcome provides empirical evidence for an alternative route to quantum advantage.

\section{Discussion}\label{sec:discussion}

    We introduce an algorithm, Q-FLAIR, to learn quantum feature-maps. These maps are utilized in the widely adopted QML models QNN and QSVM. The contributions and central elements of this algorithm are summarized briefly. Q-FLAIR iteratively probes different gates to grow the ansatz. It relies on partial analytic reconstructions of the model outputs for each gate. Because the loss function becomes efficiently evaluable on classical computers through these reconstructions, feature selection and weight parameter optimization do not require further access to the quantum computer.
    In addition to this reduced quantum resource overhead in terms of circuit executions and shot counts, Q-FLAIR learns resource-frugal quantum models by adapting to qubit topology constraints and maintaining shallow circuits due to sparse feature selection. We demonstrate strong performance in experiments across typical QML benchmarking datasets \cite{bowles_BetterClassicalSubtle_2024}, including both numerical simulations and NISQ experiments, supported by further analyses for a comprehensive interpretation.

    Building on these results, we emphasize the importance of simultaneously optimizing both the ansatz (gate and feature selection) and parameters at each iteration. Ablation studies show that removing either component degrades performance, highlighting their complementary roles. We further extend experiments to datasets with much higher feature dimensions (up to $d=784$) than typically benchmarked \cite{bowles_BetterClassicalSubtle_2024} ($d \lesssim 20$), demonstrating that Q-FLAIR scales without compromising performance by decoupling quantum resource overhead from the feature dimension. This contrasts with fixed-ansatz feature-maps, where one qubit per feature renders high-dimensional data virtually inaccessible \footnote{Due to qubit demands, we do not directly compare Q-FLAIR feature-maps with fixed-ansatz maps matching the qubit count to the feature count, e.g., angle embedding, Z-feature-map, and ZZ-feature-map \cite{havlicek_SupervisedLearningQuantumenhanced_2019}.}. Finally, the reproducibility of simulated results on NISQ hardware further supports that Q-FLAIR is both resource-efficient and viable in the near term.

    We further place the Q-FLAIR results in the broader context of prior full-resolution MNIST benchmarks on real quantum hardware. Earlier works typically relied on classical pre-\-processing: down-scaling or dimensionality reduction such as PCA or neural (autoencoder) networks \cite{kerenidis2020classification,slysz2023exploring,senokosov2024quantum,zhou2025enhanced,erkan2025quantum,chen2025exploring}, or exponentially compressed, such as amplitude-type, encodings \cite{tognini2025solving,manko2025classification,wang2025quantum,shen2024classification,kiwit2025typical}. Most demonstrations remain limited to simulations, and real-device studies are rare. Using down-scaled $4\times4$ digits, Ref.~\cite{slysz2023exploring} reached only about \SI{80}{\percent} accuracy for the \num{0} vs \num{1} task on IBM (Eagle) hardware, while otherwise exponentially deep amplitude-encoding circuits had to be approximated to shallower depth first to load a MNIST fashion item variant on earlier IBM devices \cite{shen2024classification}. Notably, Ref.~\cite{roseler2025efficient} reported \SI{96}{\percent} for full-resolution \num{0} vs \num{1} classification on the more recent IBM Heron hardware, but with a fixed ansatz derived from prior simulation-based architecture search rather than feature-maps learned directly on hardware -- an \textit{in situ} search would likely take months using traditional methods.
    By contrast, Q-FLAIR achieves over \SI{92}{\percent} test accuracy for the more challenging \num{3} vs \num{5} task \cite{bowles_BetterClassicalSubtle_2024}, fully on IBM hardware within four hours of training. Only hybrid models that employ rich classical processing report marginally higher accuracies on this task \cite{zeng2022multi,dhara2024multi,riaz2023application}.
    Moreover, for the simpler \num{0} vs \num{1} task, Q-FLAIR achieves \SI{99}{\percent} accuracy in simulation with only a single data-dependent rotation gate, which could likely be reproduced on hardware within minutes with minimal loss. To our knowledge, our IBM device experiments constitute the first demonstration of full-resolution MNIST classification on NISQ hardware with feature-maps learned entirely from scratch.

    Efficiently loading classical data into quantum states remains a central challenge in QML \cite{cerezo_ChallengesOpportunitiesQuantum_2022}, with quantum feature-maps at the core of this process. Their design is often difficult due to limited prior knowledge of effective encodings.
    We demonstrate the viability of Q-FLAIR not only on synthetic toy datasets but also on modest yet real-world datasets. This proof-of-principle is a concrete step toward overcoming the challenge of embedding classical data for QML. Q-FLAIR is particularly well-suited for near-term quantum devices, as it is adaptable to specific hardware architectures, including NISQ systems and early fault-tolerant quantum computers, and it scales even to high-dimensional datasets encountered in real-world scenarios. 
    However, learning quantum feature-maps faces not only practical resource constraints but also a key theoretical challenge in QML: the \emph{barren plateau} phenomenon \cite{mcclean_BarrenPlateausQuantum_2018,larocca_BarrenPlateausVariational_2025}. This arises primarily in fixed ansätze, where model outputs concentrate exponentially with the number of qubits, rendering learning at scale intractable. By growing the ansatz, Q-FLAIR may delay the onset of barren plateaus and identify effective quantum feature-maps before the entire gate pool suffers from concentration \cite{grimsley_AdaptiveProblemtailoredVariational_2023a}.
    Indeed, incrementally learning QNN weight layers has proven effective for mitigating barren plateaus \cite{skolik_LayerwiseLearningQuantum_2021}. However, such methods still rely on generic, fixed-ansatz feature-maps that inherently lack a dataset-specific inductive bias, requiring deep and parameter-heavy QNNs to compensate.
    More broadly, by leveraging partial analytic reconstructions, Q-FLAIR can also be seen as a contribution toward developing an interface between classical and quantum models, opening efficient pathways for their integration and helping to bridge the gap between traditional machine learning and quantum approaches. 
    For instance, enabling seamless back-propagation through reconstructions in hybrid models without additional quantum overhead.

    While Q-FLAIR relies on efficient local reconstruction of individual quantum gates, this does not imply efficient exact classical simulability via global reconstructions, provided a universal gate set \cite{nielsen_QuantumComputationQuantum_2010}. Simultaneous global reconstruction, yielding classical surrogates of the full circuits \cite{schreiber_ClassicalSurrogatesQuantum_2023}, scales exponentially with gate count, a barrier enforced by Q-FLAIR's continuous weights that induce a rich spectrum of non-canceling incommensurable frequencies.
    Although such immense expressivity typically renders fixed ansätze untrainable, Q-FLAIR’s adaptive learning successfully avoids trainability hazards. Consequently, surrogate training becomes intractable before matching the quantum model, marking an empirical demonstration of a natural advantage over classical surrogates. Such positive outcomes are rare in the literature \cite{stein_BenchmarkingQuantumSurrogate_2024}. Although we carefully adapted the benchmark to this adaptive setting, classical strategies should be considered in the future to challenge this regime. Specifically, an independent incremental search for frequencies could decouple the surrogate from the quantum learning entirely. Furthermore, applying regularization could improve generalization, while approximations via pruning similar frequencies or utilizing alternative surrogate architectures \cite{hernicht_EnhancingScalabilityClassical_2025,nair_LocalSurrogatesQuantum_2025} could delay -- though are unlikely to eliminate -- exponential scaling.

    Despite its effectiveness on QNN and QSVM, Q-FLAIR is limited to a subset of learnable quantum feature-maps, and its inherently greedy nature may restrict broader exploration.
    The first limitation stems from restricting gates to those whose generators satisfy $A^2 = I$. While this condition is necessary for a simple sine curve reconstruction, it constrains the design of the gate pool, ultimately limiting the expressivity of the quantum models and potentially hindering task-specific designs. 
    It appears that a consequence of these limitations is that Q-FLAIR learns quantum feature-maps that utilize only a few qubits, in most experiments, even though it is not explicitly constrained. Whether this is ``a bug or a feature'' is a matter of perspective: low-qubit scenarios limit the quantum advantage potential, as such models can often be efficiently simulated classically, yet a single qubit is theoretically sufficient for a universal quantum classifier via data re-uploading \cite{perez-salinas_DataReuploadingUniversal_2020}, a capability that Q-FLAIR can exploit.
    
    The second limitation is the greediness of Q-FLAIR: it biases the search towards exploitation over exploration. From a computational perspective, this favors efficiency at the potential cost of task performance. This greediness is largely due to the following algorithm design choices: deterministically selecting the gate, feature and parameter that immediately minimizes the loss, only exploring a single circuit at a time, and growing the ansatz exclusively by appending gates. The latter also requires caution in the gate pool design as certain gates may not affect the model output and loss, e.g., gates commuting with the QNN observable or data-independent gates causing the QSVM gate erasure bug \cite{paine_QuantumKernelMethods_2023,salmenpera2024impact}.

    Finally, we propose potential research directions to overcome the two main limitations identified in Q-FLAIR.
    Extending Q-FLAIR likely improves classification accuracy, e.g., on the benchmark datasets, but at the expense of extra quantum resources.
    Firstly, to increase the expressivity of quantum feature-maps, the algorithm could be extended from simple sine-curve reconstructions to higher-order truncated Fourier series. While $K$-th order reconstructions require $K$ times more circuit evaluations per gate candidate, they allow more general gate generators with $A^2 \neq I$ \cite{wierichs_GeneralParametershiftRules_2022,jager_FastGradientfreeOptimization_2025,schuld_MachineLearningQuantum_2021,schuld_EffectDataEncoding_2021} and can capture parameter repetitions, correlations, or sharing across the circuit \cite{nakanishi2020sequential}. Starting from a more expressive initial state or ansatz, e.g., applying Hadamard layers or data-dependent qubit rotations \cite{torabian_CompositionalOptimizationQuantum_2023}, may guide the algorithm to richer solutions faster. Complementary approaches include classical non-linear feature-maps ($\bm{x} \mapsto \bm{\phi}(\bm{x})$) as input, such as polynomial features for ZZ-feature-maps \cite{havlicek_SupervisedLearningQuantumenhanced_2019} in the quadratic case \footnote{The discrepancy introduced by classical non-linear pre-processing, is another reason why we refrained from directly comparing the feature-maps composed by Q-FLAIR with such fixed-ansatz quantum feature-maps as the ZZ-feature-map.}, which increase input dimension without adding quantum overhead because feature selection remains classical.

    Secondly, 
    to mitigate the greediness of selecting a single ansatz expansion (for the largest immediate loss reduction), each iteration of the algorithm could maintain a set of top-performing circuits, pruning the low-performing ones after a few steps. Alternatively, stochasticity could enhance the otherwise deterministic selection process by randomly sampling the next gate candidate, weighted by the determined loss improvements. 
    So far, stochasticity has appeared solely in the NISQ experiments, originating from mini-batch sampling, finite measurement shots, and hardware noise. E.g., improved $28 \times 28$ \textit{MNIST} performance over exact simulation provides concrete initial evidence of the benefits of stochasticity.
    Furthermore, the ansatz construction strategy could be switched from appending gates to prepending or extended to simultaneous bi-directional growth. In the QNN case, this corresponds to transitioning from data-dependent evolution of the initial state (Schrödinger picture) to reverse evolution of the observable (Heisenberg picture).
    Alternative strategies involve inserting gates at arbitrary positions in the circuit or adding new qubits through richer entanglement structures beyond linear configurations. Even though these strategies pose a risk of combinatorial blow-up, they can also facilitate gate pool and model design. For instance, enabling data-independent gates in a QSVM pool without directly triggering the \textit{gate erasure bug}, and further broadening the range of QNN observable choices, since commuting gates can have an effect at intermediate positions within the ansatz.
    Overall, combining analytic reconstructions with such compositional search strategies, including evolutionary or reinforcement learning approaches, could balance exploitation with broader exploration.

\section*{Data and Code Availability}
All data presented and code necessary to reproduce the results in this paper are publicly deposited on Zenodo \cite{zenodo} and are also openly available in a GitHub repository \url{https://github.com/phels23/Q-FLAIR}.

\begin{acknowledgments}
We thank Roman V. Krems for his advice and support on this project. Furthermore, we thank both him and Tanja Schilling for making this project possible.
We would like to thank the Plateforme d’Innovation Numerique et Quantique (PINQ\textsuperscript{2}), a non-profit organization based in Québec, Canada, for the access to the IBM QS1 machines and the computation time needed for this study. We are particularly thankful to Marie-Ève Boulanger (PINQ\textsuperscript{2}) and Paola Baca (QMI, UBC) for their dedicated efforts in facilitating this partnership. The views expressed are those of the authors and do not reflect the official policy or position of IBM or the IBM Quantum team. 
The authors acknowledge support by the state of Baden-Württemberg through bwHPC and the German Research Foundation (DFG) through grant no "INST 39/963-1 FUGG" (bwForCluster NEMO), as well as "INST 39/1232-1 FUGG" (bwForCluster NEMO 2).
This work was supported by the Natural Sciences and Engineering Research Council (NSERC) of Canada. J. J. and E. T. further acknowledge the NSERC CREATE in Quantum Computing Program, grant number 543245.
P. E. has received support by the DFG funded Research Training Group "Dynamics of Controlled Atomic and Molecular Systems" (RTG 2717).
\end{acknowledgments}

\section*{Author contributions}
J.J. conceived the project and developed the algorithm. P.E. and E.T. contributed variants and adjustments to the algorithmic framework, some of which are outlined as future work. 
P.E. implemented the algorithm, as well as the QML models, training, and evaluation code, with J.J. and E.T. assisting the process. 
E.T. contributed a literature review.
P.E. conducted the numerical simulation experiments, with J.J. performing the classical surrogate benchmarking. J.J. also carried out the quantum hardware experiments and implemented the interface specifics for the IBM platform.
All authors analyzed the data and contributed to the interpretation of the results, with P.E. producing the plots. All authors contributed to writing the manuscript, with J.J. and P.E. leading the effort.

\clearpage
\appendix

\FloatBarrier

\section{Algorithmic details}

\begin{algorithm}[H]
    \caption{{Q-FLAIR.} Labels highlight steps executed on quantum (purple Q) versus classical (green L) resources.}
    \label{alg:qflair}
    \begin{algorithmic}[1]
        \Require Training data $\mathcal{D} = \{(\bm{x}_i, y_i)\}_{i=1}^N$, gate pool $\mathcal{V} = \{V_m\}_{m=1}^M$, 
        loss function $L(\bm{y}, \bm{\hat{y}})$, loss convergence threshold $\Delta L$
        
        \CState \textbf{Initialize:} $U \gets I, \bm{\theta} \gets \emptyset, L \gets \infty,\,\text{init.~const.~model output}~z_0$
        
        \Repeat \Comment{Incremental ansatz growth}
            \CState $L_{\text{prev}} \gets L$
            \For{\textbf{each} candidate gate $V_m \in \mathcal{V}$}
                \CState Append $V_m$ gate to current circuit $U(\bm{x}_i; \bm{\theta})$
                \If{$V_m$ is a fixed gate}
                    \QState Evaluate loss $L_m^*$ over \emph{all} train.~data points $\bm{x}_i \in \mathcal{D}$
                
                \Else \Comment{$V_m(\alpha)$ is a variational gate}
                    \For{\textbf{each} training data point $\bm{x}_i \in \mathcal{D}$}
                        \CState For $\alpha_0 = 0$, reuse $z_0$ from prev.~ansatz iter.
                        \QState Estimate model output $z_\pm$ at two shifts $\alpha_0 \pm \tfrac{\pi}{2}$
                        \CState Reconstruct $f_{m,i}(\alpha) = a \sin(\alpha - b) + c$ with 
                        \CState[] \hspace{0.2cm} $a = \tfrac{1}{2}{\sqrt{(2z_0 - z_+ - z_-)^2 + (z_+ - z_-)^2}}$
                        \CState[] \hspace{0.2cm} $b = \alpha_0 -\mathrm{arctan2}( 2z_0 - z_+ - z_-, z_+ - z_- )$
                        \CState[] \hspace{0.2cm} $c = \tfrac{1}{2} (z_+ + z_-)$
                    \EndFor

                    \If{$V_m$ is weight- and data-dependent}
                        \For{feature index $k = 1, \dots, d$}
                            \CState $\alpha \gets \theta x_{i,k}$
                            \CState[] $f_{m,i,k}(\theta) \gets f_{m,i}(\alpha)$
                        \EndFor
                        \CState Candidate loss $L_m(\theta\!, k) \gets L\left(\bm{y}, [f_{m,i,k}(\!\theta\!)]_{i=1}^N\!\right)$
                        \CState[] $L_m^*, \theta_m^*, k_m^* \gets \minimize_{\theta, k} L_m(\theta, k)$ \Comment{$T_{\rm max}$}
                    \ElsIf{$V_m$ is only weight-dependent}
                        \CState $\alpha \gets \theta$
                        \CState[] $f_{m,i}(\theta) \gets f_{m,i}(\alpha)$
                        \CState[] Candidate loss $L_m(\theta) \gets L\left(\bm{y}, [f_{m,i} (\theta)]_{i=1}^N\right)$
                        \CState[] $L_m^*, \theta_m^* \gets \minimize_{\theta} L_m(\theta)$ \Comment{$T_{\rm max}$}
                    \EndIf
                
                \EndIf
                
            \EndFor
            
            \CState Select optimal candidate $m^* \gets \argmin_m \{L_m^*\}$
            \CState[] Append to ansatz $U \gets V_{m^*} U$
            \CState[] Extend $\bm{\theta} \gets \bm{\theta} \cup \{\theta_{m^*}^*\}$ and record $k_{m^*}^*$ (if applicable)
            \CState[] $L \gets L_{m^*}^*$
            
        \Until{$L_{\text{prev}} - L < \Delta L$}
        \State \Return Learned quantum feature-map $U(\bm{x}; \bm{\theta})$
    \end{algorithmic}
\end{algorithm}

Algorithm~\ref{alg:qflair} provides the pseudocode for the Q-FLAIR algorithm introduced in Sec.~\ref{sec:method}. To clarify which operations are executed on quantum versus classical hardware, we utilize the color-coding established in the schematic overview of Fig.~\ref{fig:graphical_abstract}. As is evident from the nested loop structure, the optimizations over features and weights rely solely on classical resources, while only the gate selection loop queries the quantum computer. Consequently, the runtime complexity with respect to the three optimization problem sizes per Q-FLAIR iteration decouples into a quantum complexity of $\mathcal{O}(M)$ and a classical complexity of $\mathcal{O}(d T_{\rm max})$. In contrast, traditional feature-map search methods lacking these analytical reconstructions must execute all three loops on the quantum computer, resulting in a significantly steeper quantum scaling of $\mathcal{O}(M d T_{\rm max})$. Note that in either approach, the quantum runtime inevitably scales with the training data set size $N$, making mini-batching an important strategy for practical implementation.

\subsection{Reconstruction implementation} \label{app:reconstruction}

As detailed in Sec.~\ref{sec:method}, the reconstruction of Eq.~\eqref{eq:analytic_form_rotation_gate} is determined from expectation values at three different angles, e.g., via Eqs.~\eqref{eq:coeffs:a}--\eqref{eq:coeffs:c}. 
This technique applies independent of whether the gate depends on data or not, because the weight-data-dependent gate in Eq.~\eqref{eq:def_parameter_data_gate} does not differ from the standard rotation gate definition of Eq.~\eqref{eq:def_rotation_gate} when substituting the rotation angle by the parameter times the data feature $\alpha = \theta x_{k}$. Note that if the parameter shifts are centered at $\alpha_0 = 0$, the value $z_0 = f(\alpha_0)$ at the non-shifted position $\alpha_0$ only needs to be evaluated once per data point $\bm{x}_i$. It is independent of the probed gate candidate because the gate action becomes identity when the parameter vanishes, and it suffices to evaluate only the two shifted positions $\alpha_0 \pm \pi/2$ for each gate candidate. Furthermore, when the reconstructions are computed in an iterative manner as in Q-FLAIR, the loss/expectation value for the unshifted parameter position $\alpha_0 = 0$ is known from the previous iteration (as the minimal loss when appending the last gate) and can be reused. At the initialization time of the algorithm, the loss for the unshifted parameter position, corresponding to the empty ansatz, is also usually known or efficiently computable classically. Hence, the quantum resource cost per reconstruction is effectively reduced from three to two per data point. Note that two quantum circuit evaluations exactly match the number required for the parameter-shift rule \cite{li_HybridQuantumClassicalApproach_2017,mitarai_QuantumCircuitLearning_2018,schuld_EvaluatingAnalyticGradients_2019} to determine the partial derivative of the parameter in a rotation gate at a certain parameter position.

A further reduction of evaluations can be achieved if the same gates, i.e. identical generators $A$, appear in the gate pool $\mathcal{V}$ both as weight-dependent gates and as weight-data-dependent gates. Thus, the quantum resource demand can be reduced by reconstructing with these gates only once per data point and iteration. Similarly, fixed gates that correspond to a variational gate in the pool, such as $X = R_{\rm x}(\pi)$, do not need to be evaluated on the quantum computer; instead, they can be inferred from the respective reconstruction. The same holds for purely data-dependent gates, which are a special case of the weight-data-dependent gates in which the parameter in Eq.~\eqref{eq:def_parameter_data_gate} is set to $1$ \footnote{One could further consider purely weight-dependent gates as a special case of the weight-data-dependent gates if a constant feature $\pi$ is included (considering a parameter range of $[-1, 1]$) in the input data points, which is a common approach for convenience in machine learning practice.}.

\section{Experimental setup and implementation details}

    \subsection{Hyperparameter settings}\label{app:hyperparams}

        For the training of the QNN, the logarithmic loss function defined in Eq.~\eqref{eq:QNNlog} is used for the empirical loss in Eq.~\eqref{eq:empirical_loss}. The stopping criterion is set to $\Delta L_{\mathrm{log}}<\num{e-3}$. In order to determine what threshold parameter $b$ on the QNN output is required for the decision function as defined in Eq.~\eqref{eq:qnn_output_sign}, we select the best value from the receiver operating characteristic (ROC) curve analysis. Thus, the threshold parameter $b$ provides the optimal training set separation, considering correct classifications equally important for both classes, which matches the choice of the balanced accuracy metric as in Eq.~\eqref{eq:averAcc}.
        
        The quantum feature-maps for the QSVM kernel are selected according to the target alignment of Eq.~\eqref{eq:L_TA}. We use as well a stopping criterion of $\Delta L_{\mathrm{TA}}<\num{e-3}$ between the attachment of two gates. For the QSVMs, the hyperparameter of the regularization strength $C$ is determined through a grid search (equidistant in log-space) to improve predictive performance. 
        
        In either parameter search (for $b$ or $C$ in QNNs and QSVMs, respectively), only minor changes were observed. Importantly, these searches do not involve any model evaluations on the quantum computer, but are conducted entirely classically.

    \subsection{Specifics for IBM benchmarks}\label{app:specifics_ibm}

        For both models, QNN and QSVM, for each quantum circuit evaluated for a fixed parameter binding, the all-zero state probability was estimated through computational basis measurements using 1024 shots. These measurements are implemented in the IBM hardware without basis transformations. This probability directly corresponds to the all-zero projector observable $O = \ket{0}^{\otimes n}\bra{0}^{\otimes n}$ in the QNN model and, hence, a direct realization of our numerical simulations. For the QSVM, the all-zero state probability emerges from implementing the fidelity kernel $\kappa(\bm{x}_i, \bm{x}_j)$ as the transition probability when applying the quantum feature-map circuit to the initial state $\ket{\psi_0} = \ket{0}^{\otimes n}$ once for data point $\bm{x}_i$ and once inverted for data point $\bm{x}_j$. This results in the quantum circuit
        \begin{equation}
            U^\dagger(\bm{x}_j)U(\bm{x}_i)
        \end{equation}
        to be executed starting from the all-zero computational basis state and collecting the (relative) frequencies of measuring all qubits in zero to estimate $\kappa(\bm{x}_i, \bm{x}_j)$ \cite{havlicek_SupervisedLearningQuantumenhanced_2019}. 
        Note that other realizations of the quantum state fidelity exist \cite{havlicek_SupervisedLearningQuantumenhanced_2019}, which are theoretically equivalent in the shot limit, but nevertheless can behave differently on real hardware or show different estimation guarantees under finite shots. For each pair of training data points $\bm{x}_i, \bm{x}_j$, the kernel value is only estimated in one order $\kappa(\bm{x}_i, \bm{x}_j)$ and used to populate the counterpart $\kappa(\bm{x}_j, \bm{x}_i)$ in the symmetric training kernel matrix. Furthermore, the diagonal of the training kernel matrix is never estimated using the quantum computer but set to the theoretically expected value $\kappa(\bm{x}_i, \bm{x}_i) = 1$. 
        
        Due to the (additive) shot noise present in the kernel entries, the training kernel matrix is no longer guaranteed to be positive semi-definite. The underlying optimization problem when training the SVM loses its convexity/concavity property, which is usually used by the underlying optimization routine to find the global optimum. 
        Possible corrections include shifting the kernel diagonal by the largest magnitude of the negative eigenvalues or clipping negative eigenvalues to zero in an eigendecomposition. However, these corrections did not yield any impact in terms of the model test accuracy. This finding aligns with other studies that estimate quantum fidelity kernels on real hardware \cite{havlicek_SupervisedLearningQuantumenhanced_2019}. In fact, we could observe increased overfitting in improved accuracies on training data only. Therefore, all reported results are without any kernel correction.

        A key algorithmic difference between the experiments conducted in numerical simulation and those run on the real quantum computer is the \textit{mini-batching} technique, a standard in machine learning \cite{murphy_MachineLearningProbabilistic_2012} to evaluate the (empirical) loss function on. Here, the training dataset is randomly split into smaller subsets, i.e. mini-batches, which are used individually per Q-FLAIR iteration. This approach reduces computational effort when evaluating the (empirical) loss function, albeit at the cost of a less accurate (higher variance) estimate of the true loss function. For the QNN experiments, the mini-batch size is 200 for \textit{MNIST 28 $\times$ 28} and 64 for all other datasets. The test dataset is not mini-batched but limited to the first 200 samples. For the QSVM experiments, the mini-batch size is generally 32 (mini-kernel of size $32\times32$) and 100 test data points (mini-test-kernel of size $32 \times 100$).
        Complementary to reducing quantum circuit executions by limiting the number of reconstructions to the mini-batch size, quantum resources are further saved by limiting how gate candidates are probed on new qubits. Specifically, for single-qubit gate candidates, only a single new qubit (added at the bottom of the quantum circuit) is assessed. For two-qubit gates, only two qubits (one at the top and one at the bottom of the circuit) are considered to form linear entanglement. This selective probing strategy significantly reduces the combinatorial overhead of gate placement.
        Because the gate pool for the QNN in Eq.~\eqref{eq:gateSetQNN} only contains basic operations, it is not changed for the experiments. The gate pool for the QSVM is slightly modified to further improve the runtime on the quantum hardware. It consist of
        \begin{equation}
            \begin{aligned}
            \mathcal{V} = \lbrace
                &R_{x}(\theta, x_k), R_{y}(\theta, x_k), R_{z}(\theta, x_k),\\
                &R_{xx}(\theta, x_k), R_{yy}(\theta, x_k),R_{zz}(\theta, x_k)
            \rbrace
            .
            \end{aligned}
        \end{equation}
        Recall that only a single reconstruction must be retrieved to probe the weight-data-dependent and purely weight-dependent variant of the same rotation gate.

        Each model in the IBM benchmarks was trained on a single hardware instance per dataset to guarantee the same device-specific noise characteristics and allow the optimization to tailor the quantum feature-maps to this noise specifically. While as many iterations of the algorithm as possible were performed within a single IBM runtime session, multiple sessions were needed for the number of iterations reported here. An IBM runtime session guarantees exclusive and uninterrupted access to a single hardware instance to iteratively execute quantum circuits for up to eight hours. Typically, around 10-15 Q-FLAIR iterations could be executed in a single session for the QNN model, which was only about five iterations due to the more computationally demanding kernel matrix evaluations in the QSVM case. Between sessions, it is not guaranteed that model performance will be maintained, as other computations or calibrations may run on the hardware instance, altering the noise characteristics.

        In order to mitigate hardware errors, matrix-free measurement mitigation (M3) is employed, a prominent error mitigation technique compatible with the computational basis measurement sampler in the IBM runtime \cite{nation_ScalableMitigationMeasurement_2021}. The quasi-probability distribution as a result of the mitigation is mapped to the closest probability distribution under the $L2$-norm \cite{nation_ScalableMitigationMeasurement_2021}. However, the effects of M3 were found to be negligible as they did not improve over the raw measurements to any extent that would have a noticeable impact on the model output and performance. On the other hand, the calibration (performed at the beginning of each Q-FLAIR iteration) only took a few seconds. The M3 calibration was performed using the efficient \texttt{balanced} method restricted to the physical qubits that were incorporated by the quantum feature-map candidate circuits over the gate pool in the current iteration.

        For the hyperparameter tuning in the QSVM models (regularization strength $C$), ten repetitions of the 5-fold cross-validation are now performed on the mini-batches, to improve the stability of this tuning as the (balanced) accuracy estimates on folds of the mini-batches exhibit a higher variance than on the complete training dataset.

    \subsection{Classical surrogate benchmark}\label{app:classical_surrogate_bench}

        The classical surrogate benchmark, conducted in Sec.~\ref{sec:classical_surrogate_bench}, was established in Ref.~\cite{schreiber_ClassicalSurrogatesQuantum_2023}. After a recap of the classical modeling and training procedures, we describe the key modifications necessary to accommodate the Q-FLAIR framework.

        As covered and leveraged in this work, the outputs of a quantum model (as the expectation values of an observable after executing a quantum circuit) as a function of a single rotation gate angle $\alpha$ are simple sine curves (Eq.~\eqref{eq:analytic_form_rotation_gate}). Hence, such functions have a single (unit) frequency. As per the gate definition in Eq.~\eqref{eq:def_rotation_gate}, this functional form stems from the gate generators $A$ being involutory ($A^2=I$). 
        A more generalized view is obtained through the eigenvalues $\sigma(A)$ of the generator $A$. The involution property then follows as a special case of eigenvalues being exclusively $\pm 1$. In this generalization, the functional form extends to trigonometric functions in $\alpha$ with frequencies defined by the generator eigenvalue spectra $\sigma(A)$.

        When providing data $\bm{x}$ through quantum gates encoding individual features $x_k$, the quantum model can therefore be described as a multivariate trigonometric function of $\bm{x}$, also known as a truncated Fourier series 
        \begin{equation}\label{eq:surrogate_model}
            s(\bm{x}) = a_0 + 
            \sum_{\bm{\omega} \in \Omega_+} 
            a_{\bm{\omega}} \sin\left(\bm{\omega}^\top \bm{x}\right)
            + b_{\bm{\omega}} \cos\left(\bm{\omega}^\top \bm{x}\right)
            ,
        \end{equation}
        where $\Omega_+$ contains the frequency vectors induced by the gate generators \cite{schreiber_ClassicalSurrogatesQuantum_2023}.
        This full multivariate frequency spectrum is precisely composed from gate frequency spectra of both repeated encodings of the same feature $x_k$ (re-uploading) and encodings of different features via Minkowski sums 
        \begin{equation}\label{eq:classical_surrogate_frequencies}
            \Omega
            =
            \sum_{k=1}^d 
            \sum_{\ell=1}^{L_k}
            \left\{
            \left(\lambda_i-\lambda_j\right) \boldsymbol{e}_{k}
            \mid \lambda_i, \lambda_j \in \sigma\left(A_\ell^{(k)}\right)
            \right\}
            ,
        \end{equation}
        where the $k$-th feature $x_k$ is encoded via the $L_k$ gates with generators $A_\ell^{(k)}$ \cite{caro_EncodingdependentGeneralizationBounds_2021}.
        $\Omega_+$ in Eq.~\eqref{eq:surrogate_model} constitutes a positive half-set of the full frequency spectrum $\Omega$, such that $\Omega = \Omega_+ \cup (-\Omega_+) \cup \{0\}$ and $\Omega_+ \cap (-\Omega_+) = \varnothing$. \cite{schreiber_ClassicalSurrogatesQuantum_2023}.
        It is important to emphasize that these classical surrogate models $s(\bm{x})$ are functions of all data features $x_k$ simultaneously and thus characterize the functional form of the full quantum model, not only the local effect of a single gate parameter as exploited in Q-FLAIR. Furthermore, the coefficients $\left\{ a_0 \right\} \cup \left\{a_{\bm{\omega}}, b_{\bm{\omega}} \; \middle| \; \bm{\omega} \in \Omega_+ \right\}$ is referred to as the Fourier coefficients and parameters of the classical surrogate model interchangeably with short-hand notation $\bm{a}, \bm{b}$. Clearly, the number of Fourier coefficients relates to the number of frequencies as $\left\lvert \Omega \right\rvert = 2\left\lvert \Omega_+ \right\rvert + 1$. As discussed in the main text, the parameter count of the surrogate models scales exponentially with the number of features as well as the number of encodings, which becomes clear from the two Minkowski sums in Eq.~\eqref{eq:classical_surrogate_frequencies}, constructing the full multivariate frequency spectrum.

        Reference~\cite{schreiber_ClassicalSurrogatesQuantum_2023} %
        propose the classical surrogate benchmark to assess the quality of the learned quantum model by comparing it to the classical model with its Fourier coefficients being directly fit to the training dataset instead. Conveniently, the classical surrogate model is linear in its Fourier coefficients, such that the optimal coefficients are given by the least-squares solution, which guarantees global optimality due to convexity.
        Hence, the classical surrogate model can equivalently be treated as a classical neural network with a single (fully-connected) layer (without activation function) on Fourier features $\left\{\sin\left(a_{\bm{\omega}}^\top x_k\right), \cos\left(b_{\bm{\omega}}^\top x_k\right) \; \middle| \; \bm{\omega} \in \Omega_+ , k \in \left\lbrace 1, \ldots, d\right\rbrace \right\}$.
        For large datasets and numerical stability, a stochastic gradient descent approach that samples random training batches is recommended \cite{schreiber_ClassicalSurrogatesQuantum_2023}. 

        To adapt the classical surrogate benchmark proposed by Schreiber et al.~\cite{schreiber_ClassicalSurrogatesQuantum_2023} to the Q-FLAIR framework, we introduce three careful but necessary modifications.
        First, while the original benchmark was designed for fixed-ansatz models, Q-FLAIR constructs the ansatz adaptively. Because the frequency spectrum is intrinsic to the specific gate selection (including feature and weight choices), we extend the benchmark procedure to construct a new classical surrogate model that matches the current ansatz at each iteration of the Q-FLAIR algorithm and is fully trained on the training data. 
        This approach not only offers granular assessment and insight into the progression of the algorithm's classicality, but it is also essential for computational tractability, as the spectral complexity of the final converged Q-FLAIR model typically exceeds the capacity of a tractable classical surrogate.

        Second, Q-FLAIR utilizes gates with an additional weight parameter $\theta \in [-1, 1]$ (see Eq.~\eqref{eq:def_parameter_data_gate}). These weights directly scale the generator eigenvalues, resulting in non-integer and generally incommensurate frequencies (e.g., eigenvalues become $\pm \theta$ rather than $\pm 1$). While the original benchmark \cite{schreiber_ClassicalSurrogatesQuantum_2023} focused on integer spectra, the framework remains valid for these generalized Fourier series~\cite{caro_EncodingdependentGeneralizationBounds_2021} and was already formulated as such in Eqs.~\eqref{eq:surrogate_model} and \eqref{eq:classical_surrogate_frequencies}. However, a critical consequence of incommensurability is the absence of frequency cancellations. In the fixed integer frequency case~\cite{schreiber_ClassicalSurrogatesQuantum_2023}, repeated encodings of the same feature $x_k$ result in a basis size that scales linearly with repetitions $L_k$ due to cancellations in the inner Minkowski sum in Eq.~\eqref{eq:classical_surrogate_frequencies}. In contrast, the distinct, real-valued weights in Q-FLAIR generally cause the number of unique frequencies per feature $x_k$ to grow exponentially with the number of encoding gates $L_k$.

        Finally, we adapt the loss function for binary classification tasks where $y_i \in \{-1, +1\}$. In contrast to Ref.~\cite{schreiber_ClassicalSurrogatesQuantum_2023}, which employs a squared loss for regression on continuous labels, we utilize the negative log-likelihood loss to train the classical surrogate for binary classification labels $y_i \in \{-1, +1\}$
        \begin{equation}
        \minimize_{\bm{a}, \bm{b}}
        \sum_{j=1}^{M} \log\left( 1 + \exp\left( -y_i s_{\bm{a}, \bm{b}}(\bm{x}_i) \right) \right)
        .
        \end{equation}
        Beyond being the preferred loss for discrete classification tasks~\cite{bishop_PatternRecognitionMachine_2006}, this loss aligns with the Q-FLAIR log loss (Eq.~\eqref{eq:QNNlog}) and recasts surrogate training as a logistic regression problem. Crucially, because the classical surrogate remains a linear model in the Fourier feature space defined by the frequency spectrum, this linear logistic regression is convex with respect to the trainable Fourier coefficients~\cite{bishop_PatternRecognitionMachine_2006}. Therefore, surrogate training identifies the globally optimal trigonometric function that matches the Q-FLAIR gate choice for the given iteration.

        Alternative approaches for adapting the classical surrogate benchmark of Q-FLAIR are discussed in Sec.~\ref{sec:discussion}.

\section{Additional experimental studies}
    \subsection{Expressivity-trainability benchmark with fixed ansätze} \label{sec:AccExpressZZFeature}

    The fixed-ansatz second-order Pauli-$Z$ (or $ZZ$) feature-map~\cite{havlicek_SupervisedLearningQuantumenhanced_2019} is an established choice due to its difficulty to classically simulate~\cite{havlicek_SupervisedLearningQuantumenhanced_2019}. However, as most fixed-ansatz feature-maps, the $ZZ$ feature-map frequently suffers from trainability issues~\cite{mcclean_BarrenPlateausQuantum_2018,holmes_ConnectingAnsatzExpressibility_2022,thanasilp_ExponentialConcentrationQuantum_2024}. As the data dimensionality $d$ increases, the expressivity of these fixed ansätze grows rapidly, leading directly to these expressivity-trainability issues. The quantum Fisher information matrix (QFIM) serves as a critical diagnostic tool in this context~\cite{barzen2025differential,yao2025learning}: a saturated, full-rank QFIM indicates an over-expressive model whose outputs become indistinguishable from random noise, rendering the circuit untrainable. We empirically verify that Q-FLAIR avoids these expressivity-induced trainability issues. 
    As an adaptive approach, Q-FLAIR expands circuit expressivity only as mandated by the dataset. This controlled growth is reflected in the moderate scaling of the QFIM rank, aligning with the lower \emph{effective} dimension (excluding redundant and correlated features) rather than blindly scaling with the data dimension $d$.

    The down-scaled MNIST datasets, evaluated in Sec.~\ref{ssec:featureDim} to study Q-FLAIR performance and resource dependence on the data dimension $d$, provide an ideal testbed to isolate this effect. Varying the image resolution scales the raw data dimension $d$ while maintaining the underlying image semantics (and thus, the effective dimension). Here, we compare Q-FLAIR to the $ZZ$ feature-map for short, implemented via the \texttt{ZZFeatureMap} class provided by Qiskit~\cite{qiskit2024,qiskit_zenodo}. While the accuracy of models may be obtained again from Eq.~\eqref{eq:averAcc}, quantifying differences in the expressivity of the models is more challenging. Recently, new methods have been proposed that use measures to determine the difference between distributions to quantify the expressivity \cite{monbroussou2025trainability}. 
    
    We utilize the data QFIM and take its rank as a measure of the feature-map expressivity~\cite{haug_GeneralizationQuantumMachine_2024,barzen2025differential,yao2025learning}. This rank corresponds to the number of independent data-embedding directions in the quantum state, thereby revealing which input features are strictly necessary and which transport redundant or irrelevant information. For a quantum feature-map preparing a pure state $\ket{\psi(\bm{x})}$ parameterized by a $d$-dimensional data point $\bm{x}$, the QFIM is a $d \times d$ matrix $F(\bm{x})$, defined as
    \begin{equation}
    \begin{aligned}
        F_{kl}(\bm{x}) = 
        4\Re \big[&\braket{\partial_k\psi(\bm{x})|\partial_l\psi(\bm{x})} \\
        &-\braket{\partial_k\psi(\bm{x})|\psi(\bm{x})}\braket{\psi(\bm{x})|\partial_l\psi(\bm{x})}\big],
    \end{aligned}
    \end{equation}
    with the $k$-th partial derivative with respect to the feature component $x_k$ denoted as $\ket{\partial_k\psi(\bm{x})}= \partial\ket{\psi(\bm{x})}/\partial x_k$~\cite{meyer_FisherInformationNoisy_2021,braunstein_StatisticalDistanceGeometry_1994}. 
    Rather than relying on the local geometry at a single data point, we capture the global geometry over the full data distribution by evaluating the expected data QFIM $\mathcal{F} = \mathbb{E}_{\bm{x}}\left[ F(\bm{x}) \right]$. In practice, this is estimated via the empirical data QFIM averaged over a finite sample set $\lbrace \bm{x}_i\rbrace_{i=1}^N$~\cite{haug_GeneralizationQuantumMachine_2024}, given by
    \begin{equation}
        F = \frac{1}{N} \sum_{i=1}^N F(\bm{x}_i) \approx \mathcal{F}
        .
    \end{equation}
    For the sake of brevity, we refer to this empirical expected data QFIM simply as the QFIM, denoted as $F$, henceforth.

    We have trained QSVMs using both, Q-FLAIR and $ZZ$ feature-map, on the MNIST datasets with pixel resolutions of $7\times7$, $14\times14$, and $28\times28$. For better data compatibility, the $ZZ$ feature-map is folded onto seven qubits, thus containing some structure of the data by mapping pixels of one image row on the same qubit, since the data dimensions are multiples of seven. The number of qubits is for Q-FLAIR restricted to five qubits. QSVMs are preferred over QNNs to strictly assess the intrinsic quality of fixed non-parametric feature-maps, such as the $ZZ$ feature-map. QSVMs are guaranteed to find the optimal separating hyperplane, whereas training an additional circuit or observable in QNNs risks sub-optimal separation that could obscure the true effectiveness of the feature-map \cite{schuld2021supervisedquantummachinelearning}.
    
\begin{figure}
        \centering
        \includegraphics[width=\linewidth]{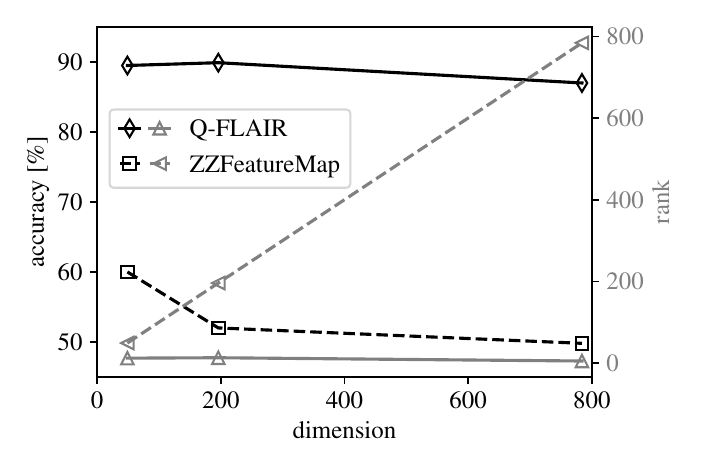}
        \caption{Classification performance and expressivity scaling with data dimension. Comparison of QSVM accuracies (black, left axis) and data QFIM ranks (gray, right axis) using the adaptive Q-FLAIR feature-map (solid lines) versus the fixed $ZZ$ feature-map (dashed lines). Models are evaluated on \textit{MNIST} dataset variants at $7 \times 7$, $14 \times 14$, and $28 \times 28$ pixel resolutions, corresponding to increasing data dimension (horizontal axis). Connecting lines are added to guide the eye.}
        \label{fig:Q-FLAIR_ZZfeatureMap}
    \end{figure}

    While the number of gates scales in the quantum circuit of the $ZZ$ feature-map with the number of features, those obtained by Q-FLAIR consist of \num{15}, respectively \num{14} data-dependent gates. Nevertheless, test datasets are classified by Q-FLAIR with an accuracy of close to \SI{90}{\percent}, independent of the dimension (see Fig.~\ref{fig:Q-FLAIR_ZZfeatureMap}).
    The predictions with the $ZZ$ feature-map reach only an accuracy of up to \SI{60}{\percent} first and decreases for higher dimensions. For the full $28\times 28$ case, there is no significant learning effect left. This difference in prediction accuracy indicates that by adapting its ansatz to each specific dataset, Q-FLAIR successfully encodes more task-relevant properties than the fixed $ZZ$ feature-map.

    That additionally the expressivity is larger can be seen from the rank of the QFIM: The maximum achievable rank of the QFIM is strictly bounded by the data dimension $d$~\cite{yao2025learning}. For the MNIST datasets evaluated here, this corresponds to theoretical upper limits of $d = 49$, $196$, and $784$ for the $7 \times 7$, $14 \times 14$, and full $28 \times 28$ pixel resolutions, respectively. Because the fixed $ZZ$ feature-map indiscriminately embeds every input feature, its QFIM rank can theoretically reach this absolute limit. In contrast, Q-FLAIR selectively incorporates only task-relevant features; therefore, its maximum attainable rank is naturally constrained by the size of this chosen feature subset rather than the full data dimension. The 15 data-dependent gates that form the Q-FLAIR circuit for the $7\times7$ depend on 12 unique features. The circuit for $14\times14$ contains 13 distinct features, and the full-resolution MNIST is described by a circuit consisting of five gates, each of which depends on a unique feature. In all three cases, the number of unique features is equivalent to the QFIM rank. This counterintuitive decrease in both the gate count and the QFIM rank at higher resolutions occurs because a sparser subset of pixels is sufficient for digit discrimination when individual pixels resolve finer structural details. In contrast, the QFIM rank of the fixed $ZZ$ feature-map increases drastically from the $14 \times 14$ to the $28 \times 28$ variant. This stark divergence explicitly validates our central claim: Simply increasing the effective dimension of the feature-map does not drive the classification performance. Instead, by adaptively tailoring the ansatz to the dataset, Q-FLAIR increases its expressivity only as required, thereby promoting generalization and averting the pitfalls of over-expressiveness and the accompanying trainability issues. 

    Beyond the mere QFIM rank, a relaxation to large vs small over non-zero vs zero eigenvalues may be more suitable as the large eigenvalues determine how many data dimensions are \emph{effectively} present in the feature-map.
    Furthermore, due to the limited precision inherent in evaluating the QFIM from numerical derivatives (finite differences) averaged over the training dataset \cite{abbas2021effectivedimensionmachinelearning}, we obtain, for all investigated cases, ranks that equal the upper limit imposed by the number of unique features encoded in the circuit.
    If we assume a (rather conservative) numerical uncertainty threshold of \num{1e-3} on the eigenvalues, we obtain a reduction to an effective rank of 674 for the $28\times28$ $ZZ$ feature-map, while the number of non-zero eigenvalues of the two lower resolution variants does not change. For Q-FLAIR, the ranks are slightly reduced in the two higher-resolution cases: 10 for $14\times14$ and 3 for $28\times28$. Nevertheless, the overall picture of the lower expressivity of $ZZ$ feature-map relative to Q-FLAIR is preserved.
    Moving beyond simple rank analysis and eigenvalue thresholding, more sophisticated metrics can be derived from QFIMs to quantify the expressivity and effective dimension of quantum feature-maps more robustly \cite{Abbas2021power,berezniuk2020scaledependentnotioneffectivedimension}.

    \subsection{Effect of parameter re-optimization}\label{app:re-optimization}
    The number of evaluations in Q-FLAIR is kept low, because every parameter gets optimized only once. This leads to local optima. By re-optimizing the parameters, in theory better results should be achievable. The re-optimization of parameters is done for the QNN of the \textit{MNIST PCA} dataset. With the limited-memory Broyden-Fletcher-Goldfarb-Shanno bounded (L-BFGS-B) algorithm \cite{byrd1995limited} and Constrained Optimization by Linear Approximation (COBYLA) \cite{powell1998direct}, two well-established optimization methods are used in the way in which they are implemented in scipy \cite{2020SciPy-NMeth}. These two methods were chosen, because they are able to respect finite definition ranges of the variables and can thus handle the restriction on the parameter space to $\theta\in(0,\pi)$.
    
\begin{figure}%
        \centering
        \includegraphics[width=\linewidth]{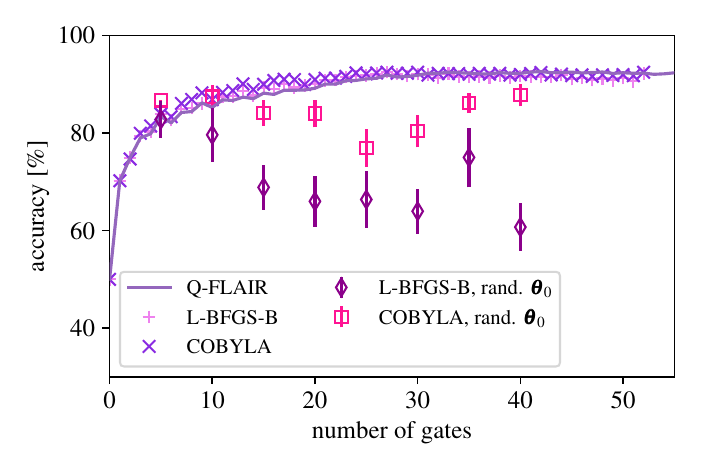}
        \caption{Comparison between different post-optimization methods. The purple curve is the native result for the \textit{MNIST PCA} testing dataset, directly obtained from the trained circuit. The `$+$' and `$\times$' are the results if the parameters in the circuit are optimized after the training with two different optimization methods (L-BFGS-B and COBYLA). In these optimizations, the optimal parameters are taken as starting points. The brown squares are the results, again for the COBYLA optimization method, but this time with random angles $\theta_0$ as initial configuration. The coral data points are the same for L-BFGS-B. The uncertainties are determined from three independent optimizations with up to \num{1000} function evaluations. 
        }
        \label{fig:ann_opt}
    \end{figure}

    In an initial attempt to increase the accuracy, the fully trained circuit is optimized with the iteratively determined parameters as starting values. The resulting mean accuracies are depicted in Fig.~\ref{fig:ann_opt} as $+$ and $\times$.
    While initially a small enhancement of the accuracy can be observed, this vanishes for larger circuits, resulting in the same performance as the circuit with no re-optimization. The behavior does not change between the two methods, which indicates that the local minima are rather stable.

    In a second optimization scheme, the starting parameters are chosen randomly, leading to initial configurations far from a minimum. This is repeated ten times with different initial parameters and the resulting accuracies are averaged. From the distribution of the results, the standard deviations are calculated. The results of these optimizations can as well be found in Fig.~\ref{fig:ann_opt}. In shallow circuits, with only up to 10 gates, it is possible to obtain slightly higher accuracies with these re-optimizations than without the additional optimization step. In none of the deeper circuits do we reach the same average accuracy through re-optimization from random starting parameters as with the iterative optimization. Throughout the optimizations, the maximally allowed number of steps towards the minimum is restricted to \num{1000}. This constraint was set in order to limit the number of explicit evaluations of the quantum circuit. COBYLA did not succeed in fully converging for circuits with more than \num{15} gates, while the optimization with L-BFGS-B was reported as converged for all investigated circuit sizes. Interestingly, the optimizations with COBYLA nevertheless outperform the L-BFGS-B results in every instance. The lowest accuracy with COBYLA is obtained for circuits with \num{25} gates. With the increasing number of parameters, it gets more difficult to find a minimum of the loss function, which leads to similar accuracy as the iterative method. By further increasing the number of gates, the accuracy increases again and approaches our previous results. This indicates that the circuit with approximately \num{40} gates has enough parameters, such that the non-uniqueness of the parameter combination allows for a high enough probability to end up in a favorable minimum. For L-BFGS-B, we do not see again such a systematic increase of the accuracy. Instead the accuracy decreases for larger gates, with the circuit consisting of \num{35} being an outlier due to statistic fluctuations. L-BFGS-B is thus more efficient in finding a minimum of the objective function, but, due to its gradient-based local search, struggles more than COBYLA in reaching minima with low values since it is incapable of reaching any minimum but one that is close to the initialization.

    The four optimizations demonstrate overall that the iterative local optimization due to the adaptive ansatz is very efficient. It leads maybe not to the global optimum, but achieves better accuracies than optimizations from random starting-parameters, which fail to reach parameter configurations that lead to similar accuracies. Also, local optimizations, starting from the parameters determined by Q-FLAIR, do not yield significantly better results. Overall, Q-FLAIR facilitates the navigation of these quantum-model-loss landscapes by optimizing parameters within an adaptive ansatz. Established optimizers, when randomly initialized in a fixed ansatz, settle in worse local optima.
    
    \subsection{Accuracy comparison of QSVM and QNN}\label{app:accuracy_depth_comp}
\begin{figure}
        \centering
        \includegraphics[width=\linewidth]{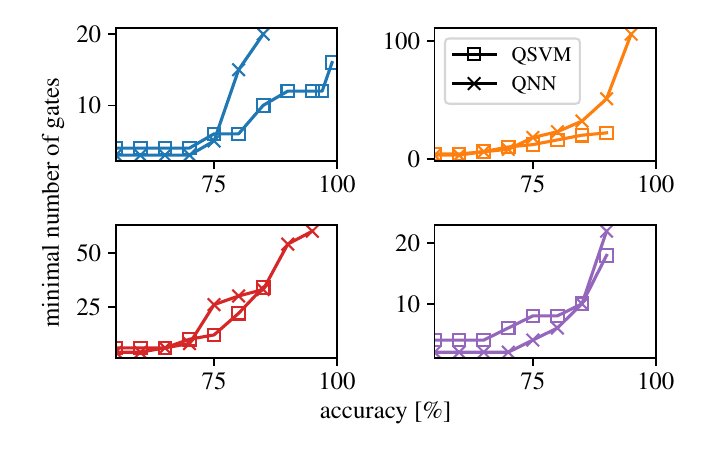}
        \caption{Minimal number of gates needed to reach an average accuracy. The number of gates for the QSVM are, in comparison with Fig.~\ref{fig:ta_acc}, multiplied by two. The colors indicate the datasets and correspond with those in Fig.~\ref{fig:ann_acc}.}
        \label{fig:minGateNeeded}
    \end{figure}

    In section~\ref{ssec:performacne_qsvm_qnn}, we concentrated mostly on the maximally achievable accuracy prior to the saturation of the loss. While this is an important quantity, it is easier to compare the performance of different models if the minimal circuit size for a given accuracy is considered. Such a comparison for the QSVMs and QNNs of \textit{two-curves}, \textit{linearly separable}, \textit{bars \& stripes}, and \textit{MNIST PCA} is depicted in Fig.~\ref{fig:minGateNeeded}.
    The number of gates in the circuit for the QNNs are counted in the same way as e.g. in Fig.~\ref{fig:ann_acc}, because this correlates directly with the number of operations that have to be calculated in Eq.~\eqref{eq:qnn_output}. For the QSVMs, we double the number of gates from Fig.~\ref{fig:ta_acc}. In this way, the two states that have to be measured for the kernel entry in Eq.~\eqref{eq:kernel} are considered.
    
    In the resulting data, we see that both models require in most cases a similar number of gates for the same accuracy. This is especially true for accuracies of up to \SI{75}{\percent}. The QNNs have a slight tendency to larger numbers of gates. While this may be a disadvantage for the usage on NISQ devices, it allows in most cases for higher accuracies, because neither the loss, nor the accuracy saturate fast. It seems thus a matter of the application, as well as the available quantum computer, which of the two models is preferable: with the exception of datasets with a similar structure to \textit{two-curves}, QNNs lead to a higher accuracy. At the same time, rather high accuracies are reached by the QSVMs with fewer gates. Overall, the two models are very similar and mediocre results can already be obtained with very few gates. The overall shallowness of the circuits is possible due to the efficiency in selecting features: for the QSVM of \textit{bars \& stripes}, this results for example in the neglect of half of the features, indicating that all the others contain redundant information.

\FloatBarrier
\bibliography{references_local}%

\begin{thebibliography}{126}%
\makeatletter
\providecommand \@ifxundefined [1]{%
 \@ifx{#1\undefined}
}%
\providecommand \@ifnum [1]{%
 \ifnum #1\expandafter \@firstoftwo
 \else \expandafter \@secondoftwo
 \fi
}%
\providecommand \@ifx [1]{%
 \ifx #1\expandafter \@firstoftwo
 \else \expandafter \@secondoftwo
 \fi
}%
\providecommand \natexlab [1]{#1}%
\providecommand \enquote  [1]{``#1''}%
\providecommand \bibnamefont  [1]{#1}%
\providecommand \bibfnamefont [1]{#1}%
\providecommand \citenamefont [1]{#1}%
\providecommand \href@noop [0]{\@secondoftwo}%
\providecommand \href [0]{\begingroup \@sanitize@url \@href}%
\providecommand \@href[1]{\@@startlink{#1}\@@href}%
\providecommand \@@href[1]{\endgroup#1\@@endlink}%
\providecommand \@sanitize@url [0]{\catcode `\\12\catcode `\$12\catcode `\&12\catcode `\#12\catcode `\^12\catcode `\_12\catcode `\%12\relax}%
\providecommand \@@startlink[1]{}%
\providecommand \@@endlink[0]{}%
\providecommand \url  [0]{\begingroup\@sanitize@url \@url }%
\providecommand \@url [1]{\endgroup\@href {#1}{\urlprefix }}%
\providecommand \urlprefix  [0]{URL }%
\providecommand \Eprint [0]{\href }%
\providecommand \doibase [0]{https://doi.org/}%
\providecommand \selectlanguage [0]{\@gobble}%
\providecommand \bibinfo  [0]{\@secondoftwo}%
\providecommand \bibfield  [0]{\@secondoftwo}%
\providecommand \translation [1]{[#1]}%
\providecommand \BibitemOpen [0]{}%
\providecommand \bibitemStop [0]{}%
\providecommand \bibitemNoStop [0]{.\EOS\space}%
\providecommand \EOS [0]{\spacefactor3000\relax}%
\providecommand \BibitemShut  [1]{\csname bibitem#1\endcsname}%
\let\auto@bib@innerbib\@empty
\bibitem [{\citenamefont {Biamonte}\ \emph {et~al.}(2017)\citenamefont {Biamonte}, \citenamefont {Wittek}, \citenamefont {Pancotti}, \citenamefont {Rebentrost}, \citenamefont {Wiebe},\ and\ \citenamefont {Lloyd}}]{biamonte_QuantumMachineLearning_2017}%
  \BibitemOpen
  \bibfield  {author} {\bibinfo {author} {\bibfnamefont {J.}~\bibnamefont {Biamonte}}, \bibinfo {author} {\bibfnamefont {P.}~\bibnamefont {Wittek}}, \bibinfo {author} {\bibfnamefont {N.}~\bibnamefont {Pancotti}}, \bibinfo {author} {\bibfnamefont {P.}~\bibnamefont {Rebentrost}}, \bibinfo {author} {\bibfnamefont {N.}~\bibnamefont {Wiebe}},\ and\ \bibinfo {author} {\bibfnamefont {S.}~\bibnamefont {Lloyd}},\ }\bibfield  {title} {{\selectlanguage {en}\bibinfo {title} {Quantum machine learning}},\ }\href {https://doi.org/10.1038/nature23474} {\bibfield  {journal} {\bibinfo  {journal} {Nature}\ }\textbf {\bibinfo {volume} {549}},\ \bibinfo {pages} {195} (\bibinfo {year} {2017})}\BibitemShut {NoStop}%
\bibitem [{\citenamefont {Dunjko}\ and\ \citenamefont {Briegel}(2018)}]{dunjko_MachineLearningArtificial_2018}%
  \BibitemOpen
  \bibfield  {author} {\bibinfo {author} {\bibfnamefont {V.}~\bibnamefont {Dunjko}}\ and\ \bibinfo {author} {\bibfnamefont {H.~J.}\ \bibnamefont {Briegel}},\ }\bibfield  {title} {{\selectlanguage {en}\bibinfo {title} {Machine learning \& artificial intelligence in the quantum domain: a review of recent progress}},\ }\href {https://doi.org/10.1088/1361-6633/aab406} {\bibfield  {journal} {\bibinfo  {journal} {Reports on Progress in Physics}\ }\textbf {\bibinfo {volume} {81}},\ \bibinfo {pages} {074001} (\bibinfo {year} {2018})}\BibitemShut {NoStop}%
\bibitem [{\citenamefont {Cerezo}\ \emph {et~al.}(2022)\citenamefont {Cerezo}, \citenamefont {Verdon}, \citenamefont {Huang}, \citenamefont {Cincio},\ and\ \citenamefont {Coles}}]{cerezo_ChallengesOpportunitiesQuantum_2022}%
  \BibitemOpen
  \bibfield  {author} {\bibinfo {author} {\bibfnamefont {M.}~\bibnamefont {Cerezo}}, \bibinfo {author} {\bibfnamefont {G.}~\bibnamefont {Verdon}}, \bibinfo {author} {\bibfnamefont {H.-Y.}\ \bibnamefont {Huang}}, \bibinfo {author} {\bibfnamefont {L.}~\bibnamefont {Cincio}},\ and\ \bibinfo {author} {\bibfnamefont {P.~J.}\ \bibnamefont {Coles}},\ }\bibfield  {title} {{\selectlanguage {en}\bibinfo {title} {Challenges and opportunities in quantum machine learning}},\ }\href {https://doi.org/10.1038/s43588-022-00311-3} {\bibfield  {journal} {\bibinfo  {journal} {Nature Computational Science}\ }\textbf {\bibinfo {volume} {2}},\ \bibinfo {pages} {567} (\bibinfo {year} {2022})}\BibitemShut {NoStop}%
\bibitem [{\citenamefont {Havlíček}\ \emph {et~al.}(2019)\citenamefont {Havlíček}, \citenamefont {Córcoles}, \citenamefont {Temme}, \citenamefont {Harrow}, \citenamefont {Kandala}, \citenamefont {Chow},\ and\ \citenamefont {Gambetta}}]{havlicek_SupervisedLearningQuantumenhanced_2019}%
  \BibitemOpen
  \bibfield  {author} {\bibinfo {author} {\bibfnamefont {V.}~\bibnamefont {Havlíček}}, \bibinfo {author} {\bibfnamefont {A.~D.}\ \bibnamefont {Córcoles}}, \bibinfo {author} {\bibfnamefont {K.}~\bibnamefont {Temme}}, \bibinfo {author} {\bibfnamefont {A.~W.}\ \bibnamefont {Harrow}}, \bibinfo {author} {\bibfnamefont {A.}~\bibnamefont {Kandala}}, \bibinfo {author} {\bibfnamefont {J.~M.}\ \bibnamefont {Chow}},\ and\ \bibinfo {author} {\bibfnamefont {J.~M.}\ \bibnamefont {Gambetta}},\ }\bibfield  {title} {\bibinfo {title} {Supervised learning with quantum-enhanced feature spaces},\ }\href {https://doi.org/10.1038/s41586-019-0980-2} {\bibfield  {journal} {\bibinfo  {journal} {Nature}\ }\textbf {\bibinfo {volume} {567}},\ \bibinfo {pages} {209} (\bibinfo {year} {2019})}\BibitemShut {NoStop}%
\bibitem [{\citenamefont {Schuld}\ and\ \citenamefont {Killoran}(2019)}]{schuld_QuantumMachineLearning_2019}%
  \BibitemOpen
  \bibfield  {author} {\bibinfo {author} {\bibfnamefont {M.}~\bibnamefont {Schuld}}\ and\ \bibinfo {author} {\bibfnamefont {N.}~\bibnamefont {Killoran}},\ }\bibfield  {title} {\bibinfo {title} {Quantum {Machine} {Learning} in {Feature} {Hilbert} {Spaces}},\ }\href {https://doi.org/10.1103/PhysRevLett.122.040504} {\bibfield  {journal} {\bibinfo  {journal} {Physical Review Letters}\ }\textbf {\bibinfo {volume} {122}},\ \bibinfo {pages} {040504} (\bibinfo {year} {2019})}\BibitemShut {NoStop}%
\bibitem [{\citenamefont {Huang}\ \emph {et~al.}(2021{\natexlab{a}})\citenamefont {Huang}, \citenamefont {Wang}, \citenamefont {Song}, \citenamefont {Xu}, \citenamefont {Li}, \citenamefont {Wang}, \citenamefont {Guo}, \citenamefont {Song}, \citenamefont {Liu}, \citenamefont {Zheng} \emph {et~al.}}]{huang2021quantum}%
  \BibitemOpen
  \bibfield  {author} {\bibinfo {author} {\bibfnamefont {K.}~\bibnamefont {Huang}}, \bibinfo {author} {\bibfnamefont {Z.-A.}\ \bibnamefont {Wang}}, \bibinfo {author} {\bibfnamefont {C.}~\bibnamefont {Song}}, \bibinfo {author} {\bibfnamefont {K.}~\bibnamefont {Xu}}, \bibinfo {author} {\bibfnamefont {H.}~\bibnamefont {Li}}, \bibinfo {author} {\bibfnamefont {Z.}~\bibnamefont {Wang}}, \bibinfo {author} {\bibfnamefont {Q.}~\bibnamefont {Guo}}, \bibinfo {author} {\bibfnamefont {Z.}~\bibnamefont {Song}}, \bibinfo {author} {\bibfnamefont {Z.-B.}\ \bibnamefont {Liu}}, \bibinfo {author} {\bibfnamefont {D.}~\bibnamefont {Zheng}}, \emph {et~al.},\ }\bibfield  {title} {\bibinfo {title} {Quantum generative adversarial networks with multiple superconducting qubits},\ }\href {https://doi.org/10.1038/s41534-021-00503-1} {\bibfield  {journal} {\bibinfo  {journal} {npj Quantum Information}\ }\textbf {\bibinfo {volume} {7}},\ \bibinfo {pages} {165} (\bibinfo {year} {2021}{\natexlab{a}})}\BibitemShut {NoStop}%
\bibitem [{\citenamefont {Sakka}\ \emph {et~al.}(2025)\citenamefont {Sakka}, \citenamefont {Mitarai},\ and\ \citenamefont {Fujii}}]{sakka2025automating}%
  \BibitemOpen
  \bibfield  {author} {\bibinfo {author} {\bibfnamefont {K.}~\bibnamefont {Sakka}}, \bibinfo {author} {\bibfnamefont {K.}~\bibnamefont {Mitarai}},\ and\ \bibinfo {author} {\bibfnamefont {K.}~\bibnamefont {Fujii}},\ }\bibfield  {title} {\bibinfo {title} {Automating quantum feature map design via large language models},\ }\href {https://doi.org/10.48550/arXiv.2504.07396} {\bibfield  {journal} {\bibinfo  {journal} {arXiv preprint arXiv:2504.07396}\ } (\bibinfo {year} {2025})}\BibitemShut {NoStop}%
\bibitem [{\citenamefont {Rist{\`e}}\ \emph {et~al.}(2017)\citenamefont {Rist{\`e}}, \citenamefont {Da~Silva}, \citenamefont {Ryan}, \citenamefont {Cross}, \citenamefont {C{\'o}rcoles}, \citenamefont {Smolin}, \citenamefont {Gambetta}, \citenamefont {Chow},\ and\ \citenamefont {Johnson}}]{riste2017demonstration}%
  \BibitemOpen
  \bibfield  {author} {\bibinfo {author} {\bibfnamefont {D.}~\bibnamefont {Rist{\`e}}}, \bibinfo {author} {\bibfnamefont {M.~P.}\ \bibnamefont {Da~Silva}}, \bibinfo {author} {\bibfnamefont {C.~A.}\ \bibnamefont {Ryan}}, \bibinfo {author} {\bibfnamefont {A.~W.}\ \bibnamefont {Cross}}, \bibinfo {author} {\bibfnamefont {A.~D.}\ \bibnamefont {C{\'o}rcoles}}, \bibinfo {author} {\bibfnamefont {J.~A.}\ \bibnamefont {Smolin}}, \bibinfo {author} {\bibfnamefont {J.~M.}\ \bibnamefont {Gambetta}}, \bibinfo {author} {\bibfnamefont {J.~M.}\ \bibnamefont {Chow}},\ and\ \bibinfo {author} {\bibfnamefont {B.~R.}\ \bibnamefont {Johnson}},\ }\bibfield  {title} {\bibinfo {title} {Demonstration of quantum advantage in machine learning},\ }\href {https://doi.org/10.1038/s41534-017-0017-3} {\bibfield  {journal} {\bibinfo  {journal} {npj Quantum Information}\ }\textbf {\bibinfo {volume} {3}},\ \bibinfo {pages} {16} (\bibinfo {year} {2017})}\BibitemShut {NoStop}%
\bibitem [{\citenamefont {Cho}\ and\ \citenamefont {Kim}(2024)}]{cho2024machine}%
  \BibitemOpen
  \bibfield  {author} {\bibinfo {author} {\bibfnamefont {G.}~\bibnamefont {Cho}}\ and\ \bibinfo {author} {\bibfnamefont {D.}~\bibnamefont {Kim}},\ }\bibfield  {title} {\bibinfo {title} {Machine learning on quantum experimental data toward solving quantum many-body problems},\ }\href {https://doi.org/10.1038/s41467-024-51932-3} {\bibfield  {journal} {\bibinfo  {journal} {Nature Communications}\ }\textbf {\bibinfo {volume} {15}},\ \bibinfo {pages} {7552} (\bibinfo {year} {2024})}\BibitemShut {NoStop}%
\bibitem [{\citenamefont {Liu}\ \emph {et~al.}(2021)\citenamefont {Liu}, \citenamefont {Arunachalam},\ and\ \citenamefont {Temme}}]{liu_RigorousRobustQuantum_2021}%
  \BibitemOpen
  \bibfield  {author} {\bibinfo {author} {\bibfnamefont {Y.}~\bibnamefont {Liu}}, \bibinfo {author} {\bibfnamefont {S.}~\bibnamefont {Arunachalam}},\ and\ \bibinfo {author} {\bibfnamefont {K.}~\bibnamefont {Temme}},\ }\bibfield  {title} {\bibinfo {title} {A rigorous and robust quantum speed-up in supervised machine learning},\ }\href {https://doi.org/10.1038/s41567-021-01287-z} {\bibfield  {journal} {\bibinfo  {journal} {Nature Physics}\ }\textbf {\bibinfo {volume} {17}},\ \bibinfo {pages} {1013} (\bibinfo {year} {2021})}\BibitemShut {NoStop}%
\bibitem [{\citenamefont {Jäger}\ and\ \citenamefont {Krems}(2023)}]{jager_UniversalExpressivenessVariational_2023}%
  \BibitemOpen
  \bibfield  {author} {\bibinfo {author} {\bibfnamefont {J.}~\bibnamefont {Jäger}}\ and\ \bibinfo {author} {\bibfnamefont {R.~V.}\ \bibnamefont {Krems}},\ }\bibfield  {title} {{\selectlanguage {en}\bibinfo {title} {Universal expressiveness of variational quantum classifiers and quantum kernels for support vector machines}},\ }\href {https://doi.org/10.1038/s41467-023-36144-5} {\bibfield  {journal} {\bibinfo  {journal} {Nature Communications}\ }\textbf {\bibinfo {volume} {14}},\ \bibinfo {pages} {576} (\bibinfo {year} {2023})}\BibitemShut {NoStop}%
\bibitem [{\citenamefont {Chang}\ and\ \citenamefont {Cerezo}(2025)}]{chang_PrimerQuantumMachine_2025}%
  \BibitemOpen
  \bibfield  {author} {\bibinfo {author} {\bibfnamefont {S.~Y.}\ \bibnamefont {Chang}}\ and\ \bibinfo {author} {\bibfnamefont {M.}~\bibnamefont {Cerezo}},\ }\href {https://doi.org/10.48550/arXiv.2511.15969} {\bibinfo {title} {A {{Primer}} on {{Quantum Machine Learning}}}} (\bibinfo {year} {2025}),\ \Eprint {https://arxiv.org/abs/2511.15969} {2511.15969 [quant-ph]} \BibitemShut {NoStop}%
\bibitem [{\citenamefont {Pira}\ and\ \citenamefont {Rebentrost}(2026)}]{pira2026fundamentalsquantummachinelearning}%
  \BibitemOpen
  \bibfield  {author} {\bibinfo {author} {\bibfnamefont {L.}~\bibnamefont {Pira}}\ and\ \bibinfo {author} {\bibfnamefont {P.}~\bibnamefont {Rebentrost}},\ }\bibfield  {title} {\bibinfo {title} {Fundamentals of quantum machine learning and robustness},\ }in\ \href@noop {} {\emph {\bibinfo {booktitle} {Quantum Robustness in Artificial Intelligence: Principles and Applications}}}\ (\bibinfo  {publisher} {Springer},\ \bibinfo {year} {2026})\ pp.\ \bibinfo {pages} {1--19}\BibitemShut {NoStop}%
\bibitem [{\citenamefont {Schuld}\ \emph {et~al.}(2021)\citenamefont {Schuld}, \citenamefont {Sweke},\ and\ \citenamefont {Meyer}}]{schuld_EffectDataEncoding_2021}%
  \BibitemOpen
  \bibfield  {author} {\bibinfo {author} {\bibfnamefont {M.}~\bibnamefont {Schuld}}, \bibinfo {author} {\bibfnamefont {R.}~\bibnamefont {Sweke}},\ and\ \bibinfo {author} {\bibfnamefont {J.~J.}\ \bibnamefont {Meyer}},\ }\bibfield  {title} {\bibinfo {title} {Effect of data encoding on the expressive power of variational quantum-machine-learning models},\ }\href {https://doi.org/10.1103/PhysRevA.103.032430} {\bibfield  {journal} {\bibinfo  {journal} {Physical Review A}\ }\textbf {\bibinfo {volume} {103}},\ \bibinfo {pages} {032430} (\bibinfo {year} {2021})}\BibitemShut {NoStop}%
\bibitem [{\citenamefont {Mitarai}\ \emph {et~al.}(2018)\citenamefont {Mitarai}, \citenamefont {Negoro}, \citenamefont {Kitagawa},\ and\ \citenamefont {Fujii}}]{mitarai_QuantumCircuitLearning_2018}%
  \BibitemOpen
  \bibfield  {author} {\bibinfo {author} {\bibfnamefont {K.}~\bibnamefont {Mitarai}}, \bibinfo {author} {\bibfnamefont {M.}~\bibnamefont {Negoro}}, \bibinfo {author} {\bibfnamefont {M.}~\bibnamefont {Kitagawa}},\ and\ \bibinfo {author} {\bibfnamefont {K.}~\bibnamefont {Fujii}},\ }\bibfield  {title} {{\selectlanguage {en}\bibinfo {title} {Quantum circuit learning}},\ }\href {https://doi.org/10.1103/PhysRevA.98.032309} {\bibfield  {journal} {\bibinfo  {journal} {Physical Review A}\ }\textbf {\bibinfo {volume} {98}},\ \bibinfo {pages} {032309} (\bibinfo {year} {2018})}\BibitemShut {NoStop}%
\bibitem [{\citenamefont {Watabe}\ \emph {et~al.}(2019)\citenamefont {Watabe}, \citenamefont {Shiba}, \citenamefont {Sogabe}, \citenamefont {Sakamoto},\ and\ \citenamefont {Sogabe}}]{watabe2019quantum}%
  \BibitemOpen
  \bibfield  {author} {\bibinfo {author} {\bibfnamefont {M.}~\bibnamefont {Watabe}}, \bibinfo {author} {\bibfnamefont {K.}~\bibnamefont {Shiba}}, \bibinfo {author} {\bibfnamefont {M.}~\bibnamefont {Sogabe}}, \bibinfo {author} {\bibfnamefont {K.}~\bibnamefont {Sakamoto}},\ and\ \bibinfo {author} {\bibfnamefont {T.}~\bibnamefont {Sogabe}},\ }\bibfield  {title} {\bibinfo {title} {Quantum circuit parameters learning with gradient descent using backpropagation},\ }\href {https://doi.org/10.48550/arXiv.1910.14266} {\bibfield  {journal} {\bibinfo  {journal} {arXiv preprint arXiv:1910.14266}\ } (\bibinfo {year} {2019})}\BibitemShut {NoStop}%
\bibitem [{\citenamefont {Qi}\ \emph {et~al.}(2023)\citenamefont {Qi}, \citenamefont {Yang},\ and\ \citenamefont {Chen}}]{qi2023qtn}%
  \BibitemOpen
  \bibfield  {author} {\bibinfo {author} {\bibfnamefont {J.}~\bibnamefont {Qi}}, \bibinfo {author} {\bibfnamefont {C.-H.}\ \bibnamefont {Yang}},\ and\ \bibinfo {author} {\bibfnamefont {P.-Y.}\ \bibnamefont {Chen}},\ }\bibfield  {title} {\bibinfo {title} {Qtn-vqc: An end-to-end learning framework for quantum neural networks},\ }\href {https://doi.org/10.1088/1402-4896/ad14d6} {\bibfield  {journal} {\bibinfo  {journal} {Physica Scripta}\ }\textbf {\bibinfo {volume} {99}},\ \bibinfo {pages} {015111} (\bibinfo {year} {2023})}\BibitemShut {NoStop}%
\bibitem [{\citenamefont {Terashi}\ \emph {et~al.}(2021)\citenamefont {Terashi}, \citenamefont {Kaneda}, \citenamefont {Kishimoto}, \citenamefont {Saito}, \citenamefont {Sawada},\ and\ \citenamefont {Tanaka}}]{terashi2021event}%
  \BibitemOpen
  \bibfield  {author} {\bibinfo {author} {\bibfnamefont {K.}~\bibnamefont {Terashi}}, \bibinfo {author} {\bibfnamefont {M.}~\bibnamefont {Kaneda}}, \bibinfo {author} {\bibfnamefont {T.}~\bibnamefont {Kishimoto}}, \bibinfo {author} {\bibfnamefont {M.}~\bibnamefont {Saito}}, \bibinfo {author} {\bibfnamefont {R.}~\bibnamefont {Sawada}},\ and\ \bibinfo {author} {\bibfnamefont {J.}~\bibnamefont {Tanaka}},\ }\bibfield  {title} {\bibinfo {title} {Event classification with quantum machine learning in high-energy physics},\ }\href {https://doi.org/10.1007/s41781-020-00047-7} {\bibfield  {journal} {\bibinfo  {journal} {Computing and Software for Big Science}\ }\textbf {\bibinfo {volume} {5}},\ \bibinfo {pages} {1} (\bibinfo {year} {2021})}\BibitemShut {NoStop}%
\bibitem [{\citenamefont {Chen}\ \emph {et~al.}(2020{\natexlab{a}})\citenamefont {Chen}, \citenamefont {Huang}, \citenamefont {Hsing},\ and\ \citenamefont {Kao}}]{chen2020hybrid}%
  \BibitemOpen
  \bibfield  {author} {\bibinfo {author} {\bibfnamefont {S.~Y.~C.}\ \bibnamefont {Chen}}, \bibinfo {author} {\bibfnamefont {C.~M.}\ \bibnamefont {Huang}}, \bibinfo {author} {\bibfnamefont {C.~W.}\ \bibnamefont {Hsing}},\ and\ \bibinfo {author} {\bibfnamefont {Y.~J.}\ \bibnamefont {Kao}},\ }\bibfield  {title} {\bibinfo {title} {Hybrid quantum-classical classifier based on tensor network and variational quantum circuit},\ }\href {https://doi.org/arXiv:2011.14651} {\bibfield  {journal} {\bibinfo  {journal} {arXiv preprint arXiv:2011.14651}\ } (\bibinfo {year} {2020}{\natexlab{a}})}\BibitemShut {NoStop}%
\bibitem [{\citenamefont {Blance}\ and\ \citenamefont {Spannowsky}(2021)}]{blance2021quantum}%
  \BibitemOpen
  \bibfield  {author} {\bibinfo {author} {\bibfnamefont {A.}~\bibnamefont {Blance}}\ and\ \bibinfo {author} {\bibfnamefont {M.}~\bibnamefont {Spannowsky}},\ }\bibfield  {title} {\bibinfo {title} {Quantum machine learning for particle physics using a variational quantum classifier},\ }\href {https://doi.org/10.1007/JHEP02(2021)212} {\bibfield  {journal} {\bibinfo  {journal} {Journal of High Energy Physics}\ }\textbf {\bibinfo {volume} {2021}},\ \bibinfo {pages} {1} (\bibinfo {year} {2021})}\BibitemShut {NoStop}%
\bibitem [{\citenamefont {Kwak}\ \emph {et~al.}(2021)\citenamefont {Kwak}, \citenamefont {Yun}, \citenamefont {Jung}, \citenamefont {Kim},\ and\ \citenamefont {Kim}}]{kwak2021introduction}%
  \BibitemOpen
  \bibfield  {author} {\bibinfo {author} {\bibfnamefont {Y.}~\bibnamefont {Kwak}}, \bibinfo {author} {\bibfnamefont {W.~J.}\ \bibnamefont {Yun}}, \bibinfo {author} {\bibfnamefont {S.}~\bibnamefont {Jung}}, \bibinfo {author} {\bibfnamefont {J.-K.}\ \bibnamefont {Kim}},\ and\ \bibinfo {author} {\bibfnamefont {J.}~\bibnamefont {Kim}},\ }\bibfield  {title} {\bibinfo {title} {Introduction to quantum reinforcement learning: Theory and pennylane-based implementation},\ }in\ \href {https://doi.org/10.1109/ICTC52510.2021.9620885} {\emph {\bibinfo {booktitle} {2021 International Conference on Information and Communication Technology Convergence (ICTC)}}}\ (\bibinfo  {publisher} {IEEE},\ \bibinfo {address} {Jeju Island},\ \bibinfo {year} {2021})\ pp.\ \bibinfo {pages} {416--420}\BibitemShut {NoStop}%
\bibitem [{\citenamefont {Sierra-Sosa}\ \emph {et~al.}(2020)\citenamefont {Sierra-Sosa}, \citenamefont {Arcila-Moreno}, \citenamefont {Garcia-Zapirain}, \citenamefont {Castillo-Olea},\ and\ \citenamefont {Elmaghraby}}]{sierra2020dementia}%
  \BibitemOpen
  \bibfield  {author} {\bibinfo {author} {\bibfnamefont {D.}~\bibnamefont {Sierra-Sosa}}, \bibinfo {author} {\bibfnamefont {J.}~\bibnamefont {Arcila-Moreno}}, \bibinfo {author} {\bibfnamefont {B.}~\bibnamefont {Garcia-Zapirain}}, \bibinfo {author} {\bibfnamefont {C.}~\bibnamefont {Castillo-Olea}},\ and\ \bibinfo {author} {\bibfnamefont {A.}~\bibnamefont {Elmaghraby}},\ }\bibfield  {title} {\bibinfo {title} {Dementia prediction applying variational quantum classifier},\ }\href {https://doi.org/10.48550/arXiv.2007.08653} {\bibfield  {journal} {\bibinfo  {journal} {arXiv preprint arXiv:2007.08653}\ } (\bibinfo {year} {2020})}\BibitemShut {NoStop}%
\bibitem [{\citenamefont {Chen}\ \emph {et~al.}(2020{\natexlab{b}})\citenamefont {Chen}, \citenamefont {Yang}, \citenamefont {Qi}, \citenamefont {Chen}, \citenamefont {Ma},\ and\ \citenamefont {Goan}}]{chen2020variational}%
  \BibitemOpen
  \bibfield  {author} {\bibinfo {author} {\bibfnamefont {S.~Y.~C.}\ \bibnamefont {Chen}}, \bibinfo {author} {\bibfnamefont {C.~H.~H.}\ \bibnamefont {Yang}}, \bibinfo {author} {\bibfnamefont {J.}~\bibnamefont {Qi}}, \bibinfo {author} {\bibfnamefont {P.~Y.}\ \bibnamefont {Chen}}, \bibinfo {author} {\bibfnamefont {X.}~\bibnamefont {Ma}},\ and\ \bibinfo {author} {\bibfnamefont {H.~S.}\ \bibnamefont {Goan}},\ }\bibfield  {title} {\bibinfo {title} {Variational quantum circuits for deep reinforcement learning},\ }\href {https://doi.org/10.1109/ACCESS.2020.3010470} {\bibfield  {journal} {\bibinfo  {journal} {IEEE Access}\ }\textbf {\bibinfo {volume} {8}},\ \bibinfo {pages} {141007} (\bibinfo {year} {2020}{\natexlab{b}})}\BibitemShut {NoStop}%
\bibitem [{\citenamefont {Lloyd}\ \emph {et~al.}(2020)\citenamefont {Lloyd}, \citenamefont {Schuld}, \citenamefont {Ijaz}, \citenamefont {Izaac},\ and\ \citenamefont {Killoran}}]{lloyd2020quantum}%
  \BibitemOpen
  \bibfield  {author} {\bibinfo {author} {\bibfnamefont {S.}~\bibnamefont {Lloyd}}, \bibinfo {author} {\bibfnamefont {M.}~\bibnamefont {Schuld}}, \bibinfo {author} {\bibfnamefont {A.}~\bibnamefont {Ijaz}}, \bibinfo {author} {\bibfnamefont {J.}~\bibnamefont {Izaac}},\ and\ \bibinfo {author} {\bibfnamefont {N.}~\bibnamefont {Killoran}},\ }\bibfield  {title} {\bibinfo {title} {Quantum embeddings for machine learning},\ }\href {https://doi.org/10.48550/arXiv.2001.03622} {\bibfield  {journal} {\bibinfo  {journal} {arXiv preprint arXiv:2001.03622}\ } (\bibinfo {year} {2020})}\BibitemShut {NoStop}%
\bibitem [{\citenamefont {Hubregtsen}\ \emph {et~al.}(2022)\citenamefont {Hubregtsen}, \citenamefont {Wierichs}, \citenamefont {Gil-Fuster}, \citenamefont {Derks}, \citenamefont {Faehrmann},\ and\ \citenamefont {Meyer}}]{hubregtsen_TrainingQuantumEmbedding_2022}%
  \BibitemOpen
  \bibfield  {author} {\bibinfo {author} {\bibfnamefont {T.}~\bibnamefont {Hubregtsen}}, \bibinfo {author} {\bibfnamefont {D.}~\bibnamefont {Wierichs}}, \bibinfo {author} {\bibfnamefont {E.}~\bibnamefont {Gil-Fuster}}, \bibinfo {author} {\bibfnamefont {P.-J. H.~S.}\ \bibnamefont {Derks}}, \bibinfo {author} {\bibfnamefont {P.~K.}\ \bibnamefont {Faehrmann}},\ and\ \bibinfo {author} {\bibfnamefont {J.~J.}\ \bibnamefont {Meyer}},\ }\bibfield  {title} {\bibinfo {title} {Training quantum embedding kernels on near-term quantum computers},\ }\href {https://doi.org/10.1103/PhysRevA.106.042431} {\bibfield  {journal} {\bibinfo  {journal} {Physical Review A}\ }\textbf {\bibinfo {volume} {106}},\ \bibinfo {pages} {042431} (\bibinfo {year} {2022})}\BibitemShut {NoStop}%
\bibitem [{\citenamefont {Henry}\ \emph {et~al.}(2021)\citenamefont {Henry}, \citenamefont {Thabet}, \citenamefont {Dalyac},\ and\ \citenamefont {Henriet}}]{henry2021quantum}%
  \BibitemOpen
  \bibfield  {author} {\bibinfo {author} {\bibfnamefont {L.~P.}\ \bibnamefont {Henry}}, \bibinfo {author} {\bibfnamefont {S.}~\bibnamefont {Thabet}}, \bibinfo {author} {\bibfnamefont {C.}~\bibnamefont {Dalyac}},\ and\ \bibinfo {author} {\bibfnamefont {L.}~\bibnamefont {Henriet}},\ }\bibfield  {title} {\bibinfo {title} {Quantum evolution kernel: Machine learning on graphs with programmable arrays of qubits},\ }\href {https://doi.org/10.1103/PhysRevA.104.032416} {\bibfield  {journal} {\bibinfo  {journal} {Physical Review A}\ }\textbf {\bibinfo {volume} {104}},\ \bibinfo {pages} {032416} (\bibinfo {year} {2021})}\BibitemShut {NoStop}%
\bibitem [{\citenamefont {Larocca}\ \emph {et~al.}(2025)\citenamefont {Larocca}, \citenamefont {Thanasilp}, \citenamefont {Wang}, \citenamefont {Sharma}, \citenamefont {Biamonte}, \citenamefont {Coles}, \citenamefont {Cincio}, \citenamefont {McClean}, \citenamefont {Holmes},\ and\ \citenamefont {Cerezo}}]{larocca_BarrenPlateausVariational_2025}%
  \BibitemOpen
  \bibfield  {author} {\bibinfo {author} {\bibfnamefont {M.}~\bibnamefont {Larocca}}, \bibinfo {author} {\bibfnamefont {S.}~\bibnamefont {Thanasilp}}, \bibinfo {author} {\bibfnamefont {S.}~\bibnamefont {Wang}}, \bibinfo {author} {\bibfnamefont {K.}~\bibnamefont {Sharma}}, \bibinfo {author} {\bibfnamefont {J.}~\bibnamefont {Biamonte}}, \bibinfo {author} {\bibfnamefont {P.~J.}\ \bibnamefont {Coles}}, \bibinfo {author} {\bibfnamefont {L.}~\bibnamefont {Cincio}}, \bibinfo {author} {\bibfnamefont {J.~R.}\ \bibnamefont {McClean}}, \bibinfo {author} {\bibfnamefont {Z.}~\bibnamefont {Holmes}},\ and\ \bibinfo {author} {\bibfnamefont {M.}~\bibnamefont {Cerezo}},\ }\bibfield  {title} {\bibinfo {title} {Barren plateaus in variational quantum computing},\ }\href {https://www.nature.com/articles/s42254-025-00813-9} {\bibfield  {journal} {\bibinfo  {journal} {Nature Reviews Physics}\ ,\ \bibinfo {pages} {1}} (\bibinfo {year} {2025})}\BibitemShut {NoStop}%
\bibitem [{\citenamefont {Du}\ \emph {et~al.}(2022)\citenamefont {Du}, \citenamefont {Huang}, \citenamefont {You}, \citenamefont {Hsieh},\ and\ \citenamefont {Tao}}]{du2022quantum}%
  \BibitemOpen
  \bibfield  {author} {\bibinfo {author} {\bibfnamefont {Y.}~\bibnamefont {Du}}, \bibinfo {author} {\bibfnamefont {T.}~\bibnamefont {Huang}}, \bibinfo {author} {\bibfnamefont {S.}~\bibnamefont {You}}, \bibinfo {author} {\bibfnamefont {M.}~\bibnamefont {Hsieh}},\ and\ \bibinfo {author} {\bibfnamefont {D.}~\bibnamefont {Tao}},\ }\bibfield  {title} {\bibinfo {title} {Quantum circuit architecture search for variational quantum algorithms},\ }\href {https://doi.org/10.1038/s41534-022-00570-y} {\bibfield  {journal} {\bibinfo  {journal} {npj Quantum Information}\ }\textbf {\bibinfo {volume} {8}},\ \bibinfo {pages} {62} (\bibinfo {year} {2022})}\BibitemShut {NoStop}%
\bibitem [{\citenamefont {Torabian}\ and\ \citenamefont {Krems}(2023)}]{torabian_CompositionalOptimizationQuantum_2023}%
  \BibitemOpen
  \bibfield  {author} {\bibinfo {author} {\bibfnamefont {E.}~\bibnamefont {Torabian}}\ and\ \bibinfo {author} {\bibfnamefont {R.~V.}\ \bibnamefont {Krems}},\ }\bibfield  {title} {\bibinfo {title} {Compositional optimization of quantum circuits for quantum kernels of support vector machines},\ }\href {https://doi.org/10.1103/PhysRevResearch.5.013211} {\bibfield  {journal} {\bibinfo  {journal} {Physical Review Research}\ }\textbf {\bibinfo {volume} {5}},\ \bibinfo {pages} {013211} (\bibinfo {year} {2023})}\BibitemShut {NoStop}%
\bibitem [{\citenamefont {Zahid}\ and\ \citenamefont {Tahir}(2024)}]{zahid2024unlocking}%
  \BibitemOpen
  \bibfield  {author} {\bibinfo {author} {\bibfnamefont {S.}~\bibnamefont {Zahid}}\ and\ \bibinfo {author} {\bibfnamefont {M.~A.}\ \bibnamefont {Tahir}},\ }\bibfield  {title} {\bibinfo {title} {Unlocking quantum svm potential: optimal feature map generation and feature selection},\ }\href {https://doi.org/10.1088/1402-4896/ad9e39} {\bibfield  {journal} {\bibinfo  {journal} {Physica Scripta}\ }\textbf {\bibinfo {volume} {100}},\ \bibinfo {pages} {015120} (\bibinfo {year} {2024})}\BibitemShut {NoStop}%
\bibitem [{\citenamefont {Fösel}\ \emph {et~al.}(2021)\citenamefont {Fösel}, \citenamefont {Niu}, \citenamefont {Marquardt},\ and\ \citenamefont {Li}}]{fosel2021quantum}%
  \BibitemOpen
  \bibfield  {author} {\bibinfo {author} {\bibfnamefont {T.}~\bibnamefont {Fösel}}, \bibinfo {author} {\bibfnamefont {M.~Y.}\ \bibnamefont {Niu}}, \bibinfo {author} {\bibfnamefont {F.}~\bibnamefont {Marquardt}},\ and\ \bibinfo {author} {\bibfnamefont {L.}~\bibnamefont {Li}},\ }\bibfield  {title} {\bibinfo {title} {Quantum circuit optimization with deep reinforcement learning},\ }\href {https://doi.org/10.48550/arXiv.2103.07585} {\bibfield  {journal} {\bibinfo  {journal} {arXiv preprint arXiv:2103.07585}\ } (\bibinfo {year} {2021})}\BibitemShut {NoStop}%
\bibitem [{\citenamefont {Chakraborty}\ \emph {et~al.}(2020)\citenamefont {Chakraborty}, \citenamefont {Shaikh}, \citenamefont {Chakrabarti},\ and\ \citenamefont {Ghosh}}]{chakraborty2020hybrid}%
  \BibitemOpen
  \bibfield  {author} {\bibinfo {author} {\bibfnamefont {S.}~\bibnamefont {Chakraborty}}, \bibinfo {author} {\bibfnamefont {S.~H.}\ \bibnamefont {Shaikh}}, \bibinfo {author} {\bibfnamefont {A.}~\bibnamefont {Chakrabarti}},\ and\ \bibinfo {author} {\bibfnamefont {R.}~\bibnamefont {Ghosh}},\ }\bibfield  {title} {\bibinfo {title} {A hybrid quantum feature selection algorithm using a quantum inspired graph theoretic approach},\ }\href {https://doi.org/10.1007/s10489-019-01604-3} {\bibfield  {journal} {\bibinfo  {journal} {Applied Intelligence}\ }\textbf {\bibinfo {volume} {50}},\ \bibinfo {pages} {1775} (\bibinfo {year} {2020})}\BibitemShut {NoStop}%
\bibitem [{\citenamefont {Lu}\ \emph {et~al.}(2021)\citenamefont {Lu}, \citenamefont {Shen},\ and\ \citenamefont {Deng}}]{lu2021markovian}%
  \BibitemOpen
  \bibfield  {author} {\bibinfo {author} {\bibfnamefont {Z.}~\bibnamefont {Lu}}, \bibinfo {author} {\bibfnamefont {P.}~\bibnamefont {Shen}},\ and\ \bibinfo {author} {\bibfnamefont {D.}~\bibnamefont {Deng}},\ }\bibfield  {title} {\bibinfo {title} {Markovian quantum neuroevolution for machine learning},\ }\href {https://doi.org/10.1103/PhysRevApplied.16.044039} {\bibfield  {journal} {\bibinfo  {journal} {Physical Review Applied}\ }\textbf {\bibinfo {volume} {16}},\ \bibinfo {pages} {044039} (\bibinfo {year} {2021})}\BibitemShut {NoStop}%
\bibitem [{\citenamefont {Wang}\ \emph {et~al.}(2022)\citenamefont {Wang}, \citenamefont {Ding}, \citenamefont {Gu}, \citenamefont {Lin}, \citenamefont {Pan}, \citenamefont {Chong},\ and\ \citenamefont {Han}}]{wang2022quantumnas}%
  \BibitemOpen
  \bibfield  {author} {\bibinfo {author} {\bibfnamefont {H.}~\bibnamefont {Wang}}, \bibinfo {author} {\bibfnamefont {Y.}~\bibnamefont {Ding}}, \bibinfo {author} {\bibfnamefont {J.}~\bibnamefont {Gu}}, \bibinfo {author} {\bibfnamefont {Y.}~\bibnamefont {Lin}}, \bibinfo {author} {\bibfnamefont {D.~Z.}\ \bibnamefont {Pan}}, \bibinfo {author} {\bibfnamefont {F.~T.}\ \bibnamefont {Chong}},\ and\ \bibinfo {author} {\bibfnamefont {S.}~\bibnamefont {Han}},\ }\bibfield  {title} {\bibinfo {title} {Quantumnas: Noise-adaptive search for robust quantum circuits},\ }in\ \href {https://doi.org/10.1109/HPCA53966.2022.00057} {\emph {\bibinfo {booktitle} {2022 IEEE International Symposium on High-Performance Computer Architecture (HPCA)}}}\ (\bibinfo  {publisher} {IEEE},\ \bibinfo {address} {Seoul},\ \bibinfo {year} {2022})\ pp.\ \bibinfo {pages} {692--708}\BibitemShut {NoStop}%
\bibitem [{\citenamefont {Altares-López}\ \emph {et~al.}(2021)\citenamefont {Altares-López}, \citenamefont {Ribeiro},\ and\ \citenamefont {García-Ripoll}}]{altares2021automatic}%
  \BibitemOpen
  \bibfield  {author} {\bibinfo {author} {\bibfnamefont {S.}~\bibnamefont {Altares-López}}, \bibinfo {author} {\bibfnamefont {A.}~\bibnamefont {Ribeiro}},\ and\ \bibinfo {author} {\bibfnamefont {J.~J.}\ \bibnamefont {García-Ripoll}},\ }\bibfield  {title} {\bibinfo {title} {Automatic design of quantum feature maps},\ }\href {https://doi.org/10.1088/2058-9565/ac1ab1} {\bibfield  {journal} {\bibinfo  {journal} {Quantum Science and Technology}\ }\textbf {\bibinfo {volume} {6}},\ \bibinfo {pages} {045015} (\bibinfo {year} {2021})}\BibitemShut {NoStop}%
\bibitem [{\citenamefont {Pellow-Jarman}\ \emph {et~al.}(2024)\citenamefont {Pellow-Jarman}, \citenamefont {Pillay}, \citenamefont {Sinayskiy},\ and\ \citenamefont {Petruccione}}]{pellow2024hybrid}%
  \BibitemOpen
  \bibfield  {author} {\bibinfo {author} {\bibfnamefont {R.}~\bibnamefont {Pellow-Jarman}}, \bibinfo {author} {\bibfnamefont {A.}~\bibnamefont {Pillay}}, \bibinfo {author} {\bibfnamefont {I.}~\bibnamefont {Sinayskiy}},\ and\ \bibinfo {author} {\bibfnamefont {F.}~\bibnamefont {Petruccione}},\ }\bibfield  {title} {\bibinfo {title} {Hybrid genetic optimization for quantum feature map design},\ }\href {https://doi.org/10.1007/s42484-024-00177-w} {\bibfield  {journal} {\bibinfo  {journal} {Quantum Machine Intelligence}\ }\textbf {\bibinfo {volume} {6}},\ \bibinfo {pages} {45} (\bibinfo {year} {2024})}\BibitemShut {NoStop}%
\bibitem [{\citenamefont {Liu}\ \emph {et~al.}(2025)\citenamefont {Liu}, \citenamefont {Meng}, \citenamefont {Wang}, \citenamefont {Hu}, \citenamefont {Li}, \citenamefont {Zhang},\ and\ \citenamefont {Yu}}]{liu2025hardware}%
  \BibitemOpen
  \bibfield  {author} {\bibinfo {author} {\bibfnamefont {Y.}~\bibnamefont {Liu}}, \bibinfo {author} {\bibfnamefont {F.}~\bibnamefont {Meng}}, \bibinfo {author} {\bibfnamefont {L.}~\bibnamefont {Wang}}, \bibinfo {author} {\bibfnamefont {Y.}~\bibnamefont {Hu}}, \bibinfo {author} {\bibfnamefont {S.}~\bibnamefont {Li}}, \bibinfo {author} {\bibfnamefont {Z.}~\bibnamefont {Zhang}},\ and\ \bibinfo {author} {\bibfnamefont {X.}~\bibnamefont {Yu}},\ }\bibfield  {title} {\bibinfo {title} {Hardware-aware quantum kernel design based on graph neural networks},\ }\href {https://doi.org/10.48550/arXiv.2506.21161} {\bibfield  {journal} {\bibinfo  {journal} {arXiv preprint arXiv:2506.21161}\ } (\bibinfo {year} {2025})}\BibitemShut {NoStop}%
\bibitem [{\citenamefont {Ostaszewski}\ \emph {et~al.}(2021)\citenamefont {Ostaszewski}, \citenamefont {Grant},\ and\ \citenamefont {Benedetti}}]{ostaszewski2021structure}%
  \BibitemOpen
  \bibfield  {author} {\bibinfo {author} {\bibfnamefont {M.}~\bibnamefont {Ostaszewski}}, \bibinfo {author} {\bibfnamefont {E.}~\bibnamefont {Grant}},\ and\ \bibinfo {author} {\bibfnamefont {M.}~\bibnamefont {Benedetti}},\ }\bibfield  {title} {\bibinfo {title} {Structure optimization for parameterized quantum circuits},\ }\href {https://doi.org/10.22331/q-2021-01-28-391} {\bibfield  {journal} {\bibinfo  {journal} {Quantum}\ }\textbf {\bibinfo {volume} {5}},\ \bibinfo {pages} {391} (\bibinfo {year} {2021})}\BibitemShut {NoStop}%
\bibitem [{\citenamefont {Nakanishi}\ \emph {et~al.}(2020)\citenamefont {Nakanishi}, \citenamefont {Fujii},\ and\ \citenamefont {Todo}}]{nakanishi2020sequential}%
  \BibitemOpen
  \bibfield  {author} {\bibinfo {author} {\bibfnamefont {K.~M.}\ \bibnamefont {Nakanishi}}, \bibinfo {author} {\bibfnamefont {K.}~\bibnamefont {Fujii}},\ and\ \bibinfo {author} {\bibfnamefont {S.}~\bibnamefont {Todo}},\ }\bibfield  {title} {\bibinfo {title} {Sequential minimal optimization for quantum-classical hybrid algorithms},\ }\href {https://doi.org/10.1103/PhysRevResearch.2.043158} {\bibfield  {journal} {\bibinfo  {journal} {Physical Review Research}\ }\textbf {\bibinfo {volume} {2}},\ \bibinfo {pages} {043158} (\bibinfo {year} {2020})}\BibitemShut {NoStop}%
\bibitem [{\citenamefont {Parrish}\ \emph {et~al.}(2019)\citenamefont {Parrish}, \citenamefont {Iosue}, \citenamefont {Ozaeta},\ and\ \citenamefont {McMahon}}]{parrish_JacobiDiagonalizationAnderson_2019}%
  \BibitemOpen
  \bibfield  {author} {\bibinfo {author} {\bibfnamefont {R.~M.}\ \bibnamefont {Parrish}}, \bibinfo {author} {\bibfnamefont {J.~T.}\ \bibnamefont {Iosue}}, \bibinfo {author} {\bibfnamefont {A.}~\bibnamefont {Ozaeta}},\ and\ \bibinfo {author} {\bibfnamefont {P.~L.}\ \bibnamefont {McMahon}},\ }\bibfield  {title} {\bibinfo {title} {A {Jacobi} {Diagonalization} and {Anderson} {Acceleration} {Algorithm} {For} {Variational} {Quantum} {Algorithm} {Parameter} {Optimization}},\ }\href {https://doi.org/10.48550/arXiv.1904.03206} {\bibfield  {journal} {\bibinfo  {journal} {arXiv preprint arXiv:1904.03206}\ } (\bibinfo {year} {2019})}\BibitemShut {NoStop}%
\bibitem [{\citenamefont {Vidal}\ and\ \citenamefont {Theis}(2018)}]{vidal_CalculusParameterizedQuantum_2018}%
  \BibitemOpen
  \bibfield  {author} {\bibinfo {author} {\bibfnamefont {J.~G.}\ \bibnamefont {Vidal}}\ and\ \bibinfo {author} {\bibfnamefont {D.~O.}\ \bibnamefont {Theis}},\ }\bibfield  {title} {\bibinfo {title} {Calculus on parameterized quantum circuits},\ }\href {http://arxiv.org/abs/1812.06323} {\bibfield  {journal} {\bibinfo  {journal} {arXiv preprint arXiv:1812.06323}\ } (\bibinfo {year} {2018})}\BibitemShut {NoStop}%
\bibitem [{\citenamefont {Feniou}\ \emph {et~al.}(2025)\citenamefont {Feniou}, \citenamefont {Hassan}, \citenamefont {Claudon}, \citenamefont {Courtat}, \citenamefont {Adjoua}, \citenamefont {Maday},\ and\ \citenamefont {Piquemal}}]{feniou2025greedy}%
  \BibitemOpen
  \bibfield  {author} {\bibinfo {author} {\bibfnamefont {C.}~\bibnamefont {Feniou}}, \bibinfo {author} {\bibfnamefont {M.}~\bibnamefont {Hassan}}, \bibinfo {author} {\bibfnamefont {B.}~\bibnamefont {Claudon}}, \bibinfo {author} {\bibfnamefont {A.}~\bibnamefont {Courtat}}, \bibinfo {author} {\bibfnamefont {O.}~\bibnamefont {Adjoua}}, \bibinfo {author} {\bibfnamefont {Y.}~\bibnamefont {Maday}},\ and\ \bibinfo {author} {\bibfnamefont {J.-P.}\ \bibnamefont {Piquemal}},\ }\bibfield  {title} {\bibinfo {title} {Greedy gradient-free adaptive variational quantum algorithms on a noisy intermediate scale quantum computer},\ }\href {https://doi.org/10.1038/s41598-025-99962-1} {\bibfield  {journal} {\bibinfo  {journal} {Scientific Reports}\ }\textbf {\bibinfo {volume} {15}},\ \bibinfo {pages} {18689} (\bibinfo {year} {2025})}\BibitemShut {NoStop}%
\bibitem [{\citenamefont {J{\"a}ger}\ \emph {et~al.}(2025)\citenamefont {J{\"a}ger}, \citenamefont {Kaldenbach}, \citenamefont {Haas},\ and\ \citenamefont {Schultheis}}]{jager_FastGradientfreeOptimization_2025}%
  \BibitemOpen
  \bibfield  {author} {\bibinfo {author} {\bibfnamefont {J.}~\bibnamefont {J{\"a}ger}}, \bibinfo {author} {\bibfnamefont {T.~N.}\ \bibnamefont {Kaldenbach}}, \bibinfo {author} {\bibfnamefont {M.}~\bibnamefont {Haas}},\ and\ \bibinfo {author} {\bibfnamefont {E.}~\bibnamefont {Schultheis}},\ }\bibfield  {title} {\bibinfo {title} {Fast gradient-free optimization of excitations in variational quantum eigensolvers},\ }\href@noop {} {\bibfield  {journal} {\bibinfo  {journal} {Communications Physics}\ }\textbf {\bibinfo {volume} {8}},\ \bibinfo {pages} {418} (\bibinfo {year} {2025})}\BibitemShut {NoStop}%
\bibitem [{\citenamefont {Grimsley}\ \emph {et~al.}(2019)\citenamefont {Grimsley}, \citenamefont {Economou}, \citenamefont {Barnes},\ and\ \citenamefont {Mayhall}}]{grimsley_AdaptiveVariationalAlgorithm_2019}%
  \BibitemOpen
  \bibfield  {author} {\bibinfo {author} {\bibfnamefont {H.~R.}\ \bibnamefont {Grimsley}}, \bibinfo {author} {\bibfnamefont {S.~E.}\ \bibnamefont {Economou}}, \bibinfo {author} {\bibfnamefont {E.}~\bibnamefont {Barnes}},\ and\ \bibinfo {author} {\bibfnamefont {N.~J.}\ \bibnamefont {Mayhall}},\ }\bibfield  {title} {{\selectlanguage {en}\bibinfo {title} {An adaptive variational algorithm for exact molecular simulations on a quantum computer}},\ }\href {https://doi.org/10.1038/s41467-019-10988-2} {\bibfield  {journal} {\bibinfo  {journal} {Nature Communications}\ }\textbf {\bibinfo {volume} {10}},\ \bibinfo {pages} {3007} (\bibinfo {year} {2019})}\BibitemShut {NoStop}%
\bibitem [{\citenamefont {Nielsen}\ and\ \citenamefont {Chuang}(2010)}]{nielsen_QuantumComputationQuantum_2010}%
  \BibitemOpen
  \bibfield  {author} {\bibinfo {author} {\bibfnamefont {M.~A.}\ \bibnamefont {Nielsen}}\ and\ \bibinfo {author} {\bibfnamefont {I.~L.}\ \bibnamefont {Chuang}},\ }\href@noop {} {{\selectlanguage {en}\emph {\bibinfo {title} {Quantum computation and quantum information}}}},\ \bibinfo {edition} {10th}\ ed.\ (\bibinfo  {publisher} {Cambridge University Press},\ \bibinfo {address} {Cambridge ; New York},\ \bibinfo {year} {2010})\BibitemShut {NoStop}%
\bibitem [{\citenamefont {Grossi}\ \emph {et~al.}(2022)\citenamefont {Grossi}, \citenamefont {Ibrahim}, \citenamefont {Radescu}, \citenamefont {Loredo}, \citenamefont {Voigt}, \citenamefont {Von~Altrock},\ and\ \citenamefont {Rudnik}}]{grossi2022mixed}%
  \BibitemOpen
  \bibfield  {author} {\bibinfo {author} {\bibfnamefont {M.}~\bibnamefont {Grossi}}, \bibinfo {author} {\bibfnamefont {N.}~\bibnamefont {Ibrahim}}, \bibinfo {author} {\bibfnamefont {V.}~\bibnamefont {Radescu}}, \bibinfo {author} {\bibfnamefont {R.}~\bibnamefont {Loredo}}, \bibinfo {author} {\bibfnamefont {K.}~\bibnamefont {Voigt}}, \bibinfo {author} {\bibfnamefont {C.}~\bibnamefont {Von~Altrock}},\ and\ \bibinfo {author} {\bibfnamefont {A.}~\bibnamefont {Rudnik}},\ }\bibfield  {title} {\bibinfo {title} {Mixed quantum--classical method for fraud detection with quantum feature selection},\ }\href {https://doi.org/10.1109/TQE.2022.3213474} {\bibfield  {journal} {\bibinfo  {journal} {IEEE Transactions on Quantum Engineering}\ }\textbf {\bibinfo {volume} {3}},\ \bibinfo {pages} {1} (\bibinfo {year} {2022})}\BibitemShut {NoStop}%
\bibitem [{\citenamefont {Albino}\ \emph {et~al.}(2023)\citenamefont {Albino}, \citenamefont {Pires}, \citenamefont {Nooblath},\ and\ \citenamefont {Nascimento}}]{albino2023evolutionary}%
  \BibitemOpen
  \bibfield  {author} {\bibinfo {author} {\bibfnamefont {A.~S.}\ \bibnamefont {Albino}}, \bibinfo {author} {\bibfnamefont {O.~M.}\ \bibnamefont {Pires}}, \bibinfo {author} {\bibfnamefont {M.~Q.}\ \bibnamefont {Nooblath}},\ and\ \bibinfo {author} {\bibfnamefont {E.~G.~S.}\ \bibnamefont {Nascimento}},\ }\bibfield  {title} {\bibinfo {title} {Evolutionary quantum feature selection},\ }\href {https://doi.org/10.48550/arXiv.2303.07131} {\bibfield  {journal} {\bibinfo  {journal} {arXiv preprint arXiv:2303.07131}\ } (\bibinfo {year} {2023})}\BibitemShut {NoStop}%
\bibitem [{\citenamefont {Wang}\ \emph {et~al.}(2023)\citenamefont {Wang}, \citenamefont {Chen}, \citenamefont {Le}, \citenamefont {Yu}, \citenamefont {Xue}, \citenamefont {Zhuang}, \citenamefont {Yan}, \citenamefont {Yang}, \citenamefont {Wu},\ and\ \citenamefont {Guo}}]{wang2023quantum}%
  \BibitemOpen
  \bibfield  {author} {\bibinfo {author} {\bibfnamefont {L.}~\bibnamefont {Wang}}, \bibinfo {author} {\bibfnamefont {Z.~Y.}\ \bibnamefont {Chen}}, \bibinfo {author} {\bibfnamefont {F.~Y.}\ \bibnamefont {Le}}, \bibinfo {author} {\bibfnamefont {Z.~Q.}\ \bibnamefont {Yu}}, \bibinfo {author} {\bibfnamefont {C.}~\bibnamefont {Xue}}, \bibinfo {author} {\bibfnamefont {X.~N.}\ \bibnamefont {Zhuang}}, \bibinfo {author} {\bibfnamefont {Q.}~\bibnamefont {Yan}}, \bibinfo {author} {\bibfnamefont {Y.}~\bibnamefont {Yang}}, \bibinfo {author} {\bibfnamefont {Y.~C.}\ \bibnamefont {Wu}},\ and\ \bibinfo {author} {\bibfnamefont {G.~P.}\ \bibnamefont {Guo}},\ }\bibfield  {title} {\bibinfo {title} {A quantum feature selection framework via ground state preparation},\ }\href {https://doi.org/10.1088/1402-4896/ad0184} {\bibfield  {journal} {\bibinfo  {journal} {Physica Scripta}\ }\textbf {\bibinfo {volume} {98}},\ \bibinfo {pages} {115121} (\bibinfo {year} {2023})}\BibitemShut {NoStop}%
\bibitem [{\citenamefont {Mücke}\ \emph {et~al.}(2023)\citenamefont {Mücke}, \citenamefont {Heese}, \citenamefont {Müller}, \citenamefont {Wolter},\ and\ \citenamefont {Piatkowski}}]{mucke2023feature}%
  \BibitemOpen
  \bibfield  {author} {\bibinfo {author} {\bibfnamefont {S.}~\bibnamefont {Mücke}}, \bibinfo {author} {\bibfnamefont {R.}~\bibnamefont {Heese}}, \bibinfo {author} {\bibfnamefont {S.}~\bibnamefont {Müller}}, \bibinfo {author} {\bibfnamefont {M.}~\bibnamefont {Wolter}},\ and\ \bibinfo {author} {\bibfnamefont {N.}~\bibnamefont {Piatkowski}},\ }\bibfield  {title} {\bibinfo {title} {Feature selection on quantum computers},\ }\href {https://doi.org/10.1007/s42484-023-00099-z} {\bibfield  {journal} {\bibinfo  {journal} {Quantum Machine Intelligence}\ }\textbf {\bibinfo {volume} {5}},\ \bibinfo {pages} {11} (\bibinfo {year} {2023})}\BibitemShut {NoStop}%
\bibitem [{\citenamefont {Battiti}(1994)}]{battiti1994using}%
  \BibitemOpen
  \bibfield  {author} {\bibinfo {author} {\bibfnamefont {R.}~\bibnamefont {Battiti}},\ }\bibfield  {title} {\bibinfo {title} {Using mutual information for selecting features in supervised neural net learning},\ }\href {https://doi.org/10.1109/72.298224} {\bibfield  {journal} {\bibinfo  {journal} {IEEE Transactions on Neural Networks and Learning Systems}\ }\textbf {\bibinfo {volume} {5}},\ \bibinfo {pages} {537} (\bibinfo {year} {1994})}\BibitemShut {NoStop}%
\bibitem [{\citenamefont {Al~Fatih Abil~Fida}\ \emph {et~al.}(2021)\citenamefont {Al~Fatih Abil~Fida}, \citenamefont {Ahmad},\ and\ \citenamefont {Ntahobari}}]{fida2021variance}%
  \BibitemOpen
  \bibfield  {author} {\bibinfo {author} {\bibfnamefont {M.}~\bibnamefont {Al~Fatih Abil~Fida}}, \bibinfo {author} {\bibfnamefont {T.}~\bibnamefont {Ahmad}},\ and\ \bibinfo {author} {\bibfnamefont {M.}~\bibnamefont {Ntahobari}},\ }\bibfield  {title} {\bibinfo {title} {Variance threshold as early screening to boruta feature selection for intrusion detection system},\ }in\ \href {https://doi.org/10.1109/ICTS52701.2021.9608852} {\emph {\bibinfo {booktitle} {2021 13th International Conference on Information \& Communication Technology and System (ICTS)}}}\ (\bibinfo  {publisher} {IEEE},\ \bibinfo {address} {Surabaya},\ \bibinfo {year} {2021})\ pp.\ \bibinfo {pages} {46--50}\BibitemShut {NoStop}%
\bibitem [{\citenamefont {Guyon}\ \emph {et~al.}(2002)\citenamefont {Guyon}, \citenamefont {Weston}, \citenamefont {Barnhill},\ and\ \citenamefont {Vapnik}}]{guyon2002gene}%
  \BibitemOpen
  \bibfield  {author} {\bibinfo {author} {\bibfnamefont {I.}~\bibnamefont {Guyon}}, \bibinfo {author} {\bibfnamefont {J.}~\bibnamefont {Weston}}, \bibinfo {author} {\bibfnamefont {S.}~\bibnamefont {Barnhill}},\ and\ \bibinfo {author} {\bibfnamefont {V.}~\bibnamefont {Vapnik}},\ }\bibfield  {title} {\bibinfo {title} {Gene selection for cancer classification using support vector machines},\ }\href {https://doi.org/10.1023/A:1012487302797} {\bibfield  {journal} {\bibinfo  {journal} {Machine Learning}\ }\textbf {\bibinfo {volume} {46}},\ \bibinfo {pages} {389} (\bibinfo {year} {2002})}\BibitemShut {NoStop}%
\bibitem [{\citenamefont {Tibshirani}(1996)}]{tibshirani1996regression}%
  \BibitemOpen
  \bibfield  {author} {\bibinfo {author} {\bibfnamefont {R.}~\bibnamefont {Tibshirani}},\ }\bibfield  {title} {\bibinfo {title} {Regression shrinkage and selection via the lasso},\ }\href {https://doi.org/10.1111/j.2517-6161.1996.tb02080.x} {\bibfield  {journal} {\bibinfo  {journal} {Journal of the Royal Statistical Society Series B: Statistical Methodology}\ }\textbf {\bibinfo {volume} {58}},\ \bibinfo {pages} {267} (\bibinfo {year} {1996})}\BibitemShut {NoStop}%
\bibitem [{\citenamefont {Bowles}\ \emph {et~al.}(2024{\natexlab{a}})\citenamefont {Bowles}, \citenamefont {Ahmed},\ and\ \citenamefont {Schuld}}]{bowles_BetterClassicalSubtle_2024}%
  \BibitemOpen
  \bibfield  {author} {\bibinfo {author} {\bibfnamefont {J.}~\bibnamefont {Bowles}}, \bibinfo {author} {\bibfnamefont {S.}~\bibnamefont {Ahmed}},\ and\ \bibinfo {author} {\bibfnamefont {M.}~\bibnamefont {Schuld}},\ }\bibfield  {title} {\bibinfo {title} {Better than classical? {The} subtle art of benchmarking quantum machine learning models},\ }\href {http://arxiv.org/abs/2403.07059} {\bibfield  {journal} {\bibinfo  {journal} {arXiv preprint arXiv:2403.07059}\ } (\bibinfo {year} {2024}{\natexlab{a}})}\BibitemShut {NoStop}%
\bibitem [{\citenamefont {Schreiber}\ \emph {et~al.}(2023)\citenamefont {Schreiber}, \citenamefont {Eisert},\ and\ \citenamefont {Meyer}}]{schreiber_ClassicalSurrogatesQuantum_2023}%
  \BibitemOpen
  \bibfield  {author} {\bibinfo {author} {\bibfnamefont {F.~J.}\ \bibnamefont {Schreiber}}, \bibinfo {author} {\bibfnamefont {J.}~\bibnamefont {Eisert}},\ and\ \bibinfo {author} {\bibfnamefont {J.~J.}\ \bibnamefont {Meyer}},\ }\bibfield  {title} {\bibinfo {title} {Classical {{Surrogates}} for {{Quantum Learning Models}}},\ }\href {https://doi.org/10.1103/PhysRevLett.131.100803} {\bibfield  {journal} {\bibinfo  {journal} {Physical Review Letters}\ }\textbf {\bibinfo {volume} {131}},\ \bibinfo {pages} {100803} (\bibinfo {year} {2023})}\BibitemShut {NoStop}%
\bibitem [{\citenamefont {Stein}\ \emph {et~al.}(2024)\citenamefont {Stein}, \citenamefont {Poppel}, \citenamefont {Adamczyk}, \citenamefont {Fabry}, \citenamefont {Wu}, \citenamefont {Kölle}, \citenamefont {Nüßlein}, \citenamefont {Schuman}, \citenamefont {Altmann}, \citenamefont {Ehmer}, \citenamefont {Narasimhan},\ and\ \citenamefont {Linnhoff{-}Popien}}]{stein_BenchmarkingQuantumSurrogate_2024}%
  \BibitemOpen
  \bibfield  {author} {\bibinfo {author} {\bibfnamefont {J.}~\bibnamefont {Stein}}, \bibinfo {author} {\bibfnamefont {M.}~\bibnamefont {Poppel}}, \bibinfo {author} {\bibfnamefont {P.}~\bibnamefont {Adamczyk}}, \bibinfo {author} {\bibfnamefont {R.}~\bibnamefont {Fabry}}, \bibinfo {author} {\bibfnamefont {Z.}~\bibnamefont {Wu}}, \bibinfo {author} {\bibfnamefont {M.}~\bibnamefont {Kölle}}, \bibinfo {author} {\bibfnamefont {J.}~\bibnamefont {Nüßlein}}, \bibinfo {author} {\bibfnamefont {D.}~\bibnamefont {Schuman}}, \bibinfo {author} {\bibfnamefont {P.}~\bibnamefont {Altmann}}, \bibinfo {author} {\bibfnamefont {T.}~\bibnamefont {Ehmer}}, \bibinfo {author} {\bibfnamefont {V.}~\bibnamefont {Narasimhan}},\ and\ \bibinfo {author} {\bibfnamefont {C.}~\bibnamefont {Linnhoff{-}Popien}},\ }\bibfield  {title} {\bibinfo {title} {Benchmarking quantum surrogate models on scarce and noisy data},\ }in\ \href {https://doi.org/10.5220/0012348900003636} {\emph {\bibinfo {booktitle} {Proceedings of the 16th International
  Conference on Agents and Artificial Intelligence - Volume 3: ICAART}}},\ \bibinfo {organization} {INSTICC}\ (\bibinfo  {publisher} {SciTePress},\ \bibinfo {address} {Rome, Italy},\ \bibinfo {year} {2024})\ pp.\ \bibinfo {pages} {352--359}\BibitemShut {NoStop}%
\bibitem [{\citenamefont {Farhi}\ and\ \citenamefont {Neven}(2018)}]{farhi_ClassificationQuantumNeural_2018}%
  \BibitemOpen
  \bibfield  {author} {\bibinfo {author} {\bibfnamefont {E.}~\bibnamefont {Farhi}}\ and\ \bibinfo {author} {\bibfnamefont {H.}~\bibnamefont {Neven}},\ }\bibfield  {title} {\bibinfo {title} {Classification with {Quantum} {Neural} {Networks} on {Near} {Term} {Processors}},\ }\href {http://arxiv.org/abs/1802.06002} {\bibfield  {journal} {\bibinfo  {journal} {arXiv preprint arXiv:1802.06002}\ } (\bibinfo {year} {2018})}\BibitemShut {NoStop}%
\bibitem [{\citenamefont {Schuld}\ \emph {et~al.}(2020)\citenamefont {Schuld}, \citenamefont {Bocharov}, \citenamefont {Svore},\ and\ \citenamefont {Wiebe}}]{schuld_CircuitcentricQuantumClassifiers_2020}%
  \BibitemOpen
  \bibfield  {author} {\bibinfo {author} {\bibfnamefont {M.}~\bibnamefont {Schuld}}, \bibinfo {author} {\bibfnamefont {A.}~\bibnamefont {Bocharov}}, \bibinfo {author} {\bibfnamefont {K.~M.}\ \bibnamefont {Svore}},\ and\ \bibinfo {author} {\bibfnamefont {N.}~\bibnamefont {Wiebe}},\ }\bibfield  {title} {\bibinfo {title} {Circuit-centric quantum classifiers},\ }\href {https://link.aps.org/doi/10.1103/PhysRevA.101.032308} {\bibfield  {journal} {\bibinfo  {journal} {Physical Review A}\ }\textbf {\bibinfo {volume} {101}},\ \bibinfo {pages} {032308} (\bibinfo {year} {2020})}\BibitemShut {NoStop}%
\bibitem [{Note1()}]{Note1}%
  \BibitemOpen
  \bibinfo {note} {This loss function draws motivation from statistics and information theory, also known as log loss, logistic loss or cross-entropy loss, and has been established as the standard choice in the classical machine learning literature -- including extensions to multi-class -- classification. \cite {bishop_PatternRecognitionMachine_2006,hastie2009elements,murphy_MachineLearningProbabilistic_2012}}\BibitemShut {NoStop}%
\bibitem [{\citenamefont {Huang}\ \emph {et~al.}(2021{\natexlab{b}})\citenamefont {Huang}, \citenamefont {Broughton}, \citenamefont {Mohseni}, \citenamefont {Babbush}, \citenamefont {Boixo}, \citenamefont {Neven},\ and\ \citenamefont {McClean}}]{huang_PowerDataQuantum_2021}%
  \BibitemOpen
  \bibfield  {author} {\bibinfo {author} {\bibfnamefont {H.-Y.}\ \bibnamefont {Huang}}, \bibinfo {author} {\bibfnamefont {M.}~\bibnamefont {Broughton}}, \bibinfo {author} {\bibfnamefont {M.}~\bibnamefont {Mohseni}}, \bibinfo {author} {\bibfnamefont {R.}~\bibnamefont {Babbush}}, \bibinfo {author} {\bibfnamefont {S.}~\bibnamefont {Boixo}}, \bibinfo {author} {\bibfnamefont {H.}~\bibnamefont {Neven}},\ and\ \bibinfo {author} {\bibfnamefont {J.~R.}\ \bibnamefont {McClean}},\ }\bibfield  {title} {{\selectlanguage {en}\bibinfo {title} {Power of data in quantum machine learning}},\ }\href {https://www.nature.com/articles/s41467-021-22539-9} {\bibfield  {journal} {\bibinfo  {journal} {Nature Communications}\ }\textbf {\bibinfo {volume} {12}},\ \bibinfo {pages} {2631} (\bibinfo {year} {2021}{\natexlab{b}})}\BibitemShut {NoStop}%
\bibitem [{\citenamefont {Boser}\ \emph {et~al.}(1992)\citenamefont {Boser}, \citenamefont {Guyon},\ and\ \citenamefont {Vapnik}}]{boser_TrainingAlgorithmOptimal_1992}%
  \BibitemOpen
  \bibfield  {author} {\bibinfo {author} {\bibfnamefont {B.~E.}\ \bibnamefont {Boser}}, \bibinfo {author} {\bibfnamefont {I.~M.}\ \bibnamefont {Guyon}},\ and\ \bibinfo {author} {\bibfnamefont {V.~N.}\ \bibnamefont {Vapnik}},\ }\bibfield  {title} {\bibinfo {title} {A training algorithm for optimal margin classifiers},\ }in\ \href {https://dl.acm.org/doi/10.1145/130385.130401} {\emph {\bibinfo {booktitle} {Proceedings of the fifth annual workshop on {Computational} learning theory}}},\ \bibinfo {series and number} {{COLT} '92}\ (\bibinfo  {publisher} {Association for Computing Machinery},\ \bibinfo {address} {New York, NY, USA},\ \bibinfo {year} {1992})\ pp.\ \bibinfo {pages} {144--152}\BibitemShut {NoStop}%
\bibitem [{Note2()}]{Note2}%
  \BibitemOpen
  \bibinfo {note} {Note that given the evaluated quantum kernel values, commonly aggregated in the kernel matrix (also referred to as Gram matrix), the training of the QSVM does not differ from that of classical SVMs \cite {bishop_PatternRecognitionMachine_2006,hastie2009elements,murphy_MachineLearningProbabilistic_2012}.}\BibitemShut {Stop}%
\bibitem [{\citenamefont {Cristianini}\ \emph {et~al.}(2001)\citenamefont {Cristianini}, \citenamefont {Shawe-Taylor}, \citenamefont {Elisseeff},\ and\ \citenamefont {Kandola}}]{cristianini2001kernel}%
  \BibitemOpen
  \bibfield  {author} {\bibinfo {author} {\bibfnamefont {N.}~\bibnamefont {Cristianini}}, \bibinfo {author} {\bibfnamefont {J.}~\bibnamefont {Shawe-Taylor}}, \bibinfo {author} {\bibfnamefont {A.}~\bibnamefont {Elisseeff}},\ and\ \bibinfo {author} {\bibfnamefont {J.}~\bibnamefont {Kandola}},\ }\bibfield  {title} {\bibinfo {title} {On kernel-target alignment},\ }\href {https://proceedings.neurips.cc/paper_files/paper/2001/file/1f71e393b3809197ed66df836fe833e5-Paper.pdf} {\bibfield  {journal} {\bibinfo  {journal} {Advances in neural information processing systems}\ }\textbf {\bibinfo {volume} {14}} (\bibinfo {year} {2001})}\BibitemShut {NoStop}%
\bibitem [{\citenamefont {Paine}\ \emph {et~al.}(2023)\citenamefont {Paine}, \citenamefont {Elfving},\ and\ \citenamefont {Kyriienko}}]{paine_QuantumKernelMethods_2023}%
  \BibitemOpen
  \bibfield  {author} {\bibinfo {author} {\bibfnamefont {A.~E.}\ \bibnamefont {Paine}}, \bibinfo {author} {\bibfnamefont {V.~E.}\ \bibnamefont {Elfving}},\ and\ \bibinfo {author} {\bibfnamefont {O.}~\bibnamefont {Kyriienko}},\ }\bibfield  {title} {\bibinfo {title} {Quantum kernel methods for solving regression problems and differential equations},\ }\href {https://link.aps.org/doi/10.1103/PhysRevA.107.032428} {\bibfield  {journal} {\bibinfo  {journal} {Physical Review A}\ }\textbf {\bibinfo {volume} {107}},\ \bibinfo {pages} {032428} (\bibinfo {year} {2023})}\BibitemShut {NoStop}%
\bibitem [{\citenamefont {Salmenperä}\ \emph {et~al.}(2024)\citenamefont {Salmenperä}, \citenamefont {Kuhtarskis}, \citenamefont {van~de Griend},\ and\ \citenamefont {Nurminen}}]{salmenpera2024impact}%
  \BibitemOpen
  \bibfield  {author} {\bibinfo {author} {\bibfnamefont {I.}~\bibnamefont {Salmenperä}}, \bibinfo {author} {\bibfnamefont {I.}~\bibnamefont {Kuhtarskis}}, \bibinfo {author} {\bibfnamefont {A.~M.}\ \bibnamefont {van~de Griend}},\ and\ \bibinfo {author} {\bibfnamefont {J.~K.}\ \bibnamefont {Nurminen}},\ }\bibfield  {title} {\bibinfo {title} {The impact of feature embedding placement in the ansatz of a quantum kernel in qsvms},\ }in\ \href {https://doi.org/10.1109/QCE60285.2024.00194} {\emph {\bibinfo {booktitle} {2024 IEEE International Conference on Quantum Computing and Engineering (QCE)}}},\ Vol.~\bibinfo {volume} {01}\ (\bibinfo  {publisher} {IEEE},\ \bibinfo {address} {Montreal},\ \bibinfo {year} {2024})\ pp.\ \bibinfo {pages} {1663--1671}\BibitemShut {NoStop}%
\bibitem [{Note3()}]{Note3}%
  \BibitemOpen
  \bibinfo {note} {The required form to apply Q-FLAIR is still maintained when appending a fixed gate along with a (weight-) data-dependent gate because the fixed gate is absorbed by the state and observable in Eq.~\protect \eqref {eq:qsvm_derivation_complex_conjugate}}\BibitemShut {NoStop}%
\bibitem [{\citenamefont {Bowles}\ \emph {et~al.}(2024{\natexlab{b}})\citenamefont {Bowles}, \citenamefont {Ahmed},\ and\ \citenamefont {Schuld}}]{bowles2024subtledata}%
  \BibitemOpen
  \bibfield  {author} {\bibinfo {author} {\bibfnamefont {J.}~\bibnamefont {Bowles}}, \bibinfo {author} {\bibfnamefont {S.}~\bibnamefont {Ahmed}},\ and\ \bibinfo {author} {\bibfnamefont {M.}~\bibnamefont {Schuld}},\ }\href@noop {} {\bibinfo {title} {Pennylane datasets for {Better} than classical? {The} subtle art of benchmarking quantum machine learning models}},\ \bibinfo {howpublished} {\url{pennylane.ai/datasets/linearly-separable}, \url{pennylane.ai/datasets/two-curves}, \url{pennylane.ai/datasets/bars-and-stripes}, \url{pennylane.ai/datasets/downscaled-mnist}} (\bibinfo {year} {2024}{\natexlab{b}}),\ \bibinfo {note} {accessed: 2024-09-22}\BibitemShut {NoStop}%
\bibitem [{\citenamefont {LeCun}(1998)}]{lecun1998}%
  \BibitemOpen
  \bibfield  {author} {\bibinfo {author} {\bibfnamefont {Y.}~\bibnamefont {LeCun}},\ }\href {http://yan.lecun.com/exdb/mnist/} {\bibinfo {title} {The {MNIST} database of handwritten digits}},\ \bibinfo {howpublished} {http://yan.lecun.com/exdb/mnist/} (\bibinfo {year} {1998})\BibitemShut {NoStop}%
\bibitem [{\citenamefont {Buchanan}\ \emph {et~al.}(2020)\citenamefont {Buchanan}, \citenamefont {Gilboa},\ and\ \citenamefont {Wright}}]{buchanan2020deep}%
  \BibitemOpen
  \bibfield  {author} {\bibinfo {author} {\bibfnamefont {S.}~\bibnamefont {Buchanan}}, \bibinfo {author} {\bibfnamefont {D.}~\bibnamefont {Gilboa}},\ and\ \bibinfo {author} {\bibfnamefont {J.}~\bibnamefont {Wright}},\ }\bibfield  {title} {\bibinfo {title} {Deep networks and the multiple manifold problem},\ }\href {https://doi.org/10.48550/arXiv.2008.11245} {\bibfield  {journal} {\bibinfo  {journal} {arXiv preprint arXiv:2008.11245}\ } (\bibinfo {year} {2020})}\BibitemShut {NoStop}%
\bibitem [{\citenamefont {Lorena}\ \emph {et~al.}(2018)\citenamefont {Lorena}, \citenamefont {Maciel}, \citenamefont {de~Miranda}, \citenamefont {Costa},\ and\ \citenamefont {Prud{\^e}ncio}}]{lorena2018data}%
  \BibitemOpen
  \bibfield  {author} {\bibinfo {author} {\bibfnamefont {A.~C.}\ \bibnamefont {Lorena}}, \bibinfo {author} {\bibfnamefont {A.~I.}\ \bibnamefont {Maciel}}, \bibinfo {author} {\bibfnamefont {P.~B.}\ \bibnamefont {de~Miranda}}, \bibinfo {author} {\bibfnamefont {I.~G.}\ \bibnamefont {Costa}},\ and\ \bibinfo {author} {\bibfnamefont {R.~B.}\ \bibnamefont {Prud{\^e}ncio}},\ }\bibfield  {title} {\bibinfo {title} {Data complexity meta-features for regression problems},\ }\href {https://doi.org/10.1007/s10994-017-5681-1} {\bibfield  {journal} {\bibinfo  {journal} {Machine Learning}\ }\textbf {\bibinfo {volume} {107}},\ \bibinfo {pages} {209} (\bibinfo {year} {2018})}\BibitemShut {NoStop}%
\bibitem [{\citenamefont {Pan}\ \emph {et~al.}(2023)\citenamefont {Pan}, \citenamefont {Cao}, \citenamefont {Wang}, \citenamefont {Hua}, \citenamefont {Cai}, \citenamefont {Li}, \citenamefont {Wang}, \citenamefont {Hu}, \citenamefont {Song}, \citenamefont {Deng} \emph {et~al.}}]{pan2023experimental}%
  \BibitemOpen
  \bibfield  {author} {\bibinfo {author} {\bibfnamefont {X.}~\bibnamefont {Pan}}, \bibinfo {author} {\bibfnamefont {X.}~\bibnamefont {Cao}}, \bibinfo {author} {\bibfnamefont {W.}~\bibnamefont {Wang}}, \bibinfo {author} {\bibfnamefont {Z.}~\bibnamefont {Hua}}, \bibinfo {author} {\bibfnamefont {W.}~\bibnamefont {Cai}}, \bibinfo {author} {\bibfnamefont {X.}~\bibnamefont {Li}}, \bibinfo {author} {\bibfnamefont {H.}~\bibnamefont {Wang}}, \bibinfo {author} {\bibfnamefont {J.}~\bibnamefont {Hu}}, \bibinfo {author} {\bibfnamefont {Y.}~\bibnamefont {Song}}, \bibinfo {author} {\bibfnamefont {D.-L.}\ \bibnamefont {Deng}}, \emph {et~al.},\ }\bibfield  {title} {\bibinfo {title} {Experimental quantum end-to-end learning on a superconducting processor},\ }\href@noop {} {\bibfield  {journal} {\bibinfo  {journal} {npj Quantum Information}\ }\textbf {\bibinfo {volume} {9}},\ \bibinfo {pages} {18} (\bibinfo {year} {2023})}\BibitemShut {NoStop}%
\bibitem [{\citenamefont {Murphy}(2012)}]{murphy_MachineLearningProbabilistic_2012}%
  \BibitemOpen
  \bibfield  {author} {\bibinfo {author} {\bibfnamefont {K.~P.}\ \bibnamefont {Murphy}},\ }\href@noop {} {\emph {\bibinfo {title} {Machine learning: a probabilistic perspective}}},\ \bibinfo {edition} {2nd}\ ed.\ (\bibinfo  {publisher} {MIT Press},\ \bibinfo {address} {Cambridge, MA},\ \bibinfo {year} {2012})\BibitemShut {NoStop}%
\bibitem [{\citenamefont {Coelho}\ \emph {et~al.}(2025)\citenamefont {Coelho}, \citenamefont {Kruse},\ and\ \citenamefont {Rosskopf}}]{coelho_QuantumEfficientKernelTarget_2025}%
  \BibitemOpen
  \bibfield  {author} {\bibinfo {author} {\bibfnamefont {R.}~\bibnamefont {Coelho}}, \bibinfo {author} {\bibfnamefont {G.}~\bibnamefont {Kruse}},\ and\ \bibinfo {author} {\bibfnamefont {A.}~\bibnamefont {Rosskopf}},\ }\bibfield  {title} {\bibinfo {title} {Quantum-{Efficient} {Kernel} {Target} {Alignment}},\ }\href {http://arxiv.org/abs/2502.08225} {\bibfield  {journal} {\bibinfo  {journal} {arXiv preprint arXiv:2502.08225}\ } (\bibinfo {year} {2025})}\BibitemShut {NoStop}%
\bibitem [{\citenamefont {Fedus}\ \emph {et~al.}(2022)\citenamefont {Fedus}, \citenamefont {Zoph},\ and\ \citenamefont {Shazeer}}]{fedus2022switch}%
  \BibitemOpen
  \bibfield  {author} {\bibinfo {author} {\bibfnamefont {W.}~\bibnamefont {Fedus}}, \bibinfo {author} {\bibfnamefont {B.}~\bibnamefont {Zoph}},\ and\ \bibinfo {author} {\bibfnamefont {N.}~\bibnamefont {Shazeer}},\ }\bibfield  {title} {\bibinfo {title} {Switch transformers: Scaling to trillion parameter models with simple and efficient sparsity},\ }\href {http://jmlr.org/papers/v23/21-0998.html} {\bibfield  {journal} {\bibinfo  {journal} {Journal of Machine Learning Research}\ }\textbf {\bibinfo {volume} {23}},\ \bibinfo {pages} {1} (\bibinfo {year} {2022})}\BibitemShut {NoStop}%
\bibitem [{\citenamefont {Reinsel}\ \emph {et~al.}(2018)\citenamefont {Reinsel}, \citenamefont {Gantz},\ and\ \citenamefont {Rydning}}]{reinsel2018digitization}%
  \BibitemOpen
  \bibfield  {author} {\bibinfo {author} {\bibfnamefont {D.}~\bibnamefont {Reinsel}}, \bibinfo {author} {\bibfnamefont {J.}~\bibnamefont {Gantz}},\ and\ \bibinfo {author} {\bibfnamefont {J.}~\bibnamefont {Rydning}},\ }\bibfield  {title} {\bibinfo {title} {The digitization of the world from edge to core},\ }\href@noop {} {\bibfield  {journal} {\bibinfo  {journal} {Framingham: International Data Corporation}\ }\textbf {\bibinfo {volume} {16}},\ \bibinfo {pages} {1} (\bibinfo {year} {2018})}\BibitemShut {NoStop}%
\bibitem [{\citenamefont {Caro}\ \emph {et~al.}(2021)\citenamefont {Caro}, \citenamefont {{Gil-Fuster}}, \citenamefont {Meyer}, \citenamefont {Eisert},\ and\ \citenamefont {Sweke}}]{caro_EncodingdependentGeneralizationBounds_2021}%
  \BibitemOpen
  \bibfield  {author} {\bibinfo {author} {\bibfnamefont {M.~C.}\ \bibnamefont {Caro}}, \bibinfo {author} {\bibfnamefont {E.}~\bibnamefont {{Gil-Fuster}}}, \bibinfo {author} {\bibfnamefont {J.~J.}\ \bibnamefont {Meyer}}, \bibinfo {author} {\bibfnamefont {J.}~\bibnamefont {Eisert}},\ and\ \bibinfo {author} {\bibfnamefont {R.}~\bibnamefont {Sweke}},\ }\bibfield  {title} {\bibinfo {title} {Encoding-dependent generalization bounds for parametrized quantum circuits},\ }\href {https://doi.org/10.22331/q-2021-11-17-582} {\bibfield  {journal} {\bibinfo  {journal} {Quantum}\ }\textbf {\bibinfo {volume} {5}},\ \bibinfo {pages} {582} (\bibinfo {year} {2021})}\BibitemShut {NoStop}%
\bibitem [{\citenamefont {McClean}\ \emph {et~al.}(2018)\citenamefont {McClean}, \citenamefont {Boixo}, \citenamefont {Smelyanskiy}, \citenamefont {Babbush},\ and\ \citenamefont {Neven}}]{mcclean_BarrenPlateausQuantum_2018}%
  \BibitemOpen
  \bibfield  {author} {\bibinfo {author} {\bibfnamefont {J.~R.}\ \bibnamefont {McClean}}, \bibinfo {author} {\bibfnamefont {S.}~\bibnamefont {Boixo}}, \bibinfo {author} {\bibfnamefont {V.~N.}\ \bibnamefont {Smelyanskiy}}, \bibinfo {author} {\bibfnamefont {R.}~\bibnamefont {Babbush}},\ and\ \bibinfo {author} {\bibfnamefont {H.}~\bibnamefont {Neven}},\ }\bibfield  {title} {{\selectlanguage {en}\bibinfo {title} {Barren plateaus in quantum neural network training landscapes}},\ }\href {https://doi.org/10.1038/s41467-018-07090-4} {\bibfield  {journal} {\bibinfo  {journal} {Nature Communications}\ }\textbf {\bibinfo {volume} {9}},\ \bibinfo {pages} {4812} (\bibinfo {year} {2018})}\BibitemShut {NoStop}%
\bibitem [{\citenamefont {Holmes}\ \emph {et~al.}(2022)\citenamefont {Holmes}, \citenamefont {Sharma}, \citenamefont {Cerezo},\ and\ \citenamefont {Coles}}]{holmes_ConnectingAnsatzExpressibility_2022}%
  \BibitemOpen
  \bibfield  {author} {\bibinfo {author} {\bibfnamefont {Z.}~\bibnamefont {Holmes}}, \bibinfo {author} {\bibfnamefont {K.}~\bibnamefont {Sharma}}, \bibinfo {author} {\bibfnamefont {M.}~\bibnamefont {Cerezo}},\ and\ \bibinfo {author} {\bibfnamefont {P.~J.}\ \bibnamefont {Coles}},\ }\bibfield  {title} {\bibinfo {title} {Connecting {Ansatz} {Expressibility} to {Gradient} {Magnitudes} and {Barren} {Plateaus}},\ }\href {https://doi.org/10.1103/PRXQuantum.3.010313} {\bibfield  {journal} {\bibinfo  {journal} {PRX Quantum}\ }\textbf {\bibinfo {volume} {3}},\ \bibinfo {pages} {010313} (\bibinfo {year} {2022})},\ \bibinfo {note} {publisher: American Physical Society}\BibitemShut {NoStop}%
\bibitem [{\citenamefont {Mittal}\ \emph {et~al.}(2025)\citenamefont {Mittal}, \citenamefont {Kumar}, \citenamefont {Chand}, \citenamefont {Krishnakumar},\ and\ \citenamefont {Kundu}}]{Mittal_ExperimentalValidationDequantization_2025}%
  \BibitemOpen
  \bibfield  {author} {\bibinfo {author} {\bibfnamefont {S.}~\bibnamefont {Mittal}}, \bibinfo {author} {\bibfnamefont {M.}~\bibnamefont {Kumar}}, \bibinfo {author} {\bibfnamefont {Y.}~\bibnamefont {Chand}}, \bibinfo {author} {\bibfnamefont {R.}~\bibnamefont {Krishnakumar}},\ and\ \bibinfo {author} {\bibfnamefont {N.~K.}\ \bibnamefont {Kundu}},\ }\bibfield  {title} {\bibinfo {title} {Experimental validation of dequantization of hybrid quantum machine learning models using classical surrogates},\ }in\ \href {https://doi.org/10.1109/AIMLSystems67835.2025.11330280} {\emph {\bibinfo {booktitle} {2025 5th International Conference on AI-ML-Systems (AIMLSystems)}}}\ (\bibinfo  {publisher} {IEEE},\ \bibinfo {address} {Bangalore},\ \bibinfo {year} {2025})\ pp.\ \bibinfo {pages} {412--417}\BibitemShut {NoStop}%
\bibitem [{\citenamefont {Hernicht}\ \emph {et~al.}(2025)\citenamefont {Hernicht}, \citenamefont {Sakhnenko}, \citenamefont {O'Meara}, \citenamefont {Cortiana},\ and\ \citenamefont {Lorenz}}]{hernicht_EnhancingScalabilityClassical_2025}%
  \BibitemOpen
  \bibfield  {author} {\bibinfo {author} {\bibfnamefont {P.~A.}\ \bibnamefont {Hernicht}}, \bibinfo {author} {\bibfnamefont {A.}~\bibnamefont {Sakhnenko}}, \bibinfo {author} {\bibfnamefont {C.}~\bibnamefont {O'Meara}}, \bibinfo {author} {\bibfnamefont {G.}~\bibnamefont {Cortiana}},\ and\ \bibinfo {author} {\bibfnamefont {J.~M.}\ \bibnamefont {Lorenz}},\ }\href {https://doi.org/10.48550/arXiv.2508.06131} {\bibinfo {title} {Enhancing the {{Scalability}} of {{Classical Surrogates}} for {{Real-World Quantum Machine Learning Applications}}}} (\bibinfo {year} {2025}),\ \Eprint {https://arxiv.org/abs/2508.06131} {arXiv:2508.06131 [quant-ph]} \BibitemShut {NoStop}%
\bibitem [{Note4()}]{Note4}%
  \BibitemOpen
  \bibinfo {note} {Due to qubit demands, we do not directly compare Q-FLAIR feature-maps with fixed-ansatz maps matching the qubit count to the feature count, e.g., angle embedding, Z-feature-map, and ZZ-feature-map \cite {havlicek_SupervisedLearningQuantumenhanced_2019}.}\BibitemShut {Stop}%
\bibitem [{\citenamefont {Kerenidis}\ and\ \citenamefont {Luongo}(2020)}]{kerenidis2020classification}%
  \BibitemOpen
  \bibfield  {author} {\bibinfo {author} {\bibfnamefont {I.}~\bibnamefont {Kerenidis}}\ and\ \bibinfo {author} {\bibfnamefont {A.}~\bibnamefont {Luongo}},\ }\bibfield  {title} {\bibinfo {title} {Classification of the mnist data set with quantum slow feature analysis},\ }\href {https://doi.org/10.1103/PhysRevA.101.062327} {\bibfield  {journal} {\bibinfo  {journal} {Physical Review A}\ }\textbf {\bibinfo {volume} {101}},\ \bibinfo {pages} {062327} (\bibinfo {year} {2020})}\BibitemShut {NoStop}%
\bibitem [{\citenamefont {Slysz}\ \emph {et~al.}(2023)\citenamefont {Slysz}, \citenamefont {Kurowski}, \citenamefont {Walig{\'o}ra},\ and\ \citenamefont {W{\k{e}}glarz}}]{slysz2023exploring}%
  \BibitemOpen
  \bibfield  {author} {\bibinfo {author} {\bibfnamefont {M.}~\bibnamefont {Slysz}}, \bibinfo {author} {\bibfnamefont {K.}~\bibnamefont {Kurowski}}, \bibinfo {author} {\bibfnamefont {G.}~\bibnamefont {Walig{\'o}ra}},\ and\ \bibinfo {author} {\bibfnamefont {J.}~\bibnamefont {W{\k{e}}glarz}},\ }\bibfield  {title} {\bibinfo {title} {Exploring the capabilities of quantum support vector machines for image classification on the mnist benchmark},\ }in\ \href@noop {} {\emph {\bibinfo {booktitle} {Computational Science -- ICCS 2023}}},\ \bibinfo {editor} {edited by\ \bibinfo {editor} {\bibfnamefont {J.}~\bibnamefont {Miky{\v{s}}ka}}, \bibinfo {editor} {\bibfnamefont {C.}~\bibnamefont {de~Mulatier}}, \bibinfo {editor} {\bibfnamefont {M.}~\bibnamefont {Paszynski}}, \bibinfo {editor} {\bibfnamefont {V.~V.}\ \bibnamefont {Krzhizhanovskaya}}, \bibinfo {editor} {\bibfnamefont {J.~J.}\ \bibnamefont {Dongarra}},\ and\ \bibinfo {editor} {\bibfnamefont {P.~M.}\ \bibnamefont {Sloot}}}\ (\bibinfo  {publisher} {Springer Nature
  Switzerland},\ \bibinfo {address} {Cham},\ \bibinfo {year} {2023})\ pp.\ \bibinfo {pages} {193--200}\BibitemShut {NoStop}%
\bibitem [{\citenamefont {Senokosov}\ \emph {et~al.}(2024)\citenamefont {Senokosov}, \citenamefont {Sedykh}, \citenamefont {Sagingalieva}, \citenamefont {Kyriacou},\ and\ \citenamefont {Melnikov}}]{senokosov2024quantum}%
  \BibitemOpen
  \bibfield  {author} {\bibinfo {author} {\bibfnamefont {A.}~\bibnamefont {Senokosov}}, \bibinfo {author} {\bibfnamefont {A.}~\bibnamefont {Sedykh}}, \bibinfo {author} {\bibfnamefont {A.}~\bibnamefont {Sagingalieva}}, \bibinfo {author} {\bibfnamefont {B.}~\bibnamefont {Kyriacou}},\ and\ \bibinfo {author} {\bibfnamefont {A.}~\bibnamefont {Melnikov}},\ }\bibfield  {title} {\bibinfo {title} {Quantum machine learning for image classification},\ }\href {https://doi.org/10.1088/2632-2153/ad2aef} {\bibfield  {journal} {\bibinfo  {journal} {Machine Learning: Science and Technology}\ }\textbf {\bibinfo {volume} {5}},\ \bibinfo {pages} {015040} (\bibinfo {year} {2024})}\BibitemShut {NoStop}%
\bibitem [{\citenamefont {Zhou}\ \emph {et~al.}(2025)\citenamefont {Zhou}, \citenamefont {Sarkar}, \citenamefont {Bose},\ and\ \citenamefont {Bayat}}]{zhou2025enhanced}%
  \BibitemOpen
  \bibfield  {author} {\bibinfo {author} {\bibfnamefont {R.}~\bibnamefont {Zhou}}, \bibinfo {author} {\bibfnamefont {S.}~\bibnamefont {Sarkar}}, \bibinfo {author} {\bibfnamefont {S.}~\bibnamefont {Bose}},\ and\ \bibinfo {author} {\bibfnamefont {A.}~\bibnamefont {Bayat}},\ }\bibfield  {title} {\bibinfo {title} {Enhanced image classification via hybridizing quantum dynamics with classical neural networks},\ }\href {https://doi.org/10.48550/arXiv.2507.13587} {\bibfield  {journal} {\bibinfo  {journal} {arXiv preprint arXiv:2507.13587}\ } (\bibinfo {year} {2025})}\BibitemShut {NoStop}%
\bibitem [{\citenamefont {Erkan}\ \emph {et~al.}(2025)\citenamefont {Erkan}, \citenamefont {Rahebi},\ and\ \citenamefont {Yelghi}}]{erkan2025quantum}%
  \BibitemOpen
  \bibfield  {author} {\bibinfo {author} {\bibfnamefont {Z.}~\bibnamefont {Erkan}}, \bibinfo {author} {\bibfnamefont {J.}~\bibnamefont {Rahebi}},\ and\ \bibinfo {author} {\bibfnamefont {A.}~\bibnamefont {Yelghi}},\ }\bibfield  {title} {\bibinfo {title} {Quantum image dataset transform (qidt) for image processing},\ }\href {https://doi.org/10.1007/s11128-025-04754-1} {\bibfield  {journal} {\bibinfo  {journal} {Quantum Information Processing}\ }\textbf {\bibinfo {volume} {24}},\ \bibinfo {pages} {156} (\bibinfo {year} {2025})}\BibitemShut {NoStop}%
\bibitem [{\citenamefont {Chen}\ \emph {et~al.}(2025)\citenamefont {Chen}, \citenamefont {Liu},\ and\ \citenamefont {Yan}}]{chen2025exploring}%
  \BibitemOpen
  \bibfield  {author} {\bibinfo {author} {\bibfnamefont {K.}~\bibnamefont {Chen}}, \bibinfo {author} {\bibfnamefont {J.}~\bibnamefont {Liu}},\ and\ \bibinfo {author} {\bibfnamefont {F.}~\bibnamefont {Yan}},\ }\bibfield  {title} {\bibinfo {title} {Exploring quantum neural networks for binary classification on mnist dataset: A swap test approach},\ }\href {https://doi.org/10.1016/j.neunet.2025.107442} {\bibfield  {journal} {\bibinfo  {journal} {Neural Networks}\ }\textbf {\bibinfo {volume} {188}},\ \bibinfo {pages} {107442} (\bibinfo {year} {2025})}\BibitemShut {NoStop}%
\bibitem [{\citenamefont {Tognini}\ \emph {et~al.}(2025)\citenamefont {Tognini}, \citenamefont {Banchi},\ and\ \citenamefont {De~Palma}}]{tognini2025solving}%
  \BibitemOpen
  \bibfield  {author} {\bibinfo {author} {\bibfnamefont {P.~A.~X.}\ \bibnamefont {Tognini}}, \bibinfo {author} {\bibfnamefont {L.}~\bibnamefont {Banchi}},\ and\ \bibinfo {author} {\bibfnamefont {G.}~\bibnamefont {De~Palma}},\ }\bibfield  {title} {\bibinfo {title} {Solving mnist with a globally trained mixture of quantum experts},\ }\href {https://doi.org/10.48550/arXiv.2505.14789} {\bibfield  {journal} {\bibinfo  {journal} {arXiv preprint arXiv:2505.14789}\ } (\bibinfo {year} {2025})}\BibitemShut {NoStop}%
\bibitem [{\citenamefont {Manko}\ and\ \citenamefont {Frolovtsev}(2025)}]{manko2025classification}%
  \BibitemOpen
  \bibfield  {author} {\bibinfo {author} {\bibfnamefont {S.}~\bibnamefont {Manko}}\ and\ \bibinfo {author} {\bibfnamefont {D.}~\bibnamefont {Frolovtsev}},\ }\bibfield  {title} {\bibinfo {title} {Classification and reconstruction for single-pixel imaging with classical and quantum neural networks},\ }\href {https://doi.org/10.1007/s11760-025-03875-5} {\bibfield  {journal} {\bibinfo  {journal} {Signal, Image and Video Processing}\ }\textbf {\bibinfo {volume} {19}},\ \bibinfo {pages} {277} (\bibinfo {year} {2025})}\BibitemShut {NoStop}%
\bibitem [{\citenamefont {Wang}\ \emph {et~al.}(2025)\citenamefont {Wang}, \citenamefont {Myers}, \citenamefont {Hollenberg},\ and\ \citenamefont {Parampalli}}]{wang2025quantum}%
  \BibitemOpen
  \bibfield  {author} {\bibinfo {author} {\bibfnamefont {P.}~\bibnamefont {Wang}}, \bibinfo {author} {\bibfnamefont {C.~R.}\ \bibnamefont {Myers}}, \bibinfo {author} {\bibfnamefont {L.~C.}\ \bibnamefont {Hollenberg}},\ and\ \bibinfo {author} {\bibfnamefont {U.}~\bibnamefont {Parampalli}},\ }\bibfield  {title} {\bibinfo {title} {Quantum hamiltonian embedding of images for data reuploading classifiers},\ }\href {https://doi.org/10.1007/s42484-025-00247-7} {\bibfield  {journal} {\bibinfo  {journal} {Quantum Machine Intelligence}\ }\textbf {\bibinfo {volume} {7}},\ \bibinfo {pages} {35} (\bibinfo {year} {2025})}\BibitemShut {NoStop}%
\bibitem [{\citenamefont {Shen}\ \emph {et~al.}(2024)\citenamefont {Shen}, \citenamefont {Jobst}, \citenamefont {Shishenina},\ and\ \citenamefont {Pollmann}}]{shen2024classification}%
  \BibitemOpen
  \bibfield  {author} {\bibinfo {author} {\bibfnamefont {K.}~\bibnamefont {Shen}}, \bibinfo {author} {\bibfnamefont {B.}~\bibnamefont {Jobst}}, \bibinfo {author} {\bibfnamefont {E.}~\bibnamefont {Shishenina}},\ and\ \bibinfo {author} {\bibfnamefont {F.}~\bibnamefont {Pollmann}},\ }\href@noop {} {\bibinfo {title} {Classification of the {{Fashion-MNIST Dataset}} on a {{Quantum Computer}}}} (\bibinfo {year} {2024}),\ \Eprint {https://arxiv.org/abs/2403.02405} {arXiv:2403.02405} \BibitemShut {NoStop}%
\bibitem [{\citenamefont {Kiwit}\ \emph {et~al.}(2025)\citenamefont {Kiwit}, \citenamefont {Jobst}, \citenamefont {Luckow}, \citenamefont {Pollmann},\ and\ \citenamefont {Riofr{\'i}o}}]{kiwit2025typical}%
  \BibitemOpen
  \bibfield  {author} {\bibinfo {author} {\bibfnamefont {F.~J.}\ \bibnamefont {Kiwit}}, \bibinfo {author} {\bibfnamefont {B.}~\bibnamefont {Jobst}}, \bibinfo {author} {\bibfnamefont {A.}~\bibnamefont {Luckow}}, \bibinfo {author} {\bibfnamefont {F.}~\bibnamefont {Pollmann}},\ and\ \bibinfo {author} {\bibfnamefont {C.~A.}\ \bibnamefont {Riofr{\'i}o}},\ }\bibfield  {title} {\bibinfo {title} {Typical machine learning datasets as low-depth quantum circuits},\ }\href {https://doi.org/10.1088/2058-9565/ae0123} {\bibfield  {journal} {\bibinfo  {journal} {Quantum Science and Technology}\ }\textbf {\bibinfo {volume} {10}},\ \bibinfo {pages} {045035} (\bibinfo {year} {2025})}\BibitemShut {NoStop}%
\bibitem [{\citenamefont {R{\"o}seler}\ \emph {et~al.}(2025)\citenamefont {R{\"o}seler}, \citenamefont {Schaudt}, \citenamefont {Berg}, \citenamefont {Bauckhage},\ and\ \citenamefont {Koch}}]{roseler2025efficient}%
  \BibitemOpen
  \bibfield  {author} {\bibinfo {author} {\bibfnamefont {P.}~\bibnamefont {R{\"o}seler}}, \bibinfo {author} {\bibfnamefont {O.}~\bibnamefont {Schaudt}}, \bibinfo {author} {\bibfnamefont {H.}~\bibnamefont {Berg}}, \bibinfo {author} {\bibfnamefont {C.}~\bibnamefont {Bauckhage}},\ and\ \bibinfo {author} {\bibfnamefont {M.}~\bibnamefont {Koch}},\ }\bibfield  {title} {\bibinfo {title} {Efficient quantum convolutional neural networks for image classification: Overcoming hardware constraints},\ }\href {https://doi.org/10.48550/arXiv.2505.05957} {\bibfield  {journal} {\bibinfo  {journal} {arXiv preprint arXiv:2505.05957}\ } (\bibinfo {year} {2025})}\BibitemShut {NoStop}%
\bibitem [{\citenamefont {Zeng}\ \emph {et~al.}(2022)\citenamefont {Zeng}, \citenamefont {Wang}, \citenamefont {He}, \citenamefont {Huang},\ and\ \citenamefont {Chang}}]{zeng2022multi}%
  \BibitemOpen
  \bibfield  {author} {\bibinfo {author} {\bibfnamefont {Y.}~\bibnamefont {Zeng}}, \bibinfo {author} {\bibfnamefont {H.}~\bibnamefont {Wang}}, \bibinfo {author} {\bibfnamefont {J.}~\bibnamefont {He}}, \bibinfo {author} {\bibfnamefont {Q.}~\bibnamefont {Huang}},\ and\ \bibinfo {author} {\bibfnamefont {S.}~\bibnamefont {Chang}},\ }\bibfield  {title} {\bibinfo {title} {A multi-classification hybrid quantum neural network using an all-qubit multi-observable measurement strategy},\ }\href@noop {} {\bibfield  {journal} {\bibinfo  {journal} {Entropy}\ }\textbf {\bibinfo {volume} {24}},\ \bibinfo {pages} {394} (\bibinfo {year} {2022})}\BibitemShut {NoStop}%
\bibitem [{\citenamefont {Dhara}\ \emph {et~al.}(2024)\citenamefont {Dhara}, \citenamefont {Agrawal},\ and\ \citenamefont {Roy}}]{dhara2024multi}%
  \BibitemOpen
  \bibfield  {author} {\bibinfo {author} {\bibfnamefont {B.}~\bibnamefont {Dhara}}, \bibinfo {author} {\bibfnamefont {M.}~\bibnamefont {Agrawal}},\ and\ \bibinfo {author} {\bibfnamefont {S.~D.}\ \bibnamefont {Roy}},\ }\bibfield  {title} {\bibinfo {title} {Multi-class classification using quantum transfer learning},\ }\href@noop {} {\bibfield  {journal} {\bibinfo  {journal} {Quantum Information Processing}\ }\textbf {\bibinfo {volume} {23}},\ \bibinfo {pages} {34} (\bibinfo {year} {2024})}\BibitemShut {NoStop}%
\bibitem [{\citenamefont {Riaz}\ \emph {et~al.}(2023)\citenamefont {Riaz}, \citenamefont {Abdulla}, \citenamefont {Suzuki}, \citenamefont {Ganguly}, \citenamefont {Deo},\ and\ \citenamefont {Hopkins}}]{riaz2023application}%
  \BibitemOpen
  \bibfield  {author} {\bibinfo {author} {\bibfnamefont {F.}~\bibnamefont {Riaz}}, \bibinfo {author} {\bibfnamefont {S.}~\bibnamefont {Abdulla}}, \bibinfo {author} {\bibfnamefont {H.}~\bibnamefont {Suzuki}}, \bibinfo {author} {\bibfnamefont {S.}~\bibnamefont {Ganguly}}, \bibinfo {author} {\bibfnamefont {R.~C.}\ \bibnamefont {Deo}},\ and\ \bibinfo {author} {\bibfnamefont {S.}~\bibnamefont {Hopkins}},\ }\bibfield  {title} {\bibinfo {title} {Application of quantum pre-processing filter for binary image classification with small samples},\ }\href@noop {} {\bibfield  {journal} {\bibinfo  {journal} {arXiv:2308.14930}\ } (\bibinfo {year} {2023})}\BibitemShut {NoStop}%
\bibitem [{\citenamefont {Grimsley}\ \emph {et~al.}(2023)\citenamefont {Grimsley}, \citenamefont {Barron}, \citenamefont {Barnes}, \citenamefont {Economou},\ and\ \citenamefont {Mayhall}}]{grimsley_AdaptiveProblemtailoredVariational_2023a}%
  \BibitemOpen
  \bibfield  {author} {\bibinfo {author} {\bibfnamefont {H.~R.}\ \bibnamefont {Grimsley}}, \bibinfo {author} {\bibfnamefont {G.~S.}\ \bibnamefont {Barron}}, \bibinfo {author} {\bibfnamefont {E.}~\bibnamefont {Barnes}}, \bibinfo {author} {\bibfnamefont {S.~E.}\ \bibnamefont {Economou}},\ and\ \bibinfo {author} {\bibfnamefont {N.~J.}\ \bibnamefont {Mayhall}},\ }\bibfield  {title} {{\selectlanguage {en}\bibinfo {title} {Adaptive, problem-tailored variational quantum eigensolver mitigates rough parameter landscapes and barren plateaus}},\ }\href {https://doi.org/10.1038/s41534-023-00681-0} {\bibfield  {journal} {\bibinfo  {journal} {npj Quantum Information}\ }\textbf {\bibinfo {volume} {9}},\ \bibinfo {pages} {1} (\bibinfo {year} {2023})}\BibitemShut {NoStop}%
\bibitem [{\citenamefont {Skolik}\ \emph {et~al.}(2021)\citenamefont {Skolik}, \citenamefont {McClean}, \citenamefont {Mohseni}, \citenamefont {van~der Smagt},\ and\ \citenamefont {Leib}}]{skolik_LayerwiseLearningQuantum_2021}%
  \BibitemOpen
  \bibfield  {author} {\bibinfo {author} {\bibfnamefont {A.}~\bibnamefont {Skolik}}, \bibinfo {author} {\bibfnamefont {J.~R.}\ \bibnamefont {McClean}}, \bibinfo {author} {\bibfnamefont {M.}~\bibnamefont {Mohseni}}, \bibinfo {author} {\bibfnamefont {P.}~\bibnamefont {van~der Smagt}},\ and\ \bibinfo {author} {\bibfnamefont {M.}~\bibnamefont {Leib}},\ }\bibfield  {title} {{\selectlanguage {en}\bibinfo {title} {Layerwise learning for quantum neural networks}},\ }\href {https://doi.org/10.1007/s42484-020-00036-4} {\bibfield  {journal} {\bibinfo  {journal} {Quantum Machine Intelligence}\ }\textbf {\bibinfo {volume} {3}},\ \bibinfo {pages} {5} (\bibinfo {year} {2021})}\BibitemShut {NoStop}%
\bibitem [{\citenamefont {Nair}\ and\ \citenamefont {Ferrie}(2025)}]{nair_LocalSurrogatesQuantum_2025}%
  \BibitemOpen
  \bibfield  {author} {\bibinfo {author} {\bibfnamefont {S.~R.}\ \bibnamefont {Nair}}\ and\ \bibinfo {author} {\bibfnamefont {C.}~\bibnamefont {Ferrie}},\ }\href {https://doi.org/10.48550/arXiv.2506.09425} {\bibinfo {title} {Local surrogates for quantum machine learning}} (\bibinfo {year} {2025}),\ \Eprint {https://arxiv.org/abs/2506.09425} {arXiv:2506.09425 [quant-ph]} \BibitemShut {NoStop}%
\bibitem [{\citenamefont {Pérez-Salinas}\ \emph {et~al.}(2020)\citenamefont {Pérez-Salinas}, \citenamefont {Cervera-Lierta}, \citenamefont {Gil-Fuster},\ and\ \citenamefont {Latorre}}]{perez-salinas_DataReuploadingUniversal_2020}%
  \BibitemOpen
  \bibfield  {author} {\bibinfo {author} {\bibfnamefont {A.}~\bibnamefont {Pérez-Salinas}}, \bibinfo {author} {\bibfnamefont {A.}~\bibnamefont {Cervera-Lierta}}, \bibinfo {author} {\bibfnamefont {E.}~\bibnamefont {Gil-Fuster}},\ and\ \bibinfo {author} {\bibfnamefont {J.~I.}\ \bibnamefont {Latorre}},\ }\bibfield  {title} {{\selectlanguage {en}\bibinfo {title} {Data re-uploading for a universal quantum classifier}},\ }\href {https://doi.org/10.22331/q-2020-02-06-226} {\bibfield  {journal} {\bibinfo  {journal} {Quantum}\ }\textbf {\bibinfo {volume} {4}},\ \bibinfo {pages} {226} (\bibinfo {year} {2020})}\BibitemShut {NoStop}%
\bibitem [{\citenamefont {Wierichs}\ \emph {et~al.}(2022)\citenamefont {Wierichs}, \citenamefont {Izaac}, \citenamefont {Wang},\ and\ \citenamefont {Lin}}]{wierichs_GeneralParametershiftRules_2022}%
  \BibitemOpen
  \bibfield  {author} {\bibinfo {author} {\bibfnamefont {D.}~\bibnamefont {Wierichs}}, \bibinfo {author} {\bibfnamefont {J.}~\bibnamefont {Izaac}}, \bibinfo {author} {\bibfnamefont {C.}~\bibnamefont {Wang}},\ and\ \bibinfo {author} {\bibfnamefont {C.~Y.-Y.}\ \bibnamefont {Lin}},\ }\bibfield  {title} {{\selectlanguage {en-GB}\bibinfo {title} {General parameter-shift rules for quantum gradients}},\ }\href {https://doi.org/10.22331/q-2022-03-30-677} {\bibfield  {journal} {\bibinfo  {journal} {Quantum}\ }\textbf {\bibinfo {volume} {6}},\ \bibinfo {pages} {677} (\bibinfo {year} {2022})}\BibitemShut {NoStop}%
\bibitem [{\citenamefont {Schuld}\ and\ \citenamefont {Petruccione}(2021)}]{schuld_MachineLearningQuantum_2021}%
  \BibitemOpen
  \bibfield  {author} {\bibinfo {author} {\bibfnamefont {M.}~\bibnamefont {Schuld}}\ and\ \bibinfo {author} {\bibfnamefont {F.}~\bibnamefont {Petruccione}},\ }\href {https://doi.org/10.1007/978-3-030-83098-4} {\emph {\bibinfo {title} {Machine {Learning} with {Quantum} {Computers}}}},\ Quantum {Science} and {Technology}\ (\bibinfo  {publisher} {Springer International Publishing},\ \bibinfo {address} {Cham},\ \bibinfo {year} {2021})\BibitemShut {NoStop}%
\bibitem [{Note5()}]{Note5}%
  \BibitemOpen
  \bibinfo {note} {The discrepancy introduced by classical non-linear pre-processing, is another reason why we refrained from directly comparing the feature-maps composed by Q-FLAIR with such fixed-ansatz quantum feature-maps as the ZZ-feature-map.}\BibitemShut {Stop}%
\bibitem [{\citenamefont {J\"{a}ger}\ \emph {et~al.}(2026)\citenamefont {J\"{a}ger}, \citenamefont {Els\"{a}sser},\ and\ \citenamefont {Torabian}}]{zenodo}%
  \BibitemOpen
  \bibfield  {author} {\bibinfo {author} {\bibfnamefont {J.}~\bibnamefont {J\"{a}ger}}, \bibinfo {author} {\bibfnamefont {P.}~\bibnamefont {Els\"{a}sser}},\ and\ \bibinfo {author} {\bibfnamefont {E.}~\bibnamefont {Torabian}},\ }\href {https://doi.org/10.5281/zenodo.20101655} {\bibinfo {title} {Q-flair: Quantum feature-map learning with reduced resource overhead}} (\bibinfo {year} {2026})\BibitemShut {NoStop}%
\bibitem [{\citenamefont {Li}\ \emph {et~al.}(2017)\citenamefont {Li}, \citenamefont {Yang}, \citenamefont {Peng},\ and\ \citenamefont {Sun}}]{li_HybridQuantumClassicalApproach_2017}%
  \BibitemOpen
  \bibfield  {author} {\bibinfo {author} {\bibfnamefont {J.}~\bibnamefont {Li}}, \bibinfo {author} {\bibfnamefont {X.}~\bibnamefont {Yang}}, \bibinfo {author} {\bibfnamefont {X.}~\bibnamefont {Peng}},\ and\ \bibinfo {author} {\bibfnamefont {C.-P.}\ \bibnamefont {Sun}},\ }\bibfield  {title} {\bibinfo {title} {Hybrid {Quantum}-{Classical} {Approach} to {Quantum} {Optimal} {Control}},\ }\href {https://doi.org/10.1103/PhysRevLett.118.150503} {\bibfield  {journal} {\bibinfo  {journal} {Physical Review Letters}\ }\textbf {\bibinfo {volume} {118}},\ \bibinfo {pages} {150503} (\bibinfo {year} {2017})}\BibitemShut {NoStop}%
\bibitem [{\citenamefont {Schuld}\ \emph {et~al.}(2019)\citenamefont {Schuld}, \citenamefont {Bergholm}, \citenamefont {Gogolin}, \citenamefont {Izaac},\ and\ \citenamefont {Killoran}}]{schuld_EvaluatingAnalyticGradients_2019}%
  \BibitemOpen
  \bibfield  {author} {\bibinfo {author} {\bibfnamefont {M.}~\bibnamefont {Schuld}}, \bibinfo {author} {\bibfnamefont {V.}~\bibnamefont {Bergholm}}, \bibinfo {author} {\bibfnamefont {C.}~\bibnamefont {Gogolin}}, \bibinfo {author} {\bibfnamefont {J.}~\bibnamefont {Izaac}},\ and\ \bibinfo {author} {\bibfnamefont {N.}~\bibnamefont {Killoran}},\ }\bibfield  {title} {{\selectlanguage {en}\bibinfo {title} {Evaluating analytic gradients on quantum hardware}},\ }\href {https://doi.org/10.1103/PhysRevA.99.032331} {\bibfield  {journal} {\bibinfo  {journal} {Physical Review A}\ }\textbf {\bibinfo {volume} {99}},\ \bibinfo {pages} {032331} (\bibinfo {year} {2019})}\BibitemShut {NoStop}%
\bibitem [{Note6()}]{Note6}%
  \BibitemOpen
  \bibinfo {note} {One could further consider purely weight-dependent gates as a special case of the weight-data-dependent gates if a constant feature $\pi $ is included (considering a parameter range of $[-1, 1]$) in the input data points, which is a common approach for convenience in machine learning practice.}\BibitemShut {Stop}%
\bibitem [{\citenamefont {Nation}\ \emph {et~al.}(2021)\citenamefont {Nation}, \citenamefont {Kang}, \citenamefont {Sundaresan},\ and\ \citenamefont {Gambetta}}]{nation_ScalableMitigationMeasurement_2021}%
  \BibitemOpen
  \bibfield  {author} {\bibinfo {author} {\bibfnamefont {P.~D.}\ \bibnamefont {Nation}}, \bibinfo {author} {\bibfnamefont {H.}~\bibnamefont {Kang}}, \bibinfo {author} {\bibfnamefont {N.}~\bibnamefont {Sundaresan}},\ and\ \bibinfo {author} {\bibfnamefont {J.~M.}\ \bibnamefont {Gambetta}},\ }\bibfield  {title} {\bibinfo {title} {Scalable {Mitigation} of {Measurement} {Errors} on {Quantum} {Computers}},\ }\href {https://doi.org/10.1103/PRXQuantum.2.040326} {\bibfield  {journal} {\bibinfo  {journal} {PRX Quantum}\ }\textbf {\bibinfo {volume} {2}},\ \bibinfo {pages} {040326} (\bibinfo {year} {2021})}\BibitemShut {NoStop}%
\bibitem [{\citenamefont {Bishop}(2006)}]{bishop_PatternRecognitionMachine_2006}%
  \BibitemOpen
  \bibfield  {author} {\bibinfo {author} {\bibfnamefont {C.~M.}\ \bibnamefont {Bishop}},\ }\href@noop {} {\emph {\bibinfo {title} {Pattern recognition and machine learning}}},\ Information science and statistics\ (\bibinfo  {publisher} {Springer},\ \bibinfo {address} {New York},\ \bibinfo {year} {2006})\BibitemShut {NoStop}%
\bibitem [{\citenamefont {Thanasilp}\ \emph {et~al.}(2024)\citenamefont {Thanasilp}, \citenamefont {Wang}, \citenamefont {Cerezo},\ and\ \citenamefont {Holmes}}]{thanasilp_ExponentialConcentrationQuantum_2024}%
  \BibitemOpen
  \bibfield  {author} {\bibinfo {author} {\bibfnamefont {S.}~\bibnamefont {Thanasilp}}, \bibinfo {author} {\bibfnamefont {S.}~\bibnamefont {Wang}}, \bibinfo {author} {\bibfnamefont {M.}~\bibnamefont {Cerezo}},\ and\ \bibinfo {author} {\bibfnamefont {Z.}~\bibnamefont {Holmes}},\ }\bibfield  {title} {\bibinfo {title} {Exponential concentration in quantum kernel methods},\ }\href {https://doi.org/10.1038/s41467-024-49287-w} {\bibfield  {journal} {\bibinfo  {journal} {Nature Communications}\ }\textbf {\bibinfo {volume} {15}},\ \bibinfo {pages} {5200} (\bibinfo {year} {2024})}\BibitemShut {NoStop}%
\bibitem [{\citenamefont {Barzen}\ and\ \citenamefont {Leymann}(2025)}]{barzen2025differential}%
  \BibitemOpen
  \bibfield  {author} {\bibinfo {author} {\bibfnamefont {J.}~\bibnamefont {Barzen}}\ and\ \bibinfo {author} {\bibfnamefont {F.}~\bibnamefont {Leymann}},\ }\bibfield  {title} {\bibinfo {title} {On the differential topology of expressivity of parameterized quantum circuits},\ }\href@noop {} {\bibfield  {journal} {\bibinfo  {journal} {AppliedMath}\ }\textbf {\bibinfo {volume} {5}},\ \bibinfo {pages} {121} (\bibinfo {year} {2025})}\BibitemShut {NoStop}%
\bibitem [{\citenamefont {Yao}(2025)}]{yao2025learning}%
  \BibitemOpen
  \bibfield  {author} {\bibinfo {author} {\bibfnamefont {J.}~\bibnamefont {Yao}},\ }\bibfield  {title} {\bibinfo {title} {Learning to maximize quantum neural network expressivity via effective rank},\ }\href@noop {} {\bibfield  {journal} {\bibinfo  {journal} {arXiv preprint arXiv:2506.15375}\ } (\bibinfo {year} {2025})}\BibitemShut {NoStop}%
\bibitem [{\citenamefont {Javadi-Abhari}\ \emph {et~al.}(2024)\citenamefont {Javadi-Abhari}, \citenamefont {Treinish}, \citenamefont {Krsulich}, \citenamefont {Wood}, \citenamefont {Lishman}, \citenamefont {Gacon}, \citenamefont {Martiel}, \citenamefont {Nation}, \citenamefont {Bishop}, \citenamefont {Cross}, \citenamefont {Johnson},\ and\ \citenamefont {Gambetta}}]{qiskit2024}%
  \BibitemOpen
  \bibfield  {author} {\bibinfo {author} {\bibfnamefont {A.}~\bibnamefont {Javadi-Abhari}}, \bibinfo {author} {\bibfnamefont {M.}~\bibnamefont {Treinish}}, \bibinfo {author} {\bibfnamefont {K.}~\bibnamefont {Krsulich}}, \bibinfo {author} {\bibfnamefont {C.~J.}\ \bibnamefont {Wood}}, \bibinfo {author} {\bibfnamefont {J.}~\bibnamefont {Lishman}}, \bibinfo {author} {\bibfnamefont {J.}~\bibnamefont {Gacon}}, \bibinfo {author} {\bibfnamefont {S.}~\bibnamefont {Martiel}}, \bibinfo {author} {\bibfnamefont {P.~D.}\ \bibnamefont {Nation}}, \bibinfo {author} {\bibfnamefont {L.~S.}\ \bibnamefont {Bishop}}, \bibinfo {author} {\bibfnamefont {A.~W.}\ \bibnamefont {Cross}}, \bibinfo {author} {\bibfnamefont {B.~R.}\ \bibnamefont {Johnson}},\ and\ \bibinfo {author} {\bibfnamefont {J.~M.}\ \bibnamefont {Gambetta}},\ }\href {https://doi.org/10.48550/arXiv.2405.08810} {\bibinfo {title} {Quantum computing with {Q}iskit}} (\bibinfo {year} {2024}),\ \Eprint {https://arxiv.org/abs/2405.08810} {arXiv:2405.08810 [quant-ph]}
  \BibitemShut {NoStop}%
\bibitem [{\citenamefont {{Qiskit contributors}}(2023)}]{qiskit_zenodo}%
  \BibitemOpen
  \bibfield  {author} {\bibinfo {author} {\bibnamefont {{Qiskit contributors}}},\ }\href {https://doi.org/10.5281/zenodo.8190968} {\bibinfo {title} {Qiskit 0.44.0}} (\bibinfo {year} {2023})\BibitemShut {NoStop}%
\bibitem [{\citenamefont {Monbroussou}\ \emph {et~al.}(2025)\citenamefont {Monbroussou}, \citenamefont {Mamon}, \citenamefont {Landman}, \citenamefont {Grilo}, \citenamefont {Kukla},\ and\ \citenamefont {Kashefi}}]{monbroussou2025trainability}%
  \BibitemOpen
  \bibfield  {author} {\bibinfo {author} {\bibfnamefont {L.}~\bibnamefont {Monbroussou}}, \bibinfo {author} {\bibfnamefont {E.~Z.}\ \bibnamefont {Mamon}}, \bibinfo {author} {\bibfnamefont {J.}~\bibnamefont {Landman}}, \bibinfo {author} {\bibfnamefont {A.~B.}\ \bibnamefont {Grilo}}, \bibinfo {author} {\bibfnamefont {R.}~\bibnamefont {Kukla}},\ and\ \bibinfo {author} {\bibfnamefont {E.}~\bibnamefont {Kashefi}},\ }\bibfield  {title} {\bibinfo {title} {Trainability and expressivity of hamming-weight preserving quantum circuits for machine learning},\ }\href@noop {} {\bibfield  {journal} {\bibinfo  {journal} {Quantum}\ }\textbf {\bibinfo {volume} {9}},\ \bibinfo {pages} {1745} (\bibinfo {year} {2025})}\BibitemShut {NoStop}%
\bibitem [{\citenamefont {Haug}\ and\ \citenamefont {Kim}(2024)}]{haug_GeneralizationQuantumMachine_2024}%
  \BibitemOpen
  \bibfield  {author} {\bibinfo {author} {\bibfnamefont {T.}~\bibnamefont {Haug}}\ and\ \bibinfo {author} {\bibfnamefont {M.~S.}\ \bibnamefont {Kim}},\ }\bibfield  {title} {\bibinfo {title} {Generalization of {{Quantum Machine Learning Models Using Quantum Fisher Information Metric}}},\ }\href {https://doi.org/10.1103/PhysRevLett.133.050603} {\bibfield  {journal} {\bibinfo  {journal} {Physical Review Letters}\ }\textbf {\bibinfo {volume} {133}},\ \bibinfo {pages} {050603} (\bibinfo {year} {2024})}\BibitemShut {NoStop}%
\bibitem [{\citenamefont {Meyer}(2021)}]{meyer_FisherInformationNoisy_2021}%
  \BibitemOpen
  \bibfield  {author} {\bibinfo {author} {\bibfnamefont {J.~J.}\ \bibnamefont {Meyer}},\ }\bibfield  {title} {\bibinfo {title} {Fisher {{Information}} in {{Noisy Intermediate-Scale Quantum Applications}}},\ }\href {https://doi.org/10.22331/q-2021-09-09-539} {\bibfield  {journal} {\bibinfo  {journal} {Quantum}\ }\textbf {\bibinfo {volume} {5}},\ \bibinfo {pages} {539} (\bibinfo {year} {2021})}\BibitemShut {NoStop}%
\bibitem [{\citenamefont {Braunstein}\ and\ \citenamefont {Caves}(1994)}]{braunstein_StatisticalDistanceGeometry_1994}%
  \BibitemOpen
  \bibfield  {author} {\bibinfo {author} {\bibfnamefont {S.~L.}\ \bibnamefont {Braunstein}}\ and\ \bibinfo {author} {\bibfnamefont {C.~M.}\ \bibnamefont {Caves}},\ }\bibfield  {title} {\bibinfo {title} {Statistical distance and the geometry of quantum states},\ }\href {https://doi.org/10.1103/PhysRevLett.72.3439} {\bibfield  {journal} {\bibinfo  {journal} {Physical Review Letters}\ }\textbf {\bibinfo {volume} {72}},\ \bibinfo {pages} {3439} (\bibinfo {year} {1994})}\BibitemShut {NoStop}%
\bibitem [{\citenamefont {Schuld}(2021)}]{schuld2021supervisedquantummachinelearning}%
  \BibitemOpen
  \bibfield  {author} {\bibinfo {author} {\bibfnamefont {M.}~\bibnamefont {Schuld}},\ }\href {https://arxiv.org/abs/2101.11020} {\bibinfo {title} {Supervised quantum machine learning models are kernel methods}} (\bibinfo {year} {2021}),\ \Eprint {https://arxiv.org/abs/2101.11020} {arXiv:2101.11020 [quant-ph]} \BibitemShut {NoStop}%
\bibitem [{\citenamefont {Abbas}\ \emph {et~al.}(2021{\natexlab{a}})\citenamefont {Abbas}, \citenamefont {Sutter}, \citenamefont {Figalli},\ and\ \citenamefont {Woerner}}]{abbas2021effectivedimensionmachinelearning}%
  \BibitemOpen
  \bibfield  {author} {\bibinfo {author} {\bibfnamefont {A.}~\bibnamefont {Abbas}}, \bibinfo {author} {\bibfnamefont {D.}~\bibnamefont {Sutter}}, \bibinfo {author} {\bibfnamefont {A.}~\bibnamefont {Figalli}},\ and\ \bibinfo {author} {\bibfnamefont {S.}~\bibnamefont {Woerner}},\ }\href {https://arxiv.org/abs/2112.04807} {\bibinfo {title} {Effective dimension of machine learning models}} (\bibinfo {year} {2021}{\natexlab{a}}),\ \Eprint {https://arxiv.org/abs/2112.04807} {arXiv:2112.04807 [cs.LG]} \BibitemShut {NoStop}%
\bibitem [{\citenamefont {Abbas}\ \emph {et~al.}(2021{\natexlab{b}})\citenamefont {Abbas}, \citenamefont {Sutter}, \citenamefont {Zoufal}, \citenamefont {Lucchi}, \citenamefont {Figalli},\ and\ \citenamefont {Woerner}}]{Abbas2021power}%
  \BibitemOpen
  \bibfield  {author} {\bibinfo {author} {\bibfnamefont {A.}~\bibnamefont {Abbas}}, \bibinfo {author} {\bibfnamefont {D.}~\bibnamefont {Sutter}}, \bibinfo {author} {\bibfnamefont {C.}~\bibnamefont {Zoufal}}, \bibinfo {author} {\bibfnamefont {A.}~\bibnamefont {Lucchi}}, \bibinfo {author} {\bibfnamefont {A.}~\bibnamefont {Figalli}},\ and\ \bibinfo {author} {\bibfnamefont {S.}~\bibnamefont {Woerner}},\ }\bibfield  {title} {\bibinfo {title} {The power of quantum neural networks},\ }\href {https://doi.org/10.1038/s43588-021-00084-1} {\bibfield  {journal} {\bibinfo  {journal} {Nature Computational Science}\ }\textbf {\bibinfo {volume} {1}},\ \bibinfo {pages} {403} (\bibinfo {year} {2021}{\natexlab{b}})}\BibitemShut {NoStop}%
\bibitem [{\citenamefont {Berezniuk}\ \emph {et~al.}(2020)\citenamefont {Berezniuk}, \citenamefont {Figalli}, \citenamefont {Ghigliazza},\ and\ \citenamefont {Musaelian}}]{berezniuk2020scaledependentnotioneffectivedimension}%
  \BibitemOpen
  \bibfield  {author} {\bibinfo {author} {\bibfnamefont {O.}~\bibnamefont {Berezniuk}}, \bibinfo {author} {\bibfnamefont {A.}~\bibnamefont {Figalli}}, \bibinfo {author} {\bibfnamefont {R.}~\bibnamefont {Ghigliazza}},\ and\ \bibinfo {author} {\bibfnamefont {K.}~\bibnamefont {Musaelian}},\ }\href {https://arxiv.org/abs/2001.10872} {\bibinfo {title} {A scale-dependent notion of effective dimension}} (\bibinfo {year} {2020}),\ \Eprint {https://arxiv.org/abs/2001.10872} {arXiv:2001.10872 [stat.ML]} \BibitemShut {NoStop}%
\bibitem [{\citenamefont {Byrd}\ \emph {et~al.}(1995)\citenamefont {Byrd}, \citenamefont {Lu}, \citenamefont {Nocedal},\ and\ \citenamefont {Zhu}}]{byrd1995limited}%
  \BibitemOpen
  \bibfield  {author} {\bibinfo {author} {\bibfnamefont {R.~H.}\ \bibnamefont {Byrd}}, \bibinfo {author} {\bibfnamefont {P.}~\bibnamefont {Lu}}, \bibinfo {author} {\bibfnamefont {J.}~\bibnamefont {Nocedal}},\ and\ \bibinfo {author} {\bibfnamefont {C.}~\bibnamefont {Zhu}},\ }\bibfield  {title} {\bibinfo {title} {A limited memory algorithm for bound constrained optimization},\ }\href {https://doi.org/10.1137/0916069} {\bibfield  {journal} {\bibinfo  {journal} {SIAM Journal on Scientific Computing}\ }\textbf {\bibinfo {volume} {16}},\ \bibinfo {pages} {1190} (\bibinfo {year} {1995})}\BibitemShut {NoStop}%
\bibitem [{\citenamefont {Powell}(1998)}]{powell1998direct}%
  \BibitemOpen
  \bibfield  {author} {\bibinfo {author} {\bibfnamefont {M.~J.}\ \bibnamefont {Powell}},\ }\bibfield  {title} {\bibinfo {title} {Direct search algorithms for optimization calculations},\ }\href {https://doi.org/10.1017/S0962492900002841} {\bibfield  {journal} {\bibinfo  {journal} {Acta Numerica}\ }\textbf {\bibinfo {volume} {7}},\ \bibinfo {pages} {287} (\bibinfo {year} {1998})}\BibitemShut {NoStop}%
\bibitem [{\citenamefont {Virtanen}\ \emph {et~al.}(2020)\citenamefont {Virtanen}, \citenamefont {Gommers}, \citenamefont {Oliphant}, \citenamefont {Haberland}, \citenamefont {Reddy}, \citenamefont {Cournapeau}, \citenamefont {Burovski}, \citenamefont {Peterson}, \citenamefont {Weckesser}, \citenamefont {Bright}, \citenamefont {{van der Walt}}, \citenamefont {Brett}, \citenamefont {Wilson}, \citenamefont {Millman}, \citenamefont {Mayorov}, \citenamefont {Nelson}, \citenamefont {Jones}, \citenamefont {Kern}, \citenamefont {Larson}, \citenamefont {Carey}, \citenamefont {Polat}, \citenamefont {Feng}, \citenamefont {Moore}, \citenamefont {{VanderPlas}}, \citenamefont {Laxalde}, \citenamefont {Perktold}, \citenamefont {Cimrman}, \citenamefont {Henriksen}, \citenamefont {Quintero}, \citenamefont {Harris}, \citenamefont {Archibald}, \citenamefont {Ribeiro}, \citenamefont {Pedregosa}, \citenamefont {{van Mulbregt}},\ and\ \citenamefont {{SciPy 1.0 Contributors}}}]{2020SciPy-NMeth}%
  \BibitemOpen
  \bibfield  {author} {\bibinfo {author} {\bibfnamefont {P.}~\bibnamefont {Virtanen}}, \bibinfo {author} {\bibfnamefont {R.}~\bibnamefont {Gommers}}, \bibinfo {author} {\bibfnamefont {T.~E.}\ \bibnamefont {Oliphant}}, \bibinfo {author} {\bibfnamefont {M.}~\bibnamefont {Haberland}}, \bibinfo {author} {\bibfnamefont {T.}~\bibnamefont {Reddy}}, \bibinfo {author} {\bibfnamefont {D.}~\bibnamefont {Cournapeau}}, \bibinfo {author} {\bibfnamefont {E.}~\bibnamefont {Burovski}}, \bibinfo {author} {\bibfnamefont {P.}~\bibnamefont {Peterson}}, \bibinfo {author} {\bibfnamefont {W.}~\bibnamefont {Weckesser}}, \bibinfo {author} {\bibfnamefont {J.}~\bibnamefont {Bright}}, \bibinfo {author} {\bibfnamefont {S.~J.}\ \bibnamefont {{van der Walt}}}, \bibinfo {author} {\bibfnamefont {M.}~\bibnamefont {Brett}}, \bibinfo {author} {\bibfnamefont {J.}~\bibnamefont {Wilson}}, \bibinfo {author} {\bibfnamefont {K.~J.}\ \bibnamefont {Millman}}, \bibinfo {author} {\bibfnamefont {N.}~\bibnamefont {Mayorov}}, \bibinfo {author} {\bibfnamefont
  {A.~R.~J.}\ \bibnamefont {Nelson}}, \bibinfo {author} {\bibfnamefont {E.}~\bibnamefont {Jones}}, \bibinfo {author} {\bibfnamefont {R.}~\bibnamefont {Kern}}, \bibinfo {author} {\bibfnamefont {E.}~\bibnamefont {Larson}}, \bibinfo {author} {\bibfnamefont {C.~J.}\ \bibnamefont {Carey}}, \bibinfo {author} {\bibfnamefont {{\.I}.}~\bibnamefont {Polat}}, \bibinfo {author} {\bibfnamefont {Y.}~\bibnamefont {Feng}}, \bibinfo {author} {\bibfnamefont {E.~W.}\ \bibnamefont {Moore}}, \bibinfo {author} {\bibfnamefont {J.}~\bibnamefont {{VanderPlas}}}, \bibinfo {author} {\bibfnamefont {D.}~\bibnamefont {Laxalde}}, \bibinfo {author} {\bibfnamefont {J.}~\bibnamefont {Perktold}}, \bibinfo {author} {\bibfnamefont {R.}~\bibnamefont {Cimrman}}, \bibinfo {author} {\bibfnamefont {I.}~\bibnamefont {Henriksen}}, \bibinfo {author} {\bibfnamefont {E.~A.}\ \bibnamefont {Quintero}}, \bibinfo {author} {\bibfnamefont {C.~R.}\ \bibnamefont {Harris}}, \bibinfo {author} {\bibfnamefont {A.~M.}\ \bibnamefont {Archibald}}, \bibinfo {author}
  {\bibfnamefont {A.~H.}\ \bibnamefont {Ribeiro}}, \bibinfo {author} {\bibfnamefont {F.}~\bibnamefont {Pedregosa}}, \bibinfo {author} {\bibfnamefont {P.}~\bibnamefont {{van Mulbregt}}},\ and\ \bibinfo {author} {\bibnamefont {{SciPy 1.0 Contributors}}},\ }\bibfield  {title} {\bibinfo {title} {{{SciPy} 1.0: Fundamental Algorithms for Scientific Computing in Python}},\ }\href {https://doi.org/10.1038/s41592-019-0686-2} {\bibfield  {journal} {\bibinfo  {journal} {Nature Methods}\ }\textbf {\bibinfo {volume} {17}},\ \bibinfo {pages} {261} (\bibinfo {year} {2020})}\BibitemShut {NoStop}%
\bibitem [{\citenamefont {Hastie}\ \emph {et~al.}(2009)\citenamefont {Hastie}, \citenamefont {Tibshirani},\ and\ \citenamefont {Friedman}}]{hastie2009elements}%
  \BibitemOpen
  \bibfield  {author} {\bibinfo {author} {\bibfnamefont {T.}~\bibnamefont {Hastie}}, \bibinfo {author} {\bibfnamefont {R.}~\bibnamefont {Tibshirani}},\ and\ \bibinfo {author} {\bibfnamefont {J.~H.}\ \bibnamefont {Friedman}},\ }\href@noop {} {\emph {\bibinfo {title} {The elements of statistical learning: data mining, inference, and prediction}}},\ Vol.~\bibinfo {volume} {2}\ (\bibinfo  {publisher} {Springer},\ \bibinfo {address} {New York},\ \bibinfo {year} {2009})\BibitemShut {NoStop}%
\end{thebibliography}%

\end{document}